\documentclass[superscriptaddress,onecolumn,pre,nofootinbib]{revtex4}
\usepackage{amsmath}
\usepackage{graphicx,color}
\usepackage{amssymb}
\usepackage{amsfonts}
\usepackage{bm}

\newcommand{\be}{\begin{equation}}
\newcommand{\ee}{\end{equation}}
\newcommand{\bea}{\begin{eqnarray}}
\newcommand{\eea}{\end{eqnarray}}

\begin{document}
\sloppy

%-title page-%

\title{A simple model of magnetic universe
without singularity associated\\
 with a quadratic equation of state}

\author{Pierre-Henri Chavanis}
\affiliation{Laboratoire de Physique Th\'eorique, Universit\'e de Toulouse,
CNRS, UPS, France}

\begin{abstract}

A model of magnetic universe based on nonlinear electrodynamics has
been introduced by Kruglov [Phys. Rev. D {\bf 92}, 123523 (2015)]. This model
describes an early inflation era followed by a radiation era. We show that this
model is related to the model of universe based on a quadratic equation of state
introduced in our previous paper [P.H. Chavanis, Universe {\bf 1}, 357 (2015)].
This correspondance may provide a more fundamental justification of our equation
of state. It may arise from quantum corrections to linear electrodynamics
when the electromagnetic field becomes very high. We discuss two
quantitatively different models of early universe. In Model
I, the
primordial density of the universe is identified with the Planck density. At
$t=0$, the universe had the characteristics of a Planck
black hole (``planckion'' particle). 
During the inflation, which takes place on a Planck timescale, the size of the
universe evolves from the Planck length to a size comparable to the Compton
wavelength of the neutrino. 
If we interpret the radius of the universe at the end of the inflation
(neutrino's Compton wavelength) as a minimum length related to quantum gravity
and use
Zeldovich's first formula of the vacuum energy, we obtain the correct value of
the cosmological constant. In Model II, the primordial density of the
universe is identified with the electron density as a consequence  of
nonlinear electrodynamics. At $t=0$, the universe had the
characteristics  of an
electron. This can be viewed as a refinement of the ``primeval atom'' of
Lema\^itre. During the inflation, which takes place on a gravitoelectronic
timescale, the size of the universe evolves
from the electron's classical radius
to a size comparable to the size of a dark energy star of the stellar mass. If
we
interpret the radius of the universe at the begining of the inflation
(electron's classical radius) as a minimum length related to quantum gravity and
use
Zeldovich's second formula of the vacuum energy, we obtain the correct value of
the cosmological constant. This provides an accurate form of Eddington's
relation between the cosmological constant and the mass of the electron. We use
these arguments to show that the present universe contains
about $10^{80}$ protons (Eddington's number). We also
introduce a
nonlinear electromagnetic Lagrangian that describes
simultaneously the early inflation, the radiation
era, and the dark energy era. It may account for a form of ``generalized
radiation'' that is present from the begining to the end of the cosmic
evolution. Baryonic and dark matter are treated as
independent noninteracting species. The dark energy era is due to
the electromagnetic vacuum energy (zero-point radiation). In this
approach, both the early inflation and the
late acceleration of the universe (dark energy) arise as a consequence of
nonlinear
electrodynamics. This leads to a simple model of magnetic universe without
singularity (aioniotic universe).

\end{abstract}

\maketitle

\section{Introduction}
\label{sec_intro}

In a series of papers
\cite{chavanisAIP,jgrav,cosmopoly1,cosmopoly2,cosmopoly3,universe,stiff,mlbec,
chavanisIOP,vacuumon}, we have developed a simple model of
universe based on a generalized equation of state of the
form
\begin{equation}
P/c^2=\alpha\rho+k\rho^{1+1/n},
\label{intro1}
\end{equation}
where $P$ is the pressure and $\rho c^2$ is the energy density. This is the sum
of a linear equation of state and a polytropic equation of state. 
When $n>0$ this equation of state describes the early universe. When $k<0$ the
energy density interpolates between a phase of primordial inflation (de Sitter)
where the scale factor increases exponentially
rapidly with time and an
$\alpha$-era where the scale factor increases algebraically. For
$\alpha=1/3$, this period corresponds to the radiation era.
When $n<0$ this equation of state describes the late universe. When $k<0$ the
energy density interpolates between an $\alpha$-era
where the scale factor increases algebraically with time and a
phase of late acceleration (de Sitter) where the scale factor increases
exponentially rapidly.\footnote{The case $k>0$ has been treated in
\cite{cosmopoly1,cosmopoly2} and leads to singular or peculiar models of
universe that we do not consider here. The case of a phantom evolution where
the density of the universe increases as the universe expands has been treated
in \cite{cosmopoly3}. We do not consider this case here neither.}

We have also introduced the quadratic equation of
state \cite{chavanisAIP,jgrav,cosmopoly1,cosmopoly2,cosmopoly3,
universe,stiff,mlbec,chavanisIOP,vacuumon}
\begin{equation}
\label{intro2}
P=-(\alpha+1)\frac{\rho^2}{\rho_P} c^2+\alpha\rho c^2-(\alpha+1)\rho_{\Lambda}
c^2,
\end{equation}
which provides a simple model of nonsingular
universe presenting an early inflation, an $\alpha$-era, and a late
acceleration (dark energy era). The density starts at early times with the
Planck density
$\rho_P=c^5/G^2\hbar=5.16\times 10^{99}\, {\rm g\, m^{-3}}$,
decreases during the $\alpha$-era, and
tends to the cosmological density $\rho_{\Lambda}=\Lambda c^2/8\pi G=5.96\times
10^{-24}\, {\rm
g\, m^{-3}}$ 
at late times (see
Fig. 16 in \cite{universe}).\footnote{The empirical value of the
cosmological constant is $\Lambda=1.11\times 10^{-52}\,
{\rm m^{-2}}$.}
Accordingly, the universe experiences a first period of exponential acceleration
(early inflation), decelerates (when $\alpha>-1/3$), and finally experiences a
second period
of exponential acceleration (late inflation), the one that we observe at
present. The two de Sitter eras are bridged by an $\alpha$-era. Therefore, this
equation of state unifies the inflation in the early universe and the dark
energy in the late universe. We argued that the
equation of state (\ref{intro2}) with $\alpha=1/3$ describes a form of
``generalized
radiation''. In order to obtain the complete evolution of the universe we have
to introduce, in addition to this generalized
radiation,  baryonic and dark matter, and possibly other
components, treated as independent noninteracting species. The
resulting cosmological 
model turns out to be equivalent to the $\Lambda$CDM model except that it
replaces the big bang singularity by a non singular (de Sitter) inflation era
with a ``graceful'' exit. A
summary of our main results is given in the introduction
of \cite{vacuumon}.

A weakness of
our model, however, is that the equation of state (\ref{intro2}) is introduced
in a rather {\it ad hoc} manner, without a ``microscopic'' derivation. Recently,
we
came accross the very interesting papers of Kruglov
\cite{kruglov2015a,kruglov2015,kruglov2020}
who
introduced a similar
model (valid in the early universe) from the viewpoint of ``magnetic cosmology''
based on nonlinear electrodynamics. Following previous authors
\cite{novello1,novello2,cgcl,novello3,vollick,novello4,garcia} he
argued that
electromagnetic fields play an important
role in cosmology and that the evolution of the early universe is fueled by a
stochastic magnetic field due to plasma fluctuations. In this sense,
electromagnetic fields are the source of
gravitational fields. When electromagnetic fields are very strong during the
early evolution of
the universe one must use nonlinear electrodynamics. This may take into account
quantum gravity corrections to linear electrodynamics.\footnote{In addition,
quantum mechanics could
be a ``classical'' process arising from the stochastic fluctuations of
the electromagnetic field (zero-point radiation)
as suggested by the theory of stochastic electrodynamics \cite{pena}.} The use
of
nonlinear
electrodynamics can remove the big
bang singularity. Instead of prescribing an equation of state as
we did,
Kruglov \cite{kruglov2015a,kruglov2015,kruglov2020} (see also
\cite{ovgun,benaoum1,benaoum2}) introduced a nonlinear
electromagnetic
Lagrangian -- a sort
of generalized Born-Infeld \cite{born1933,borninfeld} 
Lagrangian --  that produces
a
phase
of early inflation followed by a radiation era.\footnote{This  Lagrangian is
also introduced in an {\it ad hoc} manner. However, since it is connected to a
physical mechanism -- nonlinear electrodynamics -- this gives the hope to
derive this model (or more general ones) from ``microscopic'' considerations.
This is, however, beyond the scope of the present paper.}
In the present paper, we show that his nonlinear
electromagnetic Lagrangian leads precisely to the equation of state 
(\ref{intro1}) introduced
in
\cite{chavanisAIP,jgrav,cosmopoly1,cosmopoly2,cosmopoly3,universe,stiff,mlbec,
chavanisIOP,vacuumon}. We also consider the inverse problem and explain how one
can
construct more general electromagnetic  Lagrangians by prescribing an equation
of state such as Eq. (\ref{intro1}). We stress, however,
that some electromagnetic constraints
reduce the possible type of equations of state and select the indices $n=\pm 1$
among the whole polytropic family, leading in a natural manner to Eq.
(\ref{intro2}).

We discuss two quantitatively different models regarding the
early inflation. 

In Model I, the
 density of the primordial universe is identified with the Planck
density $\rho_P=5.16\times 10^{99}\,
{\rm g\, m^{-3}}$ (or a fraction of it).
During the inflation, which takes place on a Planck
timescale $t_{P}=5.39\times 10^{-44}\,
{\rm s}$, the size
of the
universe increases by $30$ orders of
magnitude. It evolves from the Planck length $l_{P}=1.62\times
10^{-35}\, {\rm m}$ to a size equal to the Compton
wavelength of the neutrino $r_{\nu}=3.91\times
10^{-5}\, {\rm m}$ with mass $m_{\nu}=5.04\times 10^{-3}\, {\rm
eV/c^2}$. This creates $10^{90}$ particles of
the Planck mass $M_P=(\hbar c/G)^{1/2}=2.18\times 10^{-5}\, {\rm g}=1.22\times
10^{19}\, {\rm GeV/c^2}$ during the inflation. This  implies  that, after the
radiation era, the mass of the universe is equal to $10^{62}M_P$. If we
interpret the
radius of the
universe at the end
of the inflation
(neutrino's Compton wavelength) as a minimum length related to quantum
gravity in the sense of Amelino-Camelia \cite{gacnature} and use
Zeldovich's first formula of the vacuum energy \cite{zeldovich,zeldovichA}, we
obtain the correct
value of
the cosmological constant \cite{ouf}:
\begin{equation}
\Lambda=\frac{G\hbar}{r_{\nu}^4c^3}=\frac{Gcm_\nu^4}{\hbar^3}
=1.11\times 10^{-52}\, {\rm m^{-2}}.
\label{an16bintro}
\end{equation}

In Model II, the density of the primordial 
universe is identified with the electron density  $\rho_e=4.07\times
10^{16}\, {\rm g\, m^{-3}}$  as a consequence  of
nonlinear electrodynamics. During the inflation, which takes place on a
gravitoelectronic timescale $t_e^*=0.0192\, {\rm s}$, the size of the
universe increase by $30$ orders of
magnitude. It evolves from the electron's classical radius $r_e=2.82\times
10^{-15}\, {\rm m}$ 
to a size of the order of the radius ${\tilde
R}_2=7.07\times 10^{15}\,
{\rm m}$ of a dark energy star of the stellar mass.
This creates $10^{90}$  particles of the electron mass $m_e=9.11\times
10^{-28}\, {\rm g}=0.511\, {\rm MeV/c^2}$ during the inflation. This implies
that, after the radiation era, the universe is made of $10^{83}$ electrons or
$10^{80}$ protons
(Eddington's number). If we
interpret the radius of the universe at the begining of the inflation
(electron's classical radius) as a minimum length related to quantum gravity
in the sense of Karolyhazy \cite{karolyhazy} and
use
Zeldovich's second formula of the vacuum energy \cite{zeldovich,zeldovichA}, we
obtain the correct
value of
the cosmological constant \cite{ouf}: 
\begin{equation}
\Lambda=\frac{G^2m_e^6}{\alpha^6\hbar^4}=\frac{G^2\hbar^2}{r_e^6c^6}=1.36\times
10^{-52}\, {\rm m^{-2}},
\label{edd3intro}
\end{equation}
where $\alpha=e^2/(\hbar c)\simeq 1/137$ is the
fine-structure constant. This provides a justification of Eddington's accurate
relation \cite{eddington1931lambda,ouf}. We note that there is no free parameter
in this
relation.

At the primordial time $t=0$ the universe had the characteristics of a Planck
black hole (``planckion'' particle) in Model I and the characteristics of an
electron in Model II. However, it is unstable and explodes, leading to a phase
of cosmological expansion. This can be viewed as a refinement of the ``primeval
atom'' of Lema\^itre \cite{lemaitreNewton}.

Finally, we introduce a
nonlinear electromagnetic Lagrangian associated with the quadratic equation
of state (\ref{intro2}) that describes simultaneously the early inflation,
the radiation
era, and the dark energy era. The dark energy era is due to the
electromagnetic vacuum energy (zero-point radiation). We
suggest that nonlinear electrodynamics must also be used when the
magnetic field is very low. Therefore, in our
approach, both the early inflation and the
late acceleration of the universe (dark energy) arise as a consequence of
nonlinear  electrodynamics.  It may account for a form of
``generalized
radiation'' that is present from the begining to the end of the cosmic
evolution.  This leads to a simple model of magnetic
universe without
singularity (aioniotic universe).

The paper is organized as follows. In Sec. \ref{sec_ne}, we recall the basic
equations of nonlinear electrodynamics and gravitation (general
relativity). In Sec. \ref{sec_baco}, we recall the basic equations of cosmology
and consider the case where the evolution of the universe is due to a
stochastic electromagnetic field. In Sec. \ref{sec_electro}, we recall the basic
equations of nonlinear electrostatics. In Sec.
\ref{sec_magnu}, we consider the case of a purely magnetic universe. In
Sec. \ref{sec_gen}, we consider the nonlinear electrodynamics associated with
a generalized polytropic equation of state and explain why the polytropic
indices $n=\pm 1$ and the equation of state parameter $\alpha=1/3$ are naturally
selected by electrodynamics. In Secs. \ref{sec_kc} and \ref{sec_kcl}, we discuss
the evolution of the early and late universe in the context of nonlinear
electrodynamics. In 
Sec. \ref{sec_total}, we introduce a nonlinear Lagrangian associated with a
generalized form of radiation that accounts both for the early inflation and for
the late acceleration of the universe (dark energy). In Sec. \ref{sec_complete},
we include baryonic and dark matter as additional species and describe the
complete evolution of the universe. In Sec. \ref{sec_nee}, we study the electric
field produced by a pointlike charge (electron) in the context of nonlinear
electrodynamics. In Sec. \ref{sec_heur}, we point out interesting heuristic
connections between nonlinear electrodynamics, quantum gravity, vacuum energy
and dark energy.

\section{Nonlinear electrodynamics and gravitation}
\label{sec_ne}

In this section, we recall the basic equations of (nonlinear) electrodynamics
and general relativity \cite{llchamps,weinbergbook}.

\subsection{The total action}
\label{sec_ta}

The geometry of spacetime in general relativity is specified by
the metric tensor $g_{\mu\nu}$ which gives the spacetime interval $ds$ between
two infinitesimally separated events, that is\footnote{The $g_{\mu\nu}$ may
also be viewed as gravitational potentials since the effect of gravity is to
modify the curvature of spacetime.}
\begin{equation}
ds^2=g_{\mu\nu}dx^{\mu}dx^{\nu}.
\label{ta1}
\end{equation}
Here and below the Greek indices $\mu,\nu$ etc run
over the spacetime coordinates (ranging from $0$ to $3$) while the Latin
indices $i,j$ etc 
run only over the space coordinates (ranging from $1$ to $3$). We assume
summation over repeated indices.

We first discuss the Maxwell equations in a general curved
spacetime and then focus on Friedmann-Lema\^itre-Roberston-Walker (FLRW) models
of cosmology. Electrodynamics
in curved spacetime is most conveniently formulated
by giving the action for the electromagnetic fields
and their interaction with charged particles. For the sake of generality, we
also include the contribution of a perfect fluid (or matter
field). The total action
of the system, which
is the sum of the
Einstein-Hilbert action of general relativity $+$ the action of the perfect
fluid  $+$ the action of the
electromagnetic field $+$ the action describing the interaction
between the charges and the electromagnetic field can be
written as 
\begin{equation}
S=\frac{c^4}{16\pi G}\int R \sqrt{-g}\,
d^4x+\int \mathcal{L}_m \sqrt{-g}\,
d^4x+\int \mathcal{L}({\cal F}) \sqrt{-g}\,
d^4x-\int A_{\mu}J^{\mu}\sqrt{-g}\, d^4x,
\label{ta2}
\end{equation}
where $R$ is the Ricci
curvature scalar, $g={\rm
det}(g_{\mu\nu})$ is the determinant of the metric tensor,
$A^{\mu}=(U/c,{\bf A})$ is the
electromagnetic quadripotential, and
$J^{\mu}$ is the quadricurrent density. We have assumed that the 
Lagrangian density $\mathcal{L}({\cal
F})$ of the electromagnetic field is an arbitrary function of the
electromagnetic invariant
\begin{equation}
{\cal
F}=\frac{1}{4\mu_0}F_{\mu\nu}F^{\mu\nu},
\label{ta3}
\end{equation}
where $F_{\mu\nu}=\partial_{\mu}A_{\nu}
-\partial_{\nu}A_{\mu}$  is
the Faraday or electromagnetic field
strength tensor (one can replace the
covariant derivative $D_{\mu}$ by the partial derivative $\partial_{\mu}$ in
this equation). In the above equation,
$\mu_0$ denotes the magnetic
permeability in vacuum. The electric
permittivity in vacuum is denoted by $\epsilon_0$. They are related by
$\mu_0\epsilon_0=1/c^2$. Ordinary (linear)
Maxwell
electrodynamics corresponds to  the Lagrangian
\begin{equation}
{\cal L}_{\rm Maxwell}=-{\cal F}.
\label{ta4}
\end{equation}

\subsection{The Einstein equations}

The Einstein equations of the gravitational field can be derived from the
principle of least action $\delta S=\delta(S_g+S_M)=0$, where $S_g$ and $S_M$
are the
actions of the gravitational field and all sources of mass-energy  (including
the electromagnetic
field), by performing
variations with respect to the metric $g_{\mu\nu}$. This yields
\begin{equation}
R_{\mu\nu}-\frac{1}{2}g_{\mu\nu}R=\frac{8\pi
G}{c^4}T_{\mu\nu},
\label{ein5}
\end{equation}
where $R_{\mu\nu}$ is the Ricci
tensor measuring
the curvature of spacetime, $R=R_{\mu}^{\mu}=g^{\mu\nu}R_{\mu\nu}$ is the
scalar curvature of spacetime, and $T_{\mu\nu}$ is the energy-momentum tensor of
the
matter  (or electromagnetic field) given by
\begin{equation}
T_{\mu\nu}=\frac{2}{\sqrt{-g}}\frac{\partial(\sqrt{-g}{\cal
L})}{\partial
g^{\mu\nu}}=2\frac{\partial {\cal
L}}{\partial
g^{\mu\nu}}-g_{\mu\nu}{\cal
L}.
\label{emt2}
\end{equation}
From this formulation, the energy-momentum tensor (\ref{emt2}) is
automatically symmetric. For a
macroscopic body, the energy-momentum tensor can be written as
\begin{equation}
T_{\mu\nu}=(\rho c^2+P)u_{\mu}u_{\nu}-P g_{\mu\nu},
\label{ein4}
\end{equation}
where $\rho c^2$ is the energy density, $P$ is the
pressure and $u^{\mu}$ is the quadrivelocity such that $u_{\mu}u^{\mu}=1$ (for
a fluid at rest $T_{00}=\rho c^2$ and
$T_{xx}=T_{yy}=T_{zz}=P$).

The conservation of energy
and momentum is
expressed by 
\begin{equation}
D_{\nu}T^{\mu\nu}=0,
\label{ein10}
\end{equation}
where $D_{\nu}=\frac{1}{\sqrt{-g}}\partial_{\nu}\sqrt{-g}$ is the covariant
derivative in a curved spacetime. According to this equation, the
divergence of the left hand side of Eq.
(\ref{ein5}) must be
zero. This is actually the case because of the contracted Bianchi identities.
As a result,  the conservation of energy
and momentum is contained in the Einstein equations (\ref{ein5}). The
equations of the gravitational field contain the equations for the matter
which
produces this field. However,  for a complete determination of the distribution
and motion of the matter (or electromagnetic field) one must still supplement to
the Einstein equations the equation of state of the matter (or electromagnetic
field), i.e., an equation relating the pressure $P$ to the energy density $\rho
c^2$.
This equation must be given along with the gravitational field equations.

\subsection{The nonlinear Maxwell equations}
\label{sec_nme}

The fundamental equations of electromagnetism are the Maxwell equations. In a
curved spacetime the first pair of Maxwell equations reads (one can replace the
covariant derivative $D_{\mu}$ by the partial derivative $\partial_{\mu}$ in
this equation)
\begin{equation}
\partial_{\lambda}F_{\mu\nu}+\partial_{\nu}F_{\lambda\mu}+\partial_{\mu}F_{
\nu\lambda } =0.
\label{nme4}
\end{equation}
The field equations
determining the second pair of Maxwell equations can be
obtained
from the principe of least action $\delta S=0$ by varying the electromagnetic
potential 
$A_{\mu}$. 
This yields
\begin{equation}
D_{\nu}\left \lbrack {\cal L}'({\cal F})F^{\mu\nu}\right
\rbrack=\mu_0 J^{\mu}.
\label{nme4b}
\end{equation}
Equations (\ref{nme4}) and (\ref{nme4b}) form the (nonlinear) Maxwell equations.
We
see that only the second pair of Maxwell equations is affected by a possible
nonlinearity of the Lagrangian.

The conservation
of charge is expressed by the equation of
continuity
\begin{equation}
D_{\mu}J^{\mu}=0.
\label{nme6}
\end{equation}
According to this equation, the
divergence of the
left hand side of Eq. (\ref{nme4b}) must be
zero. This is actually the case because of the antisymmetry of the
Faraday tensor.  The conservation of
charge is therefore included in the Maxwell equations.

\subsection{Electromagnetic energy-momentum tensor}
\label{sec_emt}

The electromagnetic energy-momentum tensor associated with the
Lagrangian ${\cal L}({\cal F})$ is 
\begin{equation}
T_{\mu\nu}=-\frac{1}{\mu_0}{\cal L}'({\cal
F})F_{\mu\alpha}F^{\alpha}_{\,\,\,\nu}-{\cal
L}({\cal F})g_{\mu\nu}.
\label{tmunu}
\end{equation}
We note that
\begin{equation}
T_{\mu}^{\mu}=4{\cal F}{\cal L}'({\cal F})-4{\cal L}({\cal F}).
\label{emt4}
\end{equation}
In linear electrodynamics (${\cal L}_{\rm Maxwell}=-{\cal F}$) the
energy-momentum tensor of
the electromagnetic
field has the property that $T=T_{\mu}^{\mu}=0$. With the identity $R=-(8\pi
G/c^4)T$ obtained from the Einstein equations (\ref{ein5}) it follows that in
the presence of an electromagnetic
field without any masses the scalar curvature of spacetime is zero:
$R=0$. On the other hand, the identity $T=\rho c^2-3P$ obtained
from Eq. (\ref{ein4}) implies that $P=\rho
c^2/3$. These relations are no more true for a nonlinear electrodynamics.

\section{Basic equations of cosmology}
\label{sec_baco}

In this section, we recapitulate the basic equations of cosmology
\cite{weinbergbook} and consider the case where the universe is filled with a
stochastic electromagnetic field.

\subsection{Friedmann equations}
\label{sec_basic}

If we consider an expanding homogeneous and isotropic cosmological spacetime
(background) with a uniform curvature, the line element is given by the
Friedmann-Lema\^itre-Roberston-Walker (FLRW) metric 
\begin{eqnarray}
\label{e1}
ds^2=c^2 dt^2-a(t)^2\left\lbrack \frac{dr^2}{1-k r^2}+r^2\,  (d\theta^2+
\sin^2\theta\,  d\phi^2)\right \rbrack,
\end{eqnarray}
where $a(t)$ represents the radius of curvature of the $3$-dimensional
space, or the scale factor. By an abuse of language, we shall sometimes call it
the ``radius of the universe''. On the other hand, $k$ determines the curvature
of space. The universe is closed if $k>0$, flat if $k=0$, or
open if $k<0$.

If the universe is isotropic and homogeneous at all points in conformity with
the line element (\ref{e1}), and contains a uniform perfect fluid of energy
density $\rho(t) c^2$ and isotropic pressure $P(t)$, the
energy-momentum tensor $T_{\mu\nu}$ is given by Eq. (\ref{ein4}) and we have
\begin{eqnarray}
\label{e2}
T^0_0=\rho c^2,\qquad T^1_1=T^2_2=T^3_3=-P.
\end{eqnarray}
The Einstein equations
\begin{eqnarray}
\label{e3}
R^{\mu}_{\nu}-\frac{1}{2}g^{\mu}_{\nu} R-\Lambda g^{\mu}_{\nu}=-\frac{8\pi
G}{c^2}T^{\mu}_{\nu}
\end{eqnarray}
relate the geometrical structure of spacetime ($g_{\mu\nu}$) to the material
content of the universe ($T_{\mu\nu}$) including the electromagnetic field. For
the sake of generality, we have included
the cosmological constant $\Lambda$ in the Einstein equations (\ref{e3}). Given
Eqs.
(\ref{e1}) and (\ref{e2}), the Einstein
equations reduce to 
\begin{eqnarray}
\label{e4}
8\pi G\rho+\Lambda=3 \frac{{\dot a}^2+kc^2}{a^2},\qquad \frac{8\pi
G}{c^2}P-\Lambda=-\frac{2 a \ddot a+{\dot a}^2+kc^2}{a^2},
\end{eqnarray}
where dots denote differentiation with respect to time. These are the well-known
Friedmann \cite{friedmann1,friedmann2} cosmological equations. The Friedmann
equations are usually written under the form
\begin{equation}
\label{e5}
H^2=\frac{8\pi
G}{3}\rho-\frac{kc^2}{a^2}+\frac{\Lambda}{3},
\end{equation}
\begin{equation}
\label{e6}
2\dot H+3H^2=-\frac{8\pi
G}{c^2} P-\frac{kc^2}{a^2}+\Lambda,
\end{equation}
where $H=\dot a/a$ is  the Hubble
parameter. From these equations, one can derive the acceleration equation
\begin{equation}
\label{e7}
\frac{\ddot a}{a}=-\frac{4\pi G}{3}\left (\rho+\frac{3P}{c^2}\right
)+\frac{\Lambda}{3}.
\end{equation}
The deceleration parameter is
defined by
\begin{equation}
q=-\frac{{\ddot a}a}{{\dot a}^2}.
\label{e8}
\end{equation}
The universe is decelerating when $q>0$ and accelerating when $q<0$.

\subsection{Energy conservation equation}
\label{sec_angr}

Combining Eqs. (\ref{e5}) and (\ref{e6}), we obtain the energy conservation
equation
\begin{equation}
\label{e9}
\frac{d\rho}{dt}+3H\left
(\rho+\frac{P}{c^2}\right )=0.
\end{equation}
This equation can be directly derived from the relation (\ref{ein10})
which is
included in the Einstein equations through the contracted Bianchi
identities. It can be written under the
form
\begin{equation}
d\rho+3\frac{da}{a}(\rho+P/c^2)=0.
\label{f8}
\end{equation}
For a given barotropic equation of state $P=P(\rho)$, we can
solve
Eq. (\ref{f8}) to obtain
\begin{equation}
\label{e10}
\ln a=-\frac{1}{3}\int \frac{d\rho}{\rho+P(\rho)/c^2}.
\end{equation}
This equation determines
the relation $\rho(a)$ between the energy density and the scale factor. We can
then solve the Friedmann equation
(\ref{e5}) with $\rho=\rho(a)$ to obtain the temporal evolution 
of the scale factor $a(t)$.

The energy conservation
equation (\ref{f8}) can be rewritten as 
\begin{equation}
\label{e11}
{d}(\rho c^2 a^3)=-P{d(a^3)}.
\end{equation}
Introducing the volume $V\propto a^3$ and the energy
$E=\rho c^2 V$, Eq. (\ref{e11}) becomes $dE=-PdV$. It can be
interpreted as the first principle of thermodynamics for an
adiabatic evolution of the universe
$dS=0$ \cite{lemaitre1927,lemaitre1931,cosmopoly1}.

The equation of state parameter is defined by
\begin{equation}
\label{e12}
w=\frac{P}{\rho c^2}.
\end{equation}
According to Eq. (\ref{e9}) the energy density decreases with the scale factor
when $w> -1$ (null
dominant energy condition) and increases with the scale factor when $w< -1$.
When
$w=-1$ the energy density is constant. The case where the energy density
increases with the scale factor
corresponds to a phantom universe
\cite{caldwell,cosmopoly3}.

{\it Remark:} It is possible to develop a useful mechanical
analogy to study the Friedmann equation (\ref{e5}). Indeed, it can be cast in
the suggestive form
\begin{equation}
E=\frac{1}{2}{\dot a}^2+V(a),
\label{ma1}
\end{equation}
where
\begin{equation}
E=-\frac{1}{2}kc^2,\qquad V(a)=-\frac{4\pi G}{3}\rho(a)a^2-\frac{\Lambda}{6}a^2.
\label{ma2}
\end{equation}
Eq. (\ref{ma1}) has the structure of the first integral of motion for a
particle of unit mass in a potential $V(a)$. In that case, $E$ represents its
conserved energy. The Friedmann equation (\ref{e5}) then has the
solution
\begin{equation}
t=\int\frac{da}{\sqrt{2[E-V(a)]}}
\label{ma3}
\end{equation}
determining $a(t)$ in reversed form.

\subsection{Flat universe}
\label{sec_flatt}

In this paper, we consider a flat universe ($k=0$) in agreement
with the inflation paradigm \cite{guthinflation} and the observations of the
cosmic
microwave
background (CMB) \cite{planck2014,planck2016}. On the other
hand, we set the cosmological constant equal to zero ($\Lambda=0$) because
dark energy
will be taken into account in the nonlinear electrodynamics. The Friedmann
equations (\ref{e5}) and (\ref{e6}) then reduce to the form
\begin{equation}
\label{e13}
H^2=\left (\frac{\dot a}{a}\right )^2=\frac{8\pi
G}{3}\rho,
\end{equation}
\begin{equation}
\label{e14}
2\dot H+3H^2=-\frac{8\pi
G}{c^2} P.
\end{equation}
The acceleration equation (\ref{e7}) becomes
\begin{equation}
\label{e15}
\frac{\ddot a}{a}=-\frac{4\pi G}{3} \left (\rho+\frac{3P}{c^2}\right ).
\end{equation}
Using Eqs. (\ref{e13}) and (\ref{e15}) we see that the deceleration parameter
(\ref{e8}) is
related to the equation of state
parameter (\ref{e12}) by
\begin{equation}
q=\frac{1+3w}{2}.
\label{e16}
\end{equation}
The universe is decelerating when $w>-1/3$ (strong
energy condition) and accelerating when $w<-1/3$; when
$w=-1/3$ the scale factor increases linearly with time \cite{cosmopoly1}.

\subsection{Tolman-Ehrenfest averaging procedure}

Let us study some general properties of nonlinear
electrodynamics in cosmology
\cite{novello1,novello2,cgcl,novello3,vollick,novello4,garcia,kruglov2015a,
kruglov2015,kruglov2020}. We assume that the universe is filled
with
electromagnetic radiation. The
electromagnetic
field that is of cosmological interest is the
cosmic microwave background (CMB). It can be considered as a random field  of
short coherent radiation wavelength  as compared
to the cosmological horizon scales. Due to the isotropy of the
spatial sections of the
FLRW model, an average procedure
is needed for compatibility reason if the electromagnetic
field is to act as a
source of gravity. Using the usual Tolman-Ehrenfest \cite{te}
procedure, we assume that the averaged electromagnetic field obeys
the equations 
\begin{equation}
\overline{E}_i=0,\quad \overline{B}_i=0,\quad \overline{E_iB_j}=0,
\label{te1}
\end{equation}
\begin{equation}
\overline{E_iE_j}=\frac{1}{3}E^2\delta_{ij},\quad
\overline{B_iB_j}=\frac{1}{3}B^2\delta_{ij},
\label{te2}
\end{equation}
where $\overline{X}$ denotes an average over a volume that is
large
compared to the radiation wavelength but small
compared to the curvature of spacetime.  In the
following, we omit the averaging bars for notational
simplicity.

With these conditions, the average value of the energy-momentum
tensor of
the electromagnetic field associated with the Lagrangian density ${\cal
L}({\cal F})$ [see Eq. (\ref{tmunu})] can be written in the
form
of the energy-momentum tensor  for a perfect fluid [see Eq. (\ref{ein4})].  The
energy
density and the pressure of the radiation are
given by \cite{novello2,novello3}
\begin{equation}
\rho c^2=-{\cal L}({\cal
F})-E^2 {\cal L}'({\cal
F}),
\label{re1}
\end{equation}
\begin{equation}
P={\cal L}({\cal
F})+\frac{E^2-2B^2}{3} {\cal L}'({\cal
F}),
\label{re2}
\end{equation}
where $E^2$ and $B^2$ are
the averaged electric and magnetic fields
squared, respectively. For convenience we have rescaled ${\bf E}$ by
$1/(\epsilon_0)^{1/2}$ and ${\bf B}$ by
$(\mu_0)^{1/2}$. The electromagnetic invariant takes the form
\begin{equation}
{\cal F}=\frac{1}{4}F_{\mu\nu}F^{\mu\nu}=\frac{B^2-E^2}{2}.
\label{re3}
\end{equation}
For Maxwell's linear electrodynamics described by the Lagrangian
(\ref{ta4}) the foregoing equations reduce to
\begin{equation}
\rho c^2=\frac{1}{2}(E^2+B^2),\qquad P=\frac{1}{6}(E^2+B^2),
\label{re4}
\end{equation}
returning the usual equation of state of radiation
\begin{equation}
P=\frac{1}{3}\rho c^2.
\label{re5}
\end{equation}

\section{Electrostatics}
\label{sec_electro}

In electrostatics, in the absence of magnetic field (${\bf B}={\bf 0}$) and if
we can neglect the expansion of the universe ($a=1$), the nonlinear Maxwell
equations can be written as 
\begin{equation}
\nabla\times
{\bf E}=0
\label{el1}
\end{equation}
and
\begin{equation}
\nabla\cdot \left\lbrack {\cal L}'({\cal F}){\bf
E}\right\rbrack=-\rho_e,
\label{el2}
\end{equation}
where $\rho_e$ is the charge density and 
\begin{equation}
{\cal F}=-\frac{1}{2} E^2.
\label{el3}
\end{equation}

The electric field ${\bf E}$ is expressed in terms of the scalar potential
$U$ by the relation
\begin{equation}
{\bf E}=-\nabla U.
\label{el4}
\end{equation}
Substituting Eq. (\ref{el4}) into Eq. (\ref{el2}), we get 
\begin{equation}
\nabla\cdot \left\lbrack {\cal L}'({\cal
F})\nabla U\right\rbrack=\rho_e.
\label{el5}
\end{equation}
In particular, in
vacuum ($\rho_e=0$), the electric potential satisfies the equation
\begin{equation}
\nabla\cdot \left\lbrack {\cal L}'({\cal
F})\nabla U\right\rbrack=0.
\label{el6}
\end{equation}
For Maxwell's linear electrodynamics, Eqs. (\ref{el5}) and (\ref{el6}) reduce to
the usual Poisson and Laplace equations $\Delta U=-\rho_e$
and $\Delta
U=0$, respectively.

Using the Gauss law (\ref{el2}), the electric
field produced by a point charge [$\rho_e=\sqrt{4\pi}
e\delta({\bf r})$] satisfies the
equation
\begin{equation}
-\nabla\cdot \left\lbrack {\cal L}'({\cal F}){\bf
E}\right\rbrack=\sqrt{4\pi} e\delta({\bf r}).
\label{el7}
\end{equation}
Its solution is 
\begin{equation}
-{\cal L}'({\cal F})E(r)=\frac{e}{\sqrt{4\pi}r^2}.
\label{el8}
\end{equation}
For Maxwell's linear electrodynamics, we recover the Coulomb
law $E(r)=e/(\sqrt{4\pi}r^2)$.

Finally, using Eq. (\ref{el3}), the electric energy density and
the
pressure [see Eqs. (\ref{re1}) and (\ref{re2})] can be written as
\begin{equation}
\rho c^2=-{\cal L}({\cal
F})+2{\cal F} {\cal L}'({\cal
F}),
\label{el9}
\end{equation}
\begin{equation}
P={\cal L}({\cal
F})-\frac{2}{3} {\cal F} {\cal L}'({\cal
F}).
\label{el10}
\end{equation}
For a given Lagrangian ${\cal L}({\cal F})$, these equations define the
equation of state $P(\rho)$ in parametric form with
parameter ${\cal
F}=-E^2/2<0$. For Maxwell's linear electrodynamics, we get $\rho c^2=E^2/2$ and
$P=E^2/6$ leading to $P=\rho c^2/3$.

\section{Magnetic universe}
\label{sec_magnu}

We assume that the universe is filled with a magnetic fluid. We suppose
that the magnetic field of nonlinear electrodynamics is the
main source of gravity.  We consider the case $E=0$ and $B\neq 0$
because only the magnetic field is important in cosmology. Indeed,
the electric field is
screened by the charged primordial plasma, while the magnetic field lines are
frozen \cite{lemoine}.
This leads to the
concept of ``magnetic universe'' \cite{novello3,novello4}. In this model, the
cosmic dynamics is
fueled by the magnetic fluid alone.

In a purely magnetic universe (${E}={0}$), the energy density and the
pressure are
given
by [see Eqs. (\ref{re1}) and (\ref{re2})]
\begin{equation}
\rho c^2=-{\cal L}({\cal
F}),
\label{f4}
\end{equation}
\begin{equation}
P={\cal L}({\cal
F})-\frac{4}{3}{\cal F} {\cal L}'({\cal
F})
\label{f5}
\end{equation}
with
\begin{equation}
{\cal F}=\frac{B^2}{2}.
\label{f5b}
\end{equation}
For a given Lagrangian ${\cal L}({\cal F})$, these equations define
the
equation of state $P(\rho)$ in parametric form with parameter ${\cal
F}=B^2/2>0$. Conversely, for a given equation of state $P(\rho)$,  we can
obtain the 
Lagrangian ${\cal L}({\cal F})$ as follows.  Combining Eqs. (\ref{f4}) and
(\ref{f5}), we find that
\begin{equation}
P/c^2=-\rho+\frac{4}{3}\frac{d\rho}{d\ln{\cal F}}.
\label{f6}
\end{equation}
For a given equation of state $P=P(\rho)$, Eq. (\ref{f6}) is a just first
order
differential equation. Its solution is
\begin{equation}
\ln{\cal F}=\frac{4}{3}\int
\frac{d\rho}{P(\rho)/c^2+\rho},
\label{f7a}
\end{equation}
which determines ${\cal F}={\cal F}(\rho)$. If this relation can be inverted,
the
electromagnetic Lagrangian is
given by ${\cal L}({\cal F})=-\rho({\cal
F})c^2$.

It is also possible to derive the electromagnetic Lagrangian from the equation
of state (or the converse) in a more direct manner by using a simple
identity and exploiting the results obtained in cosmology. Combining Eq.
(\ref{f6})
with the energy conservation equation (\ref{f8})
we get
\begin{equation}
\frac{d{\cal F}}{\cal F}=-4\frac{da}{a}.
\label{f9a}
\end{equation}
Integrating this relation, we obtain the important identity (see, e.g.
\cite{kim})
\begin{equation}
{\cal F}=\frac{{\cal F}_0}{a^4}\quad {\rm i.e.}\quad B=\frac{B_0}{a^2},
\label{f10}
\end{equation}
where $B_0$ is the present value of the magnetic field and ${\cal F}_0=B_0^2/2$.
Therefore, there exists a simple
general relation [Eq. (\ref{f10})] between the electromagnetic invariant ${\cal
F}$ and the scale
factor $a$. This identity is useful because, in cosmology, we are used to
prescribing an
equation of state $P=P(\rho)$ and, from the energy conservation
equation (\ref{f8}), derive the relation $\rho(a)c^2$ between the energy
density and
the scale factor $a$. Many equations of state have been introduced and
studied in
cosmology. To each of these models, if we know $\rho(a)c^2$ explicitly, then
using Eqs.
(\ref{f4}) and
(\ref{f10}), we can immediately associate an electromagnetic Lagrangian by
simply writing
\begin{equation}
{\cal L}({\cal F})=-\rho[({\cal F}_0/{\cal F})^{1/4}]c^2.
\label{f11a}
\end{equation}
Therefore, we can produce a great number of  electromagnetic 
Lagrangians associated with cosmological models. Some examples will be given
below. Conversely, for a given  electromagnetic Lagrangian  ${\cal L}({\cal
F})$,
we
can immediately write down the  relation $\rho(a)c^2$ between the energy
density and the scale factor $a$ without any calculation. Indeed, using Eqs.
(\ref{f4})
and (\ref{f10}), we have
\begin{equation}
\rho(a)c^2=-{\cal L}({\cal F}_0/a^4).
\label{f12}
\end{equation}
Similarly, using Eqs. (\ref{f5})
and (\ref{f10}), the evolution of the pressure with the scale factor is
immediately given by
\begin{equation}
P(a)={\cal L}({\cal F}_0/a^4)-\frac{4}{3}({\cal F}_0/a^4) {\cal L}'({\cal
F}_0/a^4).
\label{f13}
\end{equation}
Equations (\ref{f12}) and  (\ref{f13}) define the equation of state
$P(\rho)$ in parametric form with parameter $a$.

{\it Remark:} For Maxwell's linear electrodynamics, representing normal
radiation,\footnote{We call it normal radiation in order to distinguish it from
the generalized radiation introduced in Sec. \ref{sec_total}.} we have
\begin{equation}
{\cal L}=-{\cal F},\quad
\rho_{\rm rad} c^2={\cal F},\quad P_{\rm rad}=\frac{1}{3}{\cal F}, \quad P_{\rm
rad}=\frac{1}{3}\rho_{\rm rad} c^2.
\label{vo1}
\end{equation}
Since 
\begin{equation}
\rho_{\rm rad} c^2={\cal F}=\frac{{\cal F}_0}{a^4},
\label{vo2}
\end{equation}
we find that
\begin{equation}
{\cal F}_0=\rho_{\rm rad,0} c^2=\Omega_{\rm rad,0} \rho_0 c^2,
\label{vo3}
\end{equation}
where  $\rho_{\rm
rad,0}c^2$ is the present energy density of normal radiation and $\Omega_{\rm
rad,0}=\rho_{\rm rad,0}/\rho_{0}$ is the present proportion of normal
radiation ($\rho_{0}c^2$ is the present energy density of the universe).
Therefore, ${\cal F}_0$ represents the present 
energy density of normal radiation and $B_0=\sqrt{2{\cal F}_0}$ the present
magnetic field. The dimensional magnetic field is 
\begin{equation}
B_{\rm dim}=\sqrt{\mu_0} B.
\label{bdim}
\end{equation}
The vacuum permeability has the value $\mu_0=1.25663706212\times 10^{-9}\, {\rm
m \, s^2\,  T^2\,  g^{-1}}$. From the observations we
have $\Omega_{\rm rad,0}=9.23765\times 10^{-5}$ and 
$\rho_0=8.62\times 10^{-24}{\rm g\, m^{-3}}$.
This gives $B_0=4.23\times
10^{-10}\, {\rm T}$.  More generally, we
have the relations
\begin{equation}
\rho_{\rm rad} c^2=\frac{B^2}{2}={\cal
F}=\frac{\rho_{\rm rad,0}c^2}{a^4}.
\label{tsw1a}
\end{equation}
These relations have been established during the normal radiation era but,
because of the $a^{-4}$ law from Eq. (\ref{f10}) which coincides with the
$a^{-4}$ law of normal radiation \cite{cosmopoly1}, they are actually valid for
all
times, even
when $\rho\neq \rho_{\rm rad}$. In other words, ${\cal F}=B^2/2$ always
represents the energy density of the normal radiation, even when the normal
radiation is subdominant.

\section{Nonlinear electrodynamics corresponding to a generalized polytropic
equation of
state $P/c^2=\alpha\rho+k\rho^{1+1/n}$ in cosmology}
\label{sec_gen}

\subsection{Generalized polytropic equation of state and the corresponding
Lagrangian}
\label{sec_geneos}

We consider an equation of state of the form
\begin{equation}
P/c^2=\alpha\rho+k\rho^{1+1/n},
\label{gen1}
\end{equation}
where $\rho c^2$ is the energy density. This is the sum of a linear
equation of state and a polytropic equation of state. This equation of state has
been used in cosmology to describe the evolution of the early and late
universe
\cite{cosmopoly1,cosmopoly2,cosmopoly3}. Solving the
energy conservation
equation (\ref{f8}) with the equation of state (\ref{gen1}), we
obtain\footnote{Following \cite{cosmopoly1,cosmopoly2} we
consider the case where the energy density decreases with the scale factor. The
case of a phantom evolution where the energy density increases with the scale
factor is considered in \cite{cosmopoly3}.}
\begin{equation}
\rho=\frac{\rho_*}{\left \lbrack \left
(\frac{a}{a_*}\right )^{[3(1+\alpha)]/n}\mp
1\right \rbrack^n}, 
\label{gen2}
\end{equation}
where $a_*$ is an integration constant and $\rho_*=[(\alpha+1)/|k|]^n$. The
upper sign
corresponds to $k>0$ and the lower sign corresponds to
$k<0$.
The equation of state
(\ref{gen1}) can be rewritten as
\begin{equation}
P/c^2=\alpha\rho\pm (\alpha+1)\rho \left
(\frac{\rho}{\rho_*}\right )^{1/n}.
\label{gen2b}
\end{equation}
Using $\rho c^2=-{\cal
L}$ [see Eq. (\ref{f4})] and ${\cal F}={\cal F}_0/a^4$ [see Eq.
(\ref{f10})] yielding
${\cal F}/{\cal F}_*=(a_*/a)^4$
with ${\cal F}_*={\cal F}_0/a_*^4$, we find that the Lagrangian associated with
the
equation of state (\ref{gen2b}) in cosmology reads
\begin{equation}
{\cal L}=\frac{-\rho_*c^2}{\left \lbrack \left
(\frac{{\cal F}_*}{{\cal F}}\right )^{[3(1+\alpha)]/(4n)}\mp
1\right \rbrack^n}.
\label{gen3}
\end{equation}

\subsection{Conditions of validity}
\label{sec_genval}

It is important to
note that the electromagnetic invariant ${\cal F}$
has not a constant sign depending whether the electric or magnetic field
dominates. If the magnetic field dominates, we have ${\cal F}=B^2/2>0$ while 
${\cal F}=-E^2/2<0$ when the electric field dominates. The Lagrangian
(\ref{gen3})
has been derived in a cosmological context where the magnetic field dominates.
In that case, ${\cal F}$ and ${\cal F}_*$ are positive. Now, the same Lagrangian
should also describe the case where the electric field dominates. But in that
case 
${\cal F}$ is negative while ${\cal F}_*$ has the same value as before, being a
 positive constant. Therefore, ${\cal F}/{\cal F}_*>0$ in the
magnetic regime while ${\cal F}/{\cal F}_*<0$ in the electric regime. If
we want to describe the two situations with the same Lagrangian, and since the
Lagrangian (\ref{gen3}) involves a power-law $\left
({{\cal F}_*}/{{\cal F}}\right )^{[3(1+\alpha)]/(4n)}$, it
is necessary that $[3(1+\alpha)]/4n$ be an integer.\footnote{Of
course, we could put an absolute value $|{\cal F}_*/{\cal
F}|^{[3(1+\alpha)]/(4n)}$ in Eq. (\ref{gen3}) but
we find this procedure a bit artificial and prefer
selecting models where no absolute value arises.}
For simplicity, we shall require that 
\begin{equation}
\frac{3(1+\alpha)}{4n}=\pm 1.
\label{gen22}
\end{equation}
On the other hand, in order  to recover the
Maxwell Lagrangian ${\cal L}_{\rm Maxwell}=-{\cal
F}$  when $({{\cal F}}/{{\cal
F}_*})^{[3(1+\alpha)]/(4n)}\ll 1$ (correspondence principle), we
need 
\begin{equation}
\alpha=\frac{1}{3}\qquad {\rm and} \qquad {\cal F}_*=\rho_* c^2.
\label{gen23}
\end{equation}
The two conditions (\ref{gen22}) and (\ref{gen23}) imply 
\begin{equation}
n=\pm 1.
\label{gen24}
\end{equation}
Interestingly, we recover the two canonical models considered in
\cite{cosmopoly1,cosmopoly2,cosmopoly3}:

(i) The model $\alpha=1/3$ and $n=1$ corresponds
to
\begin{equation}
P/c^2=\frac{1}{3}\rho\pm \frac{4}{3}\frac{\rho^2}{\rho_*},\qquad {\cal
L}=\frac{-{\cal F}}{1\mp
\frac{{\cal F}}{{\cal F}_*}}.
\label{gen25}
\end{equation}
It describes the early universe. In practice we shall consider the lower sign
($k<0$)
corresponding
to a nonsingular (inflationary) early universe \cite{cosmopoly1}.
Interestingly, the Lagrangian from Eq. (\ref{gen25}) coincides with the one 
introduced by Kruglov \cite{kruglov2015a,kruglov2015} from other
arguments.

(ii) The model $\alpha=1/3$ and $n=-1$ corresponds to
\begin{equation}
P/c^2=\frac{1}{3}\rho\pm \frac{4}{3}\rho_*,\qquad {\cal L}=-{\cal F}\pm {\cal
F}_*.
\label{gen10}
\end{equation}
It describes the late universe. In practice we shall consider the lower
sign ($k<0$)
corresponding
to a nonsingular (inflationary) late universe \cite{cosmopoly2}. 

{\it Remark:} More generally, in the foregoing discussion, $1/n$ could be any
positive or negative
integer. However, by considering a polytropic equation of state with an
arbitrary index $n$, we have shown in \cite{vacuumon} that the values $n=\pm
1$ are selected by an extremum principle: they turn out to minimize the mass of
the real SF associated with the generalized polytropic equation of state
(\ref{gen1}). This may give them a special status, in addition to the argument
of simplicity invoked above.

\section{Nonlinear electrodynamics based on the Lagrangian
${\cal L}=-{\cal F}/(1+{\cal F}/{\cal F}_I)$: Cosmology in the
early universe}
\label{sec_kc}

In the early universe, we consider a nonlinear electrodynamics based on the
Lagrangian
\begin{equation}
{\cal L}=\frac{-{\cal F}}{1+
\frac{{\cal F}}{{\cal F}_I}},
\label{kc1}
\end{equation}
where ${\cal F}_I=\rho_I c^2$. We shall discuss later the value of the
characteristic density $\rho_I$ appearing in Eq. (\ref{kc1}).
From Eqs. (\ref{f4}) and (\ref{f5}) we find that the
energy density
and the pressure are given by 
\begin{equation}
\rho c^2=\frac{{\cal
F}}{1+\frac{{\cal F}}{\rho_I
c^2}},
\label{gr9b}
\end{equation}
\begin{equation}
P=\frac{{\cal F}\left (\frac{1}{3}-\frac{\cal F}{\rho_I c^2}\right )}{\left
(1+\frac{{\cal F}}{\rho_I
c^2}\right )^2}.
\label{gr10b}
\end{equation}
Eliminating ${\cal F}$ between these two expressions we obtain  the quadratic
equation of state
\begin{equation}
P=-\frac{4\rho^2c^2}{3\rho_I}+\frac{1}{3}\rho c^2.
\label{gen25b}
\end{equation}
This equation of state has been studied in \cite{cosmopoly1,universe} to model
the early universe. Below, we recall its main properties.

\subsection{Generalized polytropic equation of state}
\label{sec_kcsf1}

For the sake of generality, we consider a generalized polytropic equation of
state of the form
\begin{equation}
\label{kc14}
P=-(\alpha+1)\rho_I c^2 \left (\frac{\rho}{\rho_I}\right
)^{1+{1}/{n}}+\alpha\rho c^2
\end{equation}
with $n>0$ and $-1<\alpha\le 1$ \cite{cosmopoly1,universe}, even if we shall
finally take
$\alpha=1/3$ and $n=1$ for the reasons explained in Sec. \ref{sec_genval}. For
$\rho\ll\rho_I$, we obtain the linear equation of state $P\sim\alpha\rho c^2$.
For
$\rho\rightarrow\rho_I$, we get $P\rightarrow -\rho_I c^2$ corresponding to the
equation of state of vacuum energy.

\subsection{Evolution of the density, pressure and scale factor}
\label{sec_esf}

Solving
the energy
conservation equation (\ref{f8}) with the equation of state (\ref{kc14}) we
find that
the
energy density evolves with the scale factor as
\begin{equation}
\label{esf1}
\rho=\frac{\rho_I}{\left\lbrack
1+(a/a_1)^{\frac{3(1+\alpha)}{n}}\right\rbrack^n},
\end{equation}
where $a_1$ is a constant of integration.
The pressure depends on the scale
factor as
\begin{equation}
\label{esf2}
\frac{P}{\rho_I c^2}=\frac{\alpha
(a/a_1)^{\frac{3(\alpha+1)}{n}}-1}{\left\lbrack
(a/a_1)^{\frac{3(\alpha+1)}{n}}+1\right\rbrack^{n+1}}.
\end{equation}

For $a\ll a_1$, the energy
density is approximately constant
\begin{equation}
\rho\simeq \rho_I,
\label{esf3}
\end{equation}
and the pressure tends to $P\rightarrow -\rho_I c^2$ corresponding to vacuum
energy. The Hubble parameter is constant,
with value
$H=(8\pi/3)^{1/2}t_I^{-1}$,
where $t_I=1/(G\rho_I)^{1/2}$ is a characteristic time associated with
$\rho_I$. 
This leads to a
phase of early inflation during which the scale factor increases exponentially
rapidly with time as $e^{(8\pi/3)^{1/2}t/t_I}$ (early de
Sitter era).

For $a\gg a_1$, the energy
density decreases algebraically as
\begin{equation}
\rho\sim \frac{\rho_I a_1^{3(1+\alpha)}}{a^{3(1+\alpha)}}.
\label{esf4}
\end{equation}
In that
case, it behaves as an $\alpha$-fluid with a linear equation of state
$P\sim \alpha\rho c^2$. This leads to an $\alpha$-era during
which the scale factor increases algebraically rapidly with time as 
$t^{2/[3(1+\alpha)]}$ and the density decreases as $t^{-2}$. The expansion of
the universe is
decelerating if
$\alpha>-1/3$ and accelerating if $\alpha<-1/3$. We can write the
energy density of the $\alpha$-fluid as  
\begin{equation}
\rho_{\alpha}=\frac{\Omega_{\alpha,0}\rho_0}{a^{3(1+\alpha)}},
\label{esf5}
\end{equation}
where $\rho_0 c^2$ is the present
energy density of the universe and 
$\Omega_{\alpha,0}$ is the present fraction of the $\alpha$-fluid (e.g.
radiation
when $\alpha=1/3$). Comparing Eq. (\ref{esf4}) with Eq. (\ref{esf5}) and  
introducing the
convenient notation
$\Omega_{I,0}=\rho_I/\rho_0$, we get 
\begin{equation}
a_1=\left (\frac{\Omega_{\alpha,0}}{\Omega_{I,0}}\right
)^{\frac{1}{3(1+\alpha)}}.
\label{esf6a}
\end{equation}
This relation determines the constant $a_1$ (we note that its value is
independent
of $n$).
We can then rewrite Eq. (\ref{esf1}) as
\begin{equation}
\label{esf7}
\frac{\rho}{\rho_0}=\frac{\Omega_{\alpha,0}}{\left\lbrack a^{\frac{
3(\alpha+1) } { n } } +\left (\frac {\Omega_{\alpha,0}}{\Omega_{I,0}}\right
)^{1/n}\right\rbrack^n}.
\end{equation}

The
equation of state (\ref{kc14}) thus describes the smooth transition
between a
phase of
inflation and an $\alpha$-era in the early universe. The characteristic scale
$a_1$ marks the transition
between the vacuum energy (de Sitter) era and the $\alpha$-era. At $a=a_1$, we
have
$\rho_1=\rho_I/2^n$ and $P_1=[(\alpha-1)/2^{n+1}]\rho_I c^2$. The equation of
state
(\ref{kc14}) is
studied in detail in \cite{cosmopoly1,universe}. The energy density decreases
monotonically from $\rho_I$ to $0$. When $\alpha\ge 0$, the pressure increases
from $P=-\rho_I c^2$
to a maximum positive value $P_e/(\rho_I
c^2)=\alpha^{n+1}n^n/[(\alpha+1)^n(n+1)^{n+1}]$
at $a_e/a_1=[(\alpha+n+1)/(n\alpha)]^{n/[3(\alpha+1)]}$ and
$\rho_e/\rho_I=[\alpha n/(\alpha+1)(n+1)]^n$ then
decreases to zero. The pressure vanishes when
$a_w/a_1=(1/\alpha)^{n/[3(\alpha+1)]}$ and
$\rho_w/\rho_I=[\alpha/(\alpha+1)]^n$. When $\alpha<0$, the pressure
monotonically increases from $P=-\rho_I c^2$ to zero. It is always negative.
In this model, there is no initial singularity (no big bang). The universe
exists from the
infinite past and the scale factor tends to zero when $t\rightarrow -\infty$.
The temporal evolution of the scale factor $a(t)$ can be obtained  analytically
(in reversed form) in terms of hypergeometric functions [see Eq. (61) in
\cite{cosmopoly1}]. 
The evolution of the temperature is discussed in
\cite{cosmopoly1}.
We refer to Figs. 2, 4, 8 and 10 of \cite{cosmopoly1} for an illustration of the
previous results.

\subsection{Equation of state parameter, deceleration parameter and squared
speed of sound}
\label{sec_kcsf}

The equation of state parameter $w=P/(\rho c^2)$ is given by
\begin{equation}
\label{kc19}
w=-(\alpha+1)\left (\frac{\rho}{\rho_I}\right )^{1/n}+\alpha.
\end{equation}
Using Eq. (\ref{esf1}) we get
\begin{equation}
\label{kc20}
w=\frac{\alpha
(a/a_1)^{\frac{3(\alpha+1)}{n}}-1}{(a/a_1)^{\frac{3(\alpha+1)}{n}}+1}.
\end{equation}
The pressure vanishes ($w=0$) when
$a_w/a_1=(1/\alpha)^{n/[3(\alpha+1)]}$ and
$\rho_w/\rho_I=[\alpha/(\alpha+1)]^n$ (when $\alpha\ge 0$).

The deceleration parameter [see Eq. (\ref{e16})] is given
by
\begin{equation}
\label{kc22}
q=\frac{1+3\alpha}{2}-\frac{3}{2}(\alpha+1)\left (\frac{\rho}{\rho_I}\right
)^{1/n}.
\end{equation}
Using Eq. (\ref{esf1}) we get
\begin{equation}
\label{kc23}
q=\frac{(1+3\alpha) (a/a_1)^{\frac{3(\alpha+1)}{n}}-2}{2\left\lbrack
(a/a_1)^{\frac{3(\alpha+1)}{n}}+1\right\rbrack}.
\end{equation}
The universe is accelerating ($q<0$) when $a<a_c$ and
decelerating ($q>0$) when $a>a_c$ with
$a_c/a_1=[2/(1+3\alpha)]^{n/[3(\alpha+1)]}$ and
$\rho_c/\rho_I=(1+3\alpha)^n/[3(\alpha+1)]^n$ (provided
that $\alpha>-1/3$). Therefore,
$a_c$ marks the end of the early inflation.

The squared speed of sound  $c_s^2=dP/d\rho$ is given by
\begin{equation}
\label{kc19b}
\frac{c_s^2}{c^2}=-(\alpha+1)\left
(1+\frac{1}{n}\right )\left (\frac{\rho}{\rho_I}\right )^{1/n}+\alpha.
\end{equation}
Using Eq. (\ref{esf1}) we get
\begin{equation}
\label{qu13}
\frac{c_s^2}{c^2}=\frac{\alpha
(a/a_1)^{\frac{3(\alpha+1)}{n}}-\frac{\alpha+n+1}{n}}{
(a/a_1)^{\frac{3(\alpha+1)}{n}}+1}.
\end{equation}
When $\alpha\ge 0$, the speed of sound is imaginary ($c_s^2<0$) when $a<a_e$ and
real ($c_s^2>0$)
when $a>a_e$ with $a_e/a_1=[(\alpha+n+1)/(n\alpha)]^{n/[3(\alpha+1)]}$ and
$\rho_e/\rho_I=[\alpha n/(\alpha+1)(n+1)]^n$ (this is the point where the
pressure
reaches its maximum value $P_e$ so that $c_s^2=dP/d\rho=0$). When $\alpha<0$,
the speed of sound is always imaginary ($c_s^2<0$). When it is real, the speed
of sound is always less than the speed of light.

As the universe expands from $a=0$ to
$a=+\infty$, the equation of state
parameter $w$ increases from $-1$ to $\alpha$, the deceleration parameter $q$
increases from $-1$ to $(1+3\alpha)/2$ and the  ratio $(c_s/c)^2$ increases from
$-(\alpha+n+1)/n$ to $\alpha$ (see Fig. 6 in \cite{cosmopoly1}).

\subsection{Application to the radiation}
\label{sec_radearly}

In this section, we specifically apply the preceding results to the case
$n=1$ and $\alpha=1/3$ (radiation). This corresponds to the equation of state
(\ref{gen25b}). For $\rho\ll\rho_I$, we
recover the equation of state of radiation $P\sim\rho c^2/3$. For
$\rho\rightarrow\rho_I$, we get $P\rightarrow -\rho_I c^2$ corresponding to
vacuum energy. The
energy density  and the pressure evolve with the scale factor as
\begin{equation}
\label{esf1b}
\rho=\frac{\rho_I}{1+(a/a_1)^{4}}
,\qquad \frac{P}{\rho_I c^2}=\frac{\frac{1}{3}
(a/a_1)^{4}-1}{\left\lbrack
(a/a_1)^{4}+1\right\rbrack^{2}}.
\end{equation}
For $a\ll a_1$, the
energy
density is approximately constant with value $\rho\simeq \rho_I$, and the
pressure tends to $P\rightarrow -\rho_I c^2$ (vacuum
energy). This leads to a
phase of early inflation during which the
scale factor increases exponentially
rapidly with time as $e^{(8\pi/3)^{1/2}t/t_I}$ (early de
Sitter era). For $a\gg a_1$,
the energy density decreases algebraically as
\begin{equation}
\rho\sim \frac{\rho_I a_1^{4}}{a^{4}},
\label{esf4b}
\end{equation}
corresponding to the radiation with a linear equation of state
$P\sim \rho c^2/3$. During the radiation era, the scale factor increases
algebraically rapidly with time as 
$t^{1/2}$ and the density decreases as $t^{-2}$. The expansion of
the universe is
decelerating. We can write the
energy density of radiation as  
\begin{equation}
\rho_{\rm rad}=\frac{\Omega_{\rm rad,0}\rho_0}{a^{4}},
\label{esf5b}
\end{equation}
where $\rho_0 c^2$ is the present
energy density of the universe and 
$\Omega_{\rm rad,0}$ is the present fraction of radiation.

The
equation of state (\ref{gen25b}) describes the smooth transition
between a
phase of
inflation and the radiation era in the early universe. The transition between
the vacuum energy (de Sitter) era and the radiation era takes place
at
\begin{equation}
a_1=\left (\frac{\Omega_{\rm rad,0}}{\Omega_{I,0}}\right
)^{1/4}.
\label{esf6}
\end{equation}
At $a=a_1$, we
have
$\rho_1=\rho_I/2$ and $P_1=-\rho_I c^2/6$. The
equation of
state (\ref{gen25b})  is
studied in detail in \cite{cosmopoly1,universe}. The energy density
decreases
monotonically from $\rho_I$ to $0$. The pressure increases from $P=-\rho_I c^2$
to a maximum positive value $P_e/(\rho_I
c^2)=1/48$ at $a_e/a_1=7^{1/4}$ and
$\rho_e/\rho_I=1/8$   then
decreases to zero. The pressure vanishes when $a_w/a_1=3^{1/4}$ and
$\rho_w/\rho_I=1/4$.
There is no initial singularity (no big bang). The universe exists from the
infinite past and the scale factor tends to zero when $t\rightarrow -\infty$.
The
temporal evolution of the scale factor $a(t)$  is given  analytically
(in reversed form) by
\cite{chavanisAIP,jgrav,cosmopoly1,universe,chavanisIOP,vacuumon}:
\begin{equation}
\sqrt{(a/a_1)^{4}+1}-\ln\left
(\frac{1+\sqrt{(a/a_1)^{4}+1}}{(a/a_1)^{2}}\right
)=2\left (\frac{8\pi}{3}\right )^{1/2}\frac{t}{t_I}+C.
\label{esf8}
\end{equation}
The constant is given by $C\simeq 1-\ln
2+2\ln\epsilon$,
where $\epsilon\equiv a(t=0)/a_1\ll 1$. The value of $\epsilon$ will be
determined in Sec. \ref{sec_kcdur}. The evolution of the temperature is
discussed in \cite{chavanisAIP,jgrav,cosmopoly1,chavanisIOP}.
We refer to Figs. 11-17 of \cite{cosmopoly1} for an illustration of these
results.

The equation of state parameter $w=P/(\rho c^2)$ is given by
\begin{equation}
\label{kc19c}
w=-\frac{4\rho}{3\rho_I}+\frac{1}{3}=\frac{\frac{1}{3}
(a/a_1)^{4}-1}{(a/a_1)^{4}+1}.
\end{equation}
The pressure vanishes when $a_w/a_1=3^{1/4}$ and
$\rho_w/\rho_I=1/4$.

The deceleration parameter is given
by
\begin{equation}
\label{kc22b}
q=1-\frac{2\rho}{\rho_I}=\frac{
(a/a_1)^{4}-1}{(a/a_1)^{4}+1}.
\end{equation}
The universe is accelerating when $a<a_c$ and
decelerating when $a>a_c$ with
$a_c/a_1=1$ and
$\rho_c/\rho_I=1/2$. Therefore,
$a_c$ marks the end of the early inflation. We note that $a_c=a_1$.

The squared speed of sound  $c_s^2=dP/d\rho$ is given by
\begin{equation}
\label{kc19d}
\frac{c_s^2}{c^2}=-\frac{8\rho}{3\rho_I}+\frac{1}{3}=\frac{\frac{1}{3}
(a/a_1)^{4}-\frac{7}{3}}{
(a/a_1)^{4}+1}.
\end{equation}
The speed of sound is imaginary ($c_s^2<0$) when $a<a_e$ and real  ($c_s^2>0$)
when $a>a_e$ with $a_e/a_1=7^{1/4}$ and
$\rho_e/\rho_I=1/8$ (this is the point where the
pressure
reaches its maximum value $P_e$ so that $c_s^2=dP/d\rho=0$). When it is real,
the speed of sound
is
always less than the speed of light.

As the universe expands from $a=0$ to
$a=+\infty$, the equation of state
parameter $w$ increases from $-1$ to $1/3$, the deceleration parameter $q$
increases from $-1$ to $1$ and the  ratio $(c_s/c)^2$ increases from
$-7/3$ to $1/3$ (see Fig. 6 of \cite{cosmopoly1}).

{\it Remark:} The foregoing results can be directly obtained
from the Lagrangian (\ref{kc1}). We have the relation 
\begin{equation}
{\cal F}=\frac{B^2}{2}=\frac{\rho_I c^2}{(a/a_1)^4},
\label{tsw1b}
\end{equation}
which can be obtained by comparing  Eq.
(\ref{esf1b}) with
Eqs. (\ref{gr9b}) and  (\ref{gr10b}) or by comparing Eq. (\ref{esf4b})
with Eq. (\ref{tsw1a}) in the radiation era. 
The results can be therefore expressed in terms of ${\cal F}/\rho_I
c^2$
or $B^2/2\rho_I c^2$ instead of $a/a_1$ by using Eq. (\ref{tsw1b}). Applying Eq.
(\ref{tsw1b}) at the present epoch ($a=1$) we find that
\begin{equation}
a_1=\left (\frac{{\cal F}_0}{\rho_Ic^2}\right )^{1/4},
\label{gr12g}
\end{equation}
which is the same as Eq. (\ref{esf6}) according to Eq. (\ref{vo3}).

\subsection{$e$-folding number and duration of the
inflation}
\label{sec_kcdur}

The $e$-folding number before
inflation ends is defined by
\begin{equation}
N=\ln\left (\frac{a_c}{a}\right ),
\label{kc24}
\end{equation}
where $a_c$ is the scale factor at the end
of the inflation. It corresponds to $q=0$ or $w=-1/3$. In this section, we take
$n=1$ and  $\alpha=1/3$ (radiation) for the
reasons given in Sec. \ref{sec_genval}. In that case, using the results of Sec.
\ref{sec_kcsf}, we have $\rho_{c}={\rho_I}/{2}$ and $a_{c}=a_1$, where $a_1$ is
given by Eq. (\ref{esf6}). The scale factor $a_c$ at the end of the inflation
coincides with the scale factor $a_1$ marking
the transition between the inflation era and the radiation era. To determine
$a_1$
we must specify the value of $\rho_I$. At that point we can have two positions:

(i) {\it Model I:} If we follow the arguments given in our previous papers
\cite{cosmopoly1,universe}, we would
identify $\rho_I$ with the Planck density 
$\rho_P=c^5/G^2\hbar=5.16\times 10^{99}\, {\rm g\, m^{-3}}$  (or a fraction of
the Planck density). In other words,
we assume that the maximum density of the universe (corresponding to the early
inflation
era) is the Planck density. In that case, taking
$\rho_0=8.62\times 10^{-24}{\rm g\, m^{-3}}$ for the
present density of the universe, we get
$\Omega_{P,0}=5.98\times
10^{122}$ leading to
\begin{equation}
a_1=1.98\times 10^{-32}.
\label{kc30p}
\end{equation}
On the other hand, we have argued in  \cite{cosmopoly1,universe} (see also
Appendix
\ref{sec_intre}) that
\begin{equation}
\epsilon\equiv \frac{a(t=0)}{a_1}=1.71\times 10^{-30}.
\label{kc31}
\end{equation}
Therefore, $a(t=0)=3.39\times 10^{-62}$. From the above
results, we  
obtain \cite{vacuumon}
\begin{equation}
N_0=\ln\left \lbrack \frac{a_c}{a(t=0)}\right \rbrack=\ln\left
\lbrack \frac{a_1}{a(t=0)}\right \rbrack=\ln\left
(\frac{1}{\epsilon}\right )=68.5.
\label{kc32}
\end{equation}
This value is consistent with the value $N\simeq 65$ deduced from the
observations in order to solve the horizon and flatness problems
\cite{liddlelyth}. 
On the
other hand, the duration $t_c$ of the 
inflation can be evaluated by substituting $a=a_c=a_1$
into
Eq. (\ref{esf8})  giving
\begin{equation}
\sqrt{2}-\ln(1+\sqrt{2})=2\left (\frac{8\pi}{3}\right
)^{1/2}\frac{t_c}{t_P}+C,
\label{kc33}
\end{equation}
where $C=1-\ln(2)+2\ln\epsilon=-137$ and 
$t_{P}=1/(G\rho_{P})^{1/2}=(\hbar G/c^5)^{1/2}=5.39\times 10^{-44}\, {\rm
s}$ is the Planck time. This gives
$t_{c}=23.8\, t_P=1.28\times
10^{-42}\, {\rm s}$. 
Therefore, in model I, the duration of the early inflation is of the order of
the Planck time $t_P$.

(ii) {\it Model II:} If we follow the
arguments 
based on
nonlinear electrodynamics given by  Kruglov
\cite{kruglov2015a} and in Sec. \ref{sec_nee}, we should identify
$\rho_I$ with the electron density
$\rho_e={m_e}/{r_e^3}=4.07\times 10^{16}\, {\rm g\, m^{-3}}$ (a more
precise value would be $\rho_*=0.0307\rho_e=1.25\times 10^{15}\,
{\rm g\, m^{-3}}$ but it is convenient for the discussion to take $\rho_e$). In
other words, the maximum
density of the universe (corresponding to
the early inflation
era) is connected to the electron density, not to the Planck density. In that
case, we get $\Omega_{e,0}=4.72\times 10^{39}$ leading to
\begin{equation}
a_1=1.18\times 10^{-11}.
\label{kc30}
\end{equation}
On the other hand, Kruglov 
\cite{kruglov2020} takes
$\epsilon\equiv {a(t=0)}/{a_1}=6.5\times 10^{-28}$. 
This value is consistent with Eq. (\ref{kc31}). As discussed
in Appendix \ref{sec_intre}, it is convenient to determine $a(t=0)$ in model
II such that 
$\epsilon$
has the value given by Eq. (\ref{kc31}). This gives
$a(t=0)=2.10\times
10^{-41}$. In
this manner, $N_0$ has
the
same value $N_0=68.5$ in the two models and this simplifies the comparison.
Repeating the arguments given after Eq.
(\ref{kc32}), we find that the duration $t_c$ of the early 
inflation  in model II is given by $t_{c}=23.8\, t_e^*=0.457\, {\rm
s}$, where
$t_e^*=1/\sqrt{G\rho_e}=0.0192\, {\rm s}$
is a timescale constructed with the density of
the electron (instead of the Planck density).\footnote{With the value
$\epsilon\equiv {a(t=0)}/{a_1}=6.5\times 10^{-28}$,
Kruglov \cite{kruglov2020} finds that the
inflation lasts
approximately $2$ s with the reasonable $e$-folding number
$N\simeq 63$.} It
corresponds to the dynamical
time of a self-gravitating system of density $\rho_e$ \cite{btnew}. We
shall call it the gravitoelectronic time. In
model II, the duration of the early inflation is of the order of
$t_e^*$.

In conclusion, in model II based on nonlinear electrodynamics where
$\rho_I\sim \rho_e$,
the inflation is much longer ($\sim 1\, {\rm s}$) than in model I where
$\rho_I\sim \rho_P$ giving $t_{c}\sim 1.28\times 10^{-42}\, {\rm s}$. This is
because $\rho_e\sim 10^{-80}\rho_P\ll\rho_P$ \cite{ouf}. It would be interesting
to know if cosmological constraints on the duration of the inflation (or on
the density of the primordial universe) are able to
discriminate between the two models.

{\it Remark:} We can have a more direct estimate of the duration of the
inflation by applying the very accurate approximate formula
$a\simeq a(t=0)e^{(8\pi/3)^{1/2}t/t_I}$ at
$a=a_1=a_c$ \cite{cosmopoly1,universe}, giving
\begin{equation}
t_c=\left (\frac{3}{8\pi}\right )^{1/2}\ln\left (\frac{1}{\epsilon}\right
)t_I\simeq 23.7\, t_I. 
\label{qhu}
\end{equation}

\section{Nonlinear electrodynamics based on the Lagrangian ${\cal
L}=-{\cal F}-
{\cal F}_\Lambda$:
Cosmology in the late universe}
\label{sec_kcl}

In the late universe, we consider a nonlinear electrodynamics based on the
Lagrangian
\begin{equation}
{\cal L}=-{\cal F}-{\cal F}_{\Lambda},
\label{kc1z}
\end{equation}
where ${\cal F}_{\Lambda}=\rho_\Lambda c^2$. We suggest that
DE corresponds to the zero-point radiation energy that
manifests itself as a constant term in the electromagnetic Lagrangian.  We
identify this
constant with the cosmological density $\rho_{\Lambda}=\Lambda c^2/8\pi
G=5.96\times
10^{-24}\, {\rm g\, m^{-3}}$. In this model, the late
acceleration of the universe is due to the electromagnetic energy of point
zero. From Eqs. (\ref{f4}) and (\ref{f5})
we find that the
energy density
and the pressure are given by
\begin{equation}
\rho c^2={\cal
F}+\rho_{\Lambda}c^2,
\label{gr9c}
\end{equation}
\begin{equation}
P=\frac{1}{3}{\cal F}-\rho_{\Lambda}c^2.
\label{gr10c}
\end{equation}
Eliminating ${\cal F}$ between these two expressions we obtain  the affine
equation of state
\begin{equation}
P=\frac{1}{3}\rho c^2-\frac{4}{3}\rho_{\Lambda}c^2.
\label{feelcom}
\end{equation}
This equation of state has
been studied in \cite{cosmopoly2,universe} to model
the late universe. Below, we recall its main properties.

\subsection{Generalized polytropic equation of state}
\label{sec_kcsf3}

For the sake of generality, we consider a generalized polytropic equation of
state of the form
\begin{equation}
\label{gen20b}
P=\alpha\rho c^2-(\alpha+1)\rho_\Lambda c^2 \left (\frac{\rho}{\rho_I}\right
)^{1-{1}/{|n|}}
\end{equation}
with $n<0$ and $-1<\alpha\le 1$ \cite{cosmopoly2,universe}, even if we shall
finally take
$\alpha=1/3$ and $n=-1$ for the reasons explained in Sec. \ref{sec_genval}. 
For
$\rho\gg\rho_\Lambda$, we obtain the linear equation of state $P\sim\alpha\rho
c^2$. For
$\rho\rightarrow\rho_\Lambda$, we get $P\rightarrow -\rho_\Lambda c^2$
corresponding to the equation of state of dark energy.

{\it Remark:} The equation of state (\ref{gen20b})  in the late universe
can be viewed as the ``symmetric'' version of the equation of state (\ref{kc14})
in the early universe. The symmetrical structure of the equation of state in
the early ($n>0$) and late ($n<0$) universe is developed in
\cite{chavanisAIP,cosmopoly1,cosmopoly2}.

\subsection{Evolution of the density, pressure and scale factor}
\label{sec_lesf}

Solving
the energy
conservation equation (\ref{f8}) with the equation of state (\ref{gen20b}) we
find
that the
energy density evolves with the scale factor as
\begin{equation}
\label{lesf1}
\rho=\rho_\Lambda \left\lbrack
1+\frac{1}{(a/a_2)^{\frac{3(1+\alpha)}{|n|}}}\right\rbrack^{|n|},
\end{equation}
where $a_2$ is a constant of integration.  The pressure depends
on the scale
factor as
\begin{equation}
\label{lesf2}
\frac{P}{\rho_\Lambda c^2}=\left\lbrack \frac{\alpha}{
(a/a_2)^{\frac{3(\alpha+1)}{|n|}}}-1\right\rbrack \left\lbrack
\frac{1}{(a/a_2)^{\frac{3(\alpha+1)}{|n|}}}+1\right\rbrack^{|n|-1}.
\end{equation}
For $a\gg a_2$, the energy
density is approximately constant
\begin{equation}
\rho\simeq \rho_\Lambda,
\label{lesf3}
\end{equation}
and the pressure tends to $P\rightarrow -\rho_\Lambda
c^2$ corresponding to DE. Here, $\rho_{\Lambda}$ is
unambiguously associated
with the cosmological density $\rho_{\Lambda}$. The Hubble
parameter is constant, with value
$H=(8\pi/3)^{1/2}t_\Lambda^{-1}$,
where $t_\Lambda=1/(G\rho_\Lambda)^{1/2}=(8\pi/\Lambda c^2)^{1/2}=1.59\times
10^{18}\, {\rm
s}$ is a characteristic time  (cosmological time) associated with
$\rho_\Lambda$. 
This leads to a
phase of late accelerating expansion (or late inflation) during which the
scale factor increases exponentially
rapidly with time as $e^{(8\pi/3)^{1/2}t/t_\Lambda}$ (late de
Sitter era).

For $a\ll a_2$, the energy
density decreases algebraically as
\begin{equation}
\rho\sim \frac{\rho_\Lambda a_2^{3(1+\alpha)}}{a^{3(1+\alpha)}}.
\label{lesf4}
\end{equation}
In that
case, it behaves as an $\alpha$-fluid with a linear equation of state
$P\sim \alpha\rho c^2$. This leads to an $\alpha$-era during
which the scale factor increases algebraically rapidly with time as 
$t^{2/[3(1+\alpha)]}$ and the density decreases as $t^{-2}$. The expansion of
the universe is
decelerating if
$\alpha>-1/3$ and accelerating if $\alpha<-1/3$. We can write the
energy density of the $\alpha$-fluid as 
\begin{equation}
\rho_{\alpha}=\frac{\Omega_{\alpha,0}\rho_0}{a^{3(1+\alpha)}},
\label{lesf5}
\end{equation}
where $\rho_0 c^2$ is the present
energy density of the universe and 
$\Omega_{\alpha,0}$ is the present fraction of the $\alpha$-fluid (e.g.
radiation
when $\alpha=1/3$). Comparing Eq. (\ref{lesf4}) with Eq. (\ref{lesf5}) and  
introducing the present fraction of DE
$\Omega_{\Lambda,0}=\rho_\Lambda/\rho_0$, we get
\begin{equation}
a_2=\left (\frac{\Omega_{\alpha,0}}{\Omega_{\Lambda,0}}\right
)^{\frac{1}{3(1+\alpha)}}.
\label{lesf6a}
\end{equation}
This relation determines the constant $a_2$ (we note that its value is
independent
of $n$). We have the relation $a_2=a_1 \left ({\rho_I}/{\rho_{\Lambda}}\right
)^{\frac{1}{3(1+\alpha)}}$. We can then rewrite Eq.
(\ref{lesf1}) as
\begin{equation}
\label{lesf7}
\frac{\rho}{\rho_0}=\Omega_{\alpha,0}\left\lbrack \frac{1}{a^{\frac{
3(\alpha+1) } { |n| } }} +\left (\frac{\Omega_{\Lambda,0}}
{\Omega_{\alpha,0}}\right
)^{1/|n|}\right\rbrack^{|n|}.
\end{equation}

The
equation of state (\ref{gen20b}) thus describes the smooth transition
between an
$\alpha$-era and a phase of accelerating expansion (DE or
inflation) in the late universe.  The characteristic scale $a_2$ marks the
transition
between the $\alpha$-era and the dark energy (de Sitter) era. At $a=a_2$,
we
have
$\rho_2=2^{|n|}\rho_\Lambda$ and $P_2=(\alpha-1)2^{|n|-1}\rho_{\Lambda}c^2$. The
equation of
state (\ref{gen20b})  is
studied in detail in \cite{cosmopoly2,universe}. The energy density decreases
monotonically from $+\infty$ to $\rho_\Lambda$. The evolution of the pressure
depends on the sign of $\alpha$. 
When $\alpha>0$,  the pressure decreases from $+\infty$ to
$P=-\rho_\Lambda c^2$. It vanishes when
$a'_w/a_2=\alpha^{|n|/[3(\alpha+1)]}$ and
$\rho'_w/\rho_\Lambda=[(\alpha+1)/\alpha]^{|n|}$. When $\alpha<0$,  the pressure
increases from $-\infty$ to
$P=-\rho_\Lambda c^2$. When $\alpha=0$, the evolution of the pressure depends on
the value of $|n|$. When $|n|>1$ the pressure increases from
$-\infty$ to $P=-\rho_\Lambda c^2$,  when $|n|=1$ the pressure is constant
$P=-\rho_\Lambda c^2$, and when $|n|<1$ the pressure decreases 
from zero to $P=-\rho_\Lambda c^2$. The pressure may present an extremum
$P'_e/(\rho_\Lambda
c^2)=\alpha^{-|n|+1}(-|n|)^{-|n|}/[(\alpha+1)^{-|n|}(-|n|+1)^{-|n|+1}]$
at the point
$a'_e/a_2=[-(\alpha-|n|+1)/(|n|\alpha)]^{-|n|/[3(\alpha+1)]}$ and
$\rho'_e/\rho_\Lambda=[-\alpha|n|/((\alpha+1)(-|n|+1))]^{-|n|}$. Its
conditions of existence are detailed in \cite{cosmopoly2}. We refer to Figs. 1
and 2
of \cite{cosmopoly2} for an illustration of the previous results.

{\it Remark:} The temporal evolution of the scale factor $a(t)$ is given in
\cite{cosmopoly2} assuming that the late universe is described by a single
fluid with the equation of state (\ref{gen20b}). However, in general, there are
other fluids that also contribute to the density of the universe and therefore
change the temporal  evolution of the scale factor (see below).
This is why we have not given its expression here.

\subsection{Equation of state parameter and
squared
speed of sound}
\label{sec_kcsf4}

The equation of state parameter $w=P/(\rho c^2)$ is given by
\begin{equation}
\label{llq2}
w=\alpha-(\alpha+1)\left (\frac{\rho_\Lambda}{\rho}\right )^{1/|n|}.
\end{equation}
Using Eq. (\ref{lesf1}) we get
\begin{equation}
\label{llq8}
w=\frac{\alpha
(a_2/a)^{\frac{3(\alpha+1)}{|n|}}-1}{(a_2/a)^{\frac{3(\alpha+1)}{|n|}}+1}.
\end{equation}
The pressure vanishes ($w=0$) when
$a'_w/a_2=\alpha^{|n|/[3(\alpha+1)]}$ and
$\rho'_w/\rho_\Lambda=[(\alpha+1)/\alpha]^{|n|}$ (assuming $\alpha>0$).

The squared speed of sound  $c_s^2=dP/d\rho$ is given by
\begin{equation}
\label{lkc19b}
\frac{c_s^2}{c^2}=-(\alpha+1)\left
(1-\frac{1}{|n|}\right )\left (\frac{\rho_\Lambda}{\rho}\right )^{1/|n|}+\alpha.
\end{equation}
Using Eq. (\ref{lesf1}) we get
\begin{equation}
\label{lqu13}
\frac{c_s^2}{c^2}=\frac{\alpha
(a_2/a)^{\frac{3(\alpha+1)}{|n|}}+\frac{\alpha-|n|+1}{|n|}}{
(a_2/a)^{\frac{3(\alpha+1)}{|n|}}+1}.
\end{equation}
The speed of sound may vanish  at the
point $a'_e/a_2=[-(\alpha-|n|+1)/(|n|\alpha)]^{-|n|/[3(\alpha+1)]}$ and
 $\rho'_e/\rho_\Lambda=[-\alpha|n|/((\alpha+1)(-|n|+1))]^{-|n|}$.  This
is the
point where the pressure is extremum. The
speed of sound may equal
the speed of light at
the point $a'_s/a_2=[(\alpha-2|n|+1)/(|n|(1-\alpha))]^{-|n|/[3(\alpha+1)]}$
 and $\rho'_s/\rho_\Lambda=[(1-\alpha)|n|/((\alpha+1)(-|n|+1))]^{-|n|}$. The
conditions of existence of these two points are detailed in \cite{cosmopoly2}.

As the
universe expands from
$a=0$ to
$a=+\infty$,  the equation of state
parameter $w$ evolves from  $\alpha$ to $-1$  and the squared
speed of sound $c_s^2/c^2$  evolves 
from $\alpha$ to $(\alpha-|n|+1)/|n|$  (see Fig. 5 of
\cite{cosmopoly2} for an illustration).

{\it Remark:} For the reason explained previously we have not given the
deceleration parameter because its value may be affected by other fluids.

\subsection{Application to the radiation}
\label{sec_radlate}

In this section, we specifically apply the preceding results to the case
$n=-1$ and $\alpha=1/3$ (radiation). This corresponds to the equation of state
(\ref{feelcom}). It can be viewed as the ``symmetric'' version of the equation
of state
(\ref{gen25b}) in the early universe
\cite{chavanisAIP,cosmopoly1,cosmopoly2}.
For
$\rho\gg\rho_\Lambda$, we
recover the linear equation of state of radiation $P\sim \rho c^2/3$. For
$\rho\rightarrow\rho_\Lambda$, we get $P\rightarrow -\rho_\Lambda c^2$
corresponding to dark energy. The
energy density and the pressure evolve with the scale factor as
\begin{equation}
\label{rlesf1}
\rho=\rho_\Lambda \left\lbrack
1+\frac{1}{(a/a_2)^{4}}\right\rbrack,\qquad
\frac{P}{\rho_\Lambda c^2}=\frac{1}{
3(a/a_2)^{4}}-1.
\end{equation}
For $a\gg a_2$, the energy
density is approximately constant with value $\rho\simeq
\rho_\Lambda$, and the pressure tends to $P\rightarrow -\rho_\Lambda
c^2$ corresponding to DE. 
This leads to a
phase of late accelerating expansion (or late inflation) during which the
scale factor increases exponentially
rapidly with time as $e^{(8\pi/3)^{1/2}t/t_\Lambda}$ (late de
Sitter era). For $a\ll a_2$, the energy
density decreases algebraically as
\begin{equation}
\rho\sim \frac{\rho_\Lambda a_2^{4}}{a^{4}},
\label{rlesf4}
\end{equation}
corresponding to the radiation with a linear equation of state
$P\sim \rho c^2/3$. During the radiation era, the scale factor
increases algebraically rapidly with time as 
$t^{1/2}$ and the density decreases as $t^{-2}$. The expansion of
the universe is
decelerating. We can write the
energy density of the radiation as
\begin{equation}
\rho_{\rm rad}=\frac{\Omega_{\rm rad,0}\rho_0}{a^{4}},
\label{rlesf5}
\end{equation}
where $\rho_0 c^2$ is the present
energy density of the universe and 
$\Omega_{\rm rad,0}$ is the present fraction of radiation.

The
equation of state (\ref{feelcom}) describes the smooth transition
between the radiation era and a phase of accelerating expansion (DE or
inflation) in the late universe. The transition between the radiation era
and the the dark energy (de Sitter) era takes place
at
\begin{equation}
a_2=\left (\frac{\Omega_{\rm rad,0}}{\Omega_{\Lambda,0}}\right
)^{1/4}.
\label{lesf6}
\end{equation}
At $a=a_2$,
we
have
$\rho_2=2\rho_\Lambda$ and $P_2=-(2/3)\rho_{\Lambda}c^2$. The
equation of
state (\ref{gen20b})  is
studied in detail in \cite{cosmopoly2,universe}. The energy density decreases
monotonically from $+\infty$ to $\rho_\Lambda$. The pressure
decreases monotonically from $+\infty$ to $P=-\rho_\Lambda c^2$. It vanishes
when $a'_w/a_2=1/81$ and $\rho'_w/\rho_\Lambda=4$.

The equation of state parameter $w=P/(\rho c^2)$ is given by
\begin{equation}
\label{llq2b}
w=\frac{1}{3}-\frac{4\rho_\Lambda}{3\rho}=\frac{\frac{1}{3}
(a_2/a)^{4}-1}{(a_2/a)^{4}+1}.
\end{equation}
The pressure vanishes when $a'_w/a_2=1/81$ and
$\rho'_w/\rho_\Lambda=4$. As the universe expands from
$a=0$ to
$a=+\infty$,  the equation of state
parameter $w$ decreases from  $1/3$ to $-1$.

The squared speed of sound  $c_s^2=dP/d\rho$ is given by
\begin{equation}
\label{lkc19}
\frac{c_s^2}{c^2}=\frac{1}{3}.
\end{equation}
The speed of sound is constant and equal to
${c_s}/{c}=1/\sqrt{3}$. The speed of sound is less than the speed of light.

We will see in Sec. \ref{sec_total} that $\rho$ given by Eq.
(\ref{rlesf1}) represents the
density of generalized radiation in the late universe. For $a\ll a_2$ it
corresponds to the ordinary
radiation. For $a_{\rm eq}\ll a\ll a_2$ (where $a_{\rm eq}$ is the value of the
scale factor at
radiation-matter equality) it is subdominant with respect to baryonic and dark
matter viewed as
different species. This is why we have not given the temporal evolution of
the
scale factor $a(t)$ nor the deceleration parameter $q(t)$ in this period because
we have to take into account the contribution of
matter. This is done in Sec. \ref{sec_complete} where we
present the complete
model. By
contrast, for $a_1\ll a\ll a_{\rm eq}$ we are in the radiation era and for 
$a\gg
a'_{2}$ we are in the dark energy era where
the asymptotic
results for $a(t)$ given above are valid. In these limits, the deceleration
parameter is given by Eq. (\ref{e16}).

{\it Remark:} The foregoing results can be directly obtained
from the Lagrangian (\ref{kc1z}). We have the relation 
\begin{equation}
{\cal
F}=\frac{B^2}{2}=\frac{\rho_\Lambda c^2}{(a/a_2)^4},
\label{tswc}
\end{equation}
which can be obtained by comparing  Eq.
(\ref{rlesf1}) with
Eqs.  (\ref{gr9c}) and  (\ref{gr10c}) or by comparing Eq.
(\ref{rlesf4}) with Eq. (\ref{tsw1a})  in the radiation era. 
The results can be therefore expressed in terms of ${\cal F}/\rho_\Lambda
c^2$
or $B^2/2\rho_\Lambda c^2$ instead of $a/a_2$ by using Eq. (\ref{tswc}).
Applying Eq.
(\ref{tswc}) at the present epoch ($a=1$) we find that
\begin{equation}
a_2=\left (\frac{{\cal F}_0}{\rho_\Lambda c^2}\right )^{1/4},
\label{gr12gb}
\end{equation}
which is the same as Eq. (\ref{lesf6}) according to Eq. (\ref{vo3}).

\section{The Lagrangian of the generalized radiation}
\label{sec_total}

In Sec. \ref{sec_kc} we have considered a nonlinear electrodynamics based
on a Lagrangian of the form
\begin{equation}
{\cal
L} =\frac{-{\cal
F}}{1+\frac{{\cal F}}{\rho_I
c^2}}
\label{gr6}
\end{equation}
with $\rho_I=\rho_P$ (model I) or $\rho_I=\rho_e$ (model II). We have shown that
this Lagrangian could describe the evolution of the early
universe. In Sec. \ref{sec_kcl} we have considered a nonlinear electrodynamics
based
on a Lagrangian of the form
\begin{equation}
{\cal L}=-{\cal
F}-\rho_{\Lambda}c^2.
\label{gr7}
\end{equation}
We have shown that this Lagrangian could describe the evolution of the
late
universe. We now want to connect these two
Lagrangians in
order to describe the complete evolution of the universe (see
Sec. \ref{sec_complete}). We see that
there is a common period
corresponding to ${\cal F}\ll {\cal F}_I=\rho_I c^2$ in the early universe and 
${\cal
F}\gg {\cal F}_\Lambda=\rho_\Lambda c^2$ in the late universe. In this common
period,  corresponding to the normal radiation era, the Lagrangians
(\ref{gr6}) and
(\ref{gr7}) reduce to Maxwell's linear electrodynamics
characterized by the Lagrangian
\begin{equation}
{\cal L}=-{\cal F}.
\label{gr5}
\end{equation}
The Lagrangian  of the nonlinear electrodynamics valid during the whole
evolution
of the universe is therefore
\begin{equation}
{\cal L}=\frac{-{\cal
F}}{1+\frac{{\cal F}}{\rho_I
c^2}}-\rho_{\Lambda}c^2.
\label{gr8}
\end{equation}
In the early universe, it reduces to Eq. (\ref{gr6}) and in the late universe it
reduces to Eq. (\ref{gr7}). In the intermediate period (radiation era), it
returns the ordinary
Maxwell electrodynamics (\ref{gr5}).

\begin{figure}[!h]
\begin{center}
\includegraphics[clip,scale=0.3]{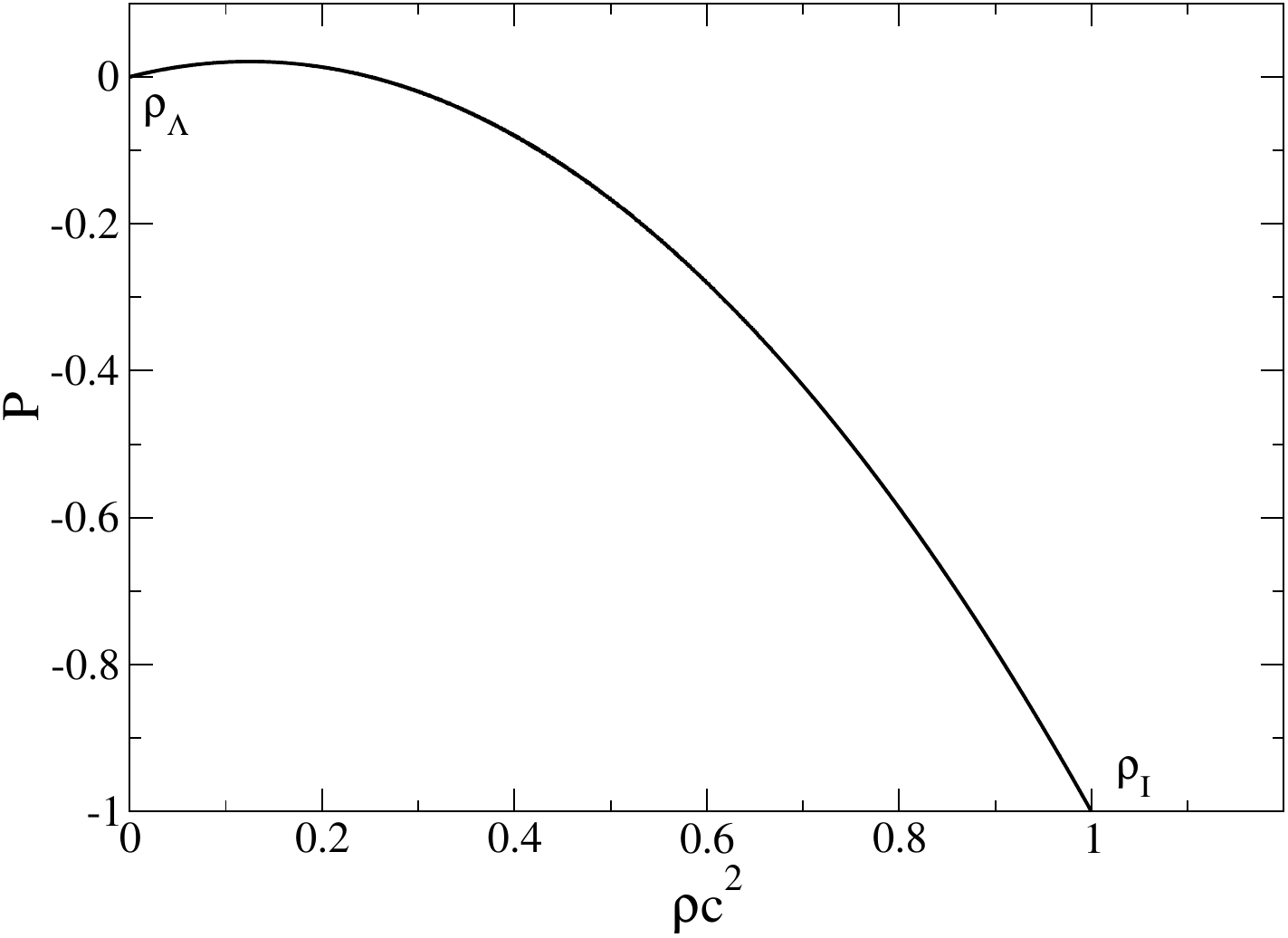}
\caption{Equation of state of the generalized radiation [see
Eq. (\ref{gr11})] associated with the Lagrangian (\ref{gr8}). The pressure $P$
and the energy density $\rho c^2$ are normalized by
$\rho_I c^2$.}
\label{eos}
\end{center}
\end{figure}

From Eqs. (\ref{f4}) and (\ref{f5}) we find that the energy density
and the pressure are given by
\begin{equation}
\rho c^2=\frac{{\cal
F}}{1+\frac{{\cal F}}{\rho_I
c^2}}+\rho_{\Lambda}c^2,
\label{gr9}
\end{equation}
\begin{equation}
P=\frac{{\cal F}\left (\frac{1}{3}-\frac{\cal F}{\rho_I c^2}\right )}{\left
(1+\frac{{\cal F}}{\rho_I
c^2}\right )^2}-\rho_{\Lambda}c^2.
\label{gr10}
\end{equation}
Eliminating ${\cal F}$ between these two expressions we obtain in excellent
approximation the quadratic
equation of state\footnote{To obtain this equation we have used the fact that
$\rho_P/\rho_\Lambda\sim 10^{120}\gg 1$ and $\rho_e/\rho_\Lambda\sim
10^{40}\gg 1$. Since these dimensionless numbers are huge \cite{ouf}, the
approximation is quasi perfect.}
\begin{equation}
P=-\frac{4\rho^2}{3\rho_I}c^2+\frac{1}{3}\rho
c^2-\frac{4}{3}\rho_{\Lambda}c^2.
\label{gr11}
\end{equation}
This equation of state (see Fig. \ref{eos}) was introduced and studied
in
\cite{chavanisAIP,jgrav,cosmopoly1,cosmopoly2,cosmopoly3,universe,stiff,mlbec,
chavanisIOP,vacuumon}. In the early
universe, we recover the quadratic equation of state (\ref{gen25b})
associated with the Lagrangian (\ref{gr6}) and in the
late
universe we recover the affine equation of state  (\ref{feelcom}) associated
with the Lagrangian (\ref{gr7}). In the
intermediate period, we recover the linear equation of state  
$P=\frac{1}{3}\rho c^2$ of the ordinary radiation associated with Maxwell's
electrodynamics. Following the interpretation given in our previous works
\cite{chavanisAIP,jgrav,cosmopoly1,cosmopoly2,cosmopoly3,universe,stiff,mlbec,
chavanisIOP,vacuumon}, the equation of state
(\ref{gr11}) describes a form of {\it
generalized radiation} which is responsible for the early inflation, the normal
radiation era and the present and late acceleration (dark
energy)
of the universe. In the present paper, we have connected this ``generalized
radiation'' to a form of nonlinear electrodynamics. Solving the energy
conservation equation (\ref{f8}) with the
equation of state (\ref{gr11}) we obtain in excellent approximation (see
footnote 10) the
following relation between the density of the generalized radiation and the
scale factor
\cite{chavanisAIP,jgrav,cosmopoly1,cosmopoly2,cosmopoly3,universe,stiff,mlbec,
chavanisIOP,vacuumon}:
\begin{equation}
\rho_{\rm Rad}=\frac{\rho_I}{1+\left
({a}/{a_1}\right )^{4}}+\rho_{\Lambda}.
\label{gr12}
\end{equation}
This is also the exact relation obtained from the
Lagrangian (\ref{gr8}) with Eq. (\ref{tsw1b}) and (\ref{gr9}).
In the
early universe, we recover
the results of Sec. \ref{sec_radearly}. In the late
universe
we recover the results of  Sec. \ref{sec_radlate}. We stress that  $\rho_{\rm
Rad}\neq \rho_{\rm
rad}$ in general. It is only in the intermediate period corresponding to
Maxwell's linear
electrodynamics that the energy density of the generalized radiation $\rho_{\rm
Rad}$ reduces to the energy density of the normal radiation $\rho_{\rm
rad}$.

{\it Remark:} It is possible to find the exact relation $\rho_{\rm Rad}(a)$
determined by the equation of state (\ref{gr11}). It is given in Appendix C
of \cite{universe}. Then, using Eqs. (\ref{f4}) and (\ref{tsw1b}) we
can obtain
the exact Lagrangian ${\cal L}({\cal F})$ corresponding to the equation of
 state
(\ref{gr11}). Conversely, eliminating ${\cal F}$ between Eqs. (\ref{gr9}) and
(\ref{gr10}) it is possible to obtain the exact equation of state corresponding
to the Lagrangian (\ref{gr8}). This equation of state exactly leads to the
relation
$\rho_{\rm Rad}(a)$ from Eq. (\ref{gr12}).  However, the approximate expressions
given above are so accurate (see
footnote 10) that this refinement is not necessary.

\section{Complete evolution of the universe}
\label{sec_complete}

The Lagrangian (\ref{gr8}) 
describes the
generalized radiation. It
accounts for the early inflation, the radiation era and the late acceleration
of the universe. Baryonic matter and dark matter must be treated
independently, as additional components. As a
result, the total Lagrangian describing the mass-energy content of the
universe
is
\begin{equation}
\mathcal{L}=\mathcal{L}_{\rm Rad}+\mathcal{L}_{\rm matter},
\label{gr13}
\end{equation}
where $\mathcal{L}_{\rm Rad}$ is the Lagrangian of the generalized
radiation\footnote{It is possible that there exist different
forms of radiation. For example, a complex SF with a
repulsive $|\varphi|^4$ self-interaction can behave like radiation
\cite{becas}. In that case, we should describe each form of radiation by a
specific Lagrangian. However, for simplicity, we shall regroup all forms
of
radiation in the Lagrangian (\ref{gr8}).}  and $\mathcal{L}_{\rm
matter}={\cal
L}_{\rm b}+{\cal L}_{\rm dm}$
is the Lagrangian of the baryonic and dark matter. The energy density of
the generalized radiation is given by Eq. (\ref{gr12}) with
$a_1=(\rho_{\rm rad,0}/\rho_I)^{1/4}$ [see Eq. (\ref{esf6})]. If we assume
for
simplicity that baryonic matter and dark matter are
pressureless ($P_{\rm b}=P_{\rm dm}=0$),\footnote{We could also consider a
more general 
equation of state of the form $P=\alpha\rho c^2$ with $\alpha\simeq 0$
\cite{cosmopoly2,universe,vacuumon}.} we find that their
densities evolve with the scale factor as
\begin{equation}
\rho_{\rm b}=\frac{\rho_{\rm b,0}}{a^3},\qquad \rho_{\rm dm}=\frac{\rho_{\rm
dm,0}}{a^3}.
\label{gr14}
\end{equation}
Therefore, the total energy density of the
universe is
\begin{equation}
\rho=\frac{\rho_I}{1+({a}/{a_1})^{4}}+\frac{\rho_{\rm
b,0}}{a^3}+\frac{\rho_{\rm dm,0}}{a^3}+\rho_{\Lambda}. 
\label{gr15}
\end{equation}
It can be rewritten as
\begin{equation}
\rho=\frac{\rho_I}{1+\frac{\rho_I a^4}{\rho_{\rm rad,0}}}+\frac{\rho_{\rm
b,0}}{a^3}+\frac{\rho_{\rm dm,0}}{a^3}+\rho_{\Lambda}. 
\label{gr19}
\end{equation}
This model can account for the whole evolution of the universe from the early
inflation to its late accelerating expansion. It exhibits two de Sitter
eras connected by a radiation era and a matter era.  We detail
these different periods below.

{\it Remark:} In this model there is no past singularity (no big bang) nor
future singularity  (no little or big rip).\footnote{We have assumed a
non-phantom evolution. See Ref. \cite{cosmopoly3} for cosmological models
presenting a singular or peculiar late evolution (phantom models).} The universe
exists eternally in the past and in the future. The
scale factor tends to zero when $t\rightarrow -\infty$ and to infinity when
$t\rightarrow +\infty$. It has been called the ``aioniotic'' universe
\cite{cosmopoly1,cosmopoly2}.

\subsection{Early universe: inflation + radiation}

In the inflation + radiation era (early universe), the energy density of the
universe is given by
\begin{equation}
\rho=\frac{\rho_I}{1+({a}/{a_1})^{4}}
\label{gr16}
\end{equation}
with $a_1=(\Omega_{\rm rad,0}/\Omega_{P,0})^{1/4}=1.98\times 10^{-32}$ if
$\rho_I=\rho_P$ and $a_1=(\Omega_{\rm rad,0}/\Omega_{e,0})^{1/4}=1.18\times
10^{-11}$ if
$\rho_I=\rho_e$. It can
be rewritten as
\begin{equation}
\label{gr19b}
\rho=\frac{\rho_I}{1+\frac{\rho_I a^{4}}{\rho_{\rm rad,0}}}.
\end{equation}
We recover the results detailed in Sec. \ref{sec_kc}. 
When $a\ll a_1$, we are in the
inflation era. The density is approximately 
constant $\rho\simeq \rho_I$.
This leads
to a phase of early accelerating expansion (or early inflation) where the scale
factor increases exponentially rapidly with time as 
$e^{(8\pi/3)^{1/2}t/t_I}$ (early de Sitter era). The universe is
accelerating. The transition between the radiation
era and the matter era takes place at  $a_{\rm
eq}=\Omega_{\rm rad,0}/\Omega_{\rm m,0}=3.00\times 10^{-4}$ (see Sec.
\ref{sec_int}). When $a_{1}\ll a\ll a_{\rm
eq}$, we are in
the ordinary radiation era described by the linear equation of state
$P_{\rm rad}=\rho c^2/3$. The density decreases algebraically
as $\rho=\rho_{\rm rad,0}/a^{4}$. The scale factor increases
algebraically with time as $t^{1/2}$ and the density decreases as $t^{-2}$. The
universe
is decelerating. We thus have a
transition between a phase of accelerating
expansion (vacuum energy/de Sitter) in the early
universe and a phase of decelerating expansion in the radiation
era. The transition takes place at
$a\simeq a_1$. This transition between the inflation era and the radiation era
is studied in detail
in \cite{cosmopoly1,universe}. The temporal evolution of the scale factor $a(t)$
is given analytically (in reversed form)
by 
\begin{equation}
\sqrt{(a/a_1)^{4}+1}-\ln\left
(\frac{1+\sqrt{(a/a_1)^{4}+1}}{(a/a_1)^{2}}\right
)=2\left (\frac{8\pi}{3}\right )^{1/2}\frac{t}{t_I}+C
\label{nesf8}
\end{equation}
with $C\simeq 1-\ln
2+2\ln\epsilon$,
where $\epsilon=a(t=0)/a_1=1.71\times 10^{-30}$ (see Sec. \ref{sec_kcdur}).
In the inflation era ($a\ll a_1$):
\begin{equation}
a\sim a(t=0) e^{(8\pi/3)^{1/2}t/t_I},\qquad 
\rho\simeq \rho_{I}.
\label{nesf8b}
\end{equation}
In the radiation era ($a_{1}\ll a\ll a_{\rm
eq}$):
\begin{equation}
a\sim  \left
(2\sqrt{\Omega_{\rm rad,0}}H_0 t\right
)^{1/2},\qquad \frac{\rho}{\rho_{0}}\sim \frac{1}{\left
(2H_0 t\right )^2},
\label{nesf8c}
\end{equation}
where $H_0=(8\pi G\rho_0/3)^{1/2}=2.195\times 10^{-18}\, {\rm s}^{-1}$ is the
present value of the Hubble parameter.

\subsection{Intermediate universe: radiation + matter}
\label{sec_int}

In the radiation + matter era  (intermediate universe), the energy density of
the universe evolves with the scale factor as
\begin{equation}
\rho=\frac{\rho_{\rm
rad,0}}{a^4}+\frac{\rho_{\rm m,0}}{a^3}.
\label{gr17}
\end{equation}
If we normalize the density by its present value, we get
\begin{equation}
\frac{\rho}{\rho_0}=\frac{\Omega_{\rm rad,0}
}{a^4}+\frac{\Omega_{\rm m,0}}{a^3}.
\label{radd0}
\end{equation}
When $a\ll a_{\rm eq}$, we are in
the ordinary radiation era described by a linear equation of state
$P_{\rm rad}=\rho c^2/3$. The density decreases algebraically
as $\rho=\rho_{\rm rad,0}/a^{4}$. The scale factor increases
algebraically with time as $t^{1/2}$ and the density decreases as $t^{-2}$. The
universe
is decelerating. When $a\gg
a_{\rm eq}$, we are in the matter era (Einstein-de Sitter) with a vanishing
pressure $P_{\rm
m}=0$. The density decreases algebraically
as $\rho=\rho_{\rm m,0}/a^{3}$. The
scale factor increases
algebraically with time as $t^{2/3}$ and the density decreases as $t^{-2}$. The
universe
is decelerating. We thus have a
transition between two phases of decelerating
expansion (radiation and matter). The transition takes place at
$a\simeq a_{\rm eq}$ with
\begin{equation}
a_{\rm eq}=\frac{\Omega_{\rm rad,0}}{\Omega_{\rm m,0}}=3.00\times 10^{-4}.
\label{aeq}
\end{equation}
We have taken $\Omega_{\rm rad,0}=9.23765\times
10^{-5}$ and $\Omega_{\rm m,0}=0.3075$. This transition between the radiation
era and the matter era is
studied
in detail in \cite{stiff}. The
temporal evolution of the scale factor  is
given analytically (in reversed form)
by
\begin{eqnarray}
H_0 t=-\frac{2}{3}\frac{1}{(\Omega_{\rm m,0})^{1/2}}\left
(\frac{2\Omega_{\rm rad,0}}{\Omega_{\rm m,0}}-a\right
)\sqrt{\frac{\Omega_{\rm rad,0}}{\Omega_{\rm m,0}}+a}
+\frac{4}{3}\frac{(\Omega_{\rm rad,0})^{3/2}}{(\Omega_{\rm m,0})^2}.
\label{radd1}
\end{eqnarray}
It can also be written as
\begin{equation}
a^3-3\frac{\Omega_{\rm rad,0}}{\Omega_{\rm m,0}}a^2=\frac{9}{4}\Omega_{\rm
m,0}H_0^2t^2-6\frac{\Omega_{\rm rad,0}^{3/2}}{\Omega_{\rm m,0} }
H_0 t.
\label{radd2}
\end{equation}
This is a cubic equation for $a$ which can be solved by
standard methods. However, in order to plot the curve $a(t)$, it is
simpler to compute $t(a)$ and represent $a$ versus $t$. In the radiation era we
recover Eq. (\ref{nesf8c}) and in the matter era we recover Eq. (\ref{gr21b})
given below.

\subsection{Late universe: matter + dark energy ($\Lambda$CDM)}

In the matter + dark energy era (late universe), the energy density is given by
\begin{equation}
\rho=\frac{\rho_{\rm
m,0}}{a^3}+\rho_{\Lambda}.
\label{gr18}
\end{equation}
If we normalize the density by its present value, we get
\begin{equation}
\label{gr19c}
\frac{\rho}{\rho_0}=\frac{\Omega_{\rm m,0}}{a^{3}
} +\Omega_{\Lambda,0}.
\end{equation}
We recover the $\Lambda$CDM model. We note
that, in the present model, dark energy comes
from the generalized
radiation associated with nonlinear electrodynamics. It arises from the
constant term in the Lagrangian (\ref{gr8}) taking into account the
electromagnetic
energy of point zero. Introducing $a'_2=(\Omega_{\rm
m,0}/\Omega_{\Lambda,0})^{1/3}=0.7634$ (we have taken
$\Omega_{\rm m,0}=0.3075$ and 
$\Omega_{\Lambda,0}=0.6911$), the relation between the energy density
and the scale factor can be rewritten as
\begin{equation}
\label{gr20}
\frac{\rho}{\rho_\Lambda}=\frac{1}{(a/a'_2)^{3}}+1.
\end{equation}
When $a_{\rm eq}\ll a\ll
a'_{2}$, we are in the matter era (Einstein-de Sitter)
with a vanishing
pressure $P_{\rm
m}=0$. The density decreases algebraically
as $\rho=\rho_{\rm m,0}/a^{3}$. The
scale factor increases
algebraically with time  as $t^{2/3}$ and the density decreases as $t^{-2}$. The
universe
is decelerating. When $a\gg a'_2$,
we are in the dark energy era. The density is approximately 
constant $\rho\simeq \rho_\Lambda$, equal to the cosmological density. This
leads
to a phase of late accelerating expansion (or late inflation) where the scale
factor increases exponentially rapidly with time as
$e^{(8\pi/3)^{1/2}t/t_\Lambda}$ (late de Sitter era). The universe is
accelerating. We thus have a
transition between a phase of decelerating expansion in the matter era
(Einstein-de Sitter) and
a phase of accelerating
expansion (vacuum energy/de Sitter) in the late universe. The transition takes
place at
$a\simeq a'_2$. This transition between the matter era and the dark energy era
is studied
in detail in \cite{cosmopoly2,universe}. 
The
temporal evolution of the scale factor $a(t)$ and density $\rho(t)$ is
given analytically
by
\begin{equation}
\frac{a}{a'_2}=\sinh^{{2}/{3}}\left\lbrack
\frac{3}{2}\left (\frac{8\pi}{3}\right
)^{1/2}\frac{t}{t_{\Lambda}}\right\rbrack,\qquad 
\frac{\rho}{\rho_{\Lambda}}=\frac{1}{\tanh^{2}\left\lbrack
\frac{3}{2}\left (\frac{8\pi}{3}\right
)^{1/2}\frac{t}{t_{\Lambda}}\right\rbrack},
\label{gr21}
\end{equation}
with $t_{\Lambda}=1/\sqrt{G\rho_{\Lambda}}=(8\pi/\Lambda c^2)^{1/2}=1.59\times
10^{18}\, {\rm s}$ (cosmological time).  It can also be written
as
\begin{equation}
a=\left(\frac{\Omega_{m,0}}{\Omega_{\Lambda,0}}\right )^{1/3}\sinh^{{2}/{3}}
\left (\frac{3}{2}\sqrt{\Omega_{\Lambda,0}}H_0 t\right ),\qquad 
\frac{\rho}{\rho_{0}}=\frac{\Omega_{\Lambda,0}}{\tanh^{2}\left
(\frac{3}{2}\sqrt{\Omega_{\Lambda,0}}H_0 t\right )}.
\label{gr21by}
\end{equation}
In the matter era ($a_{\rm eq}\ll a\ll a'_{2}$):
\begin{equation}
a\sim  \left
(\frac{3}{2}\sqrt{\Omega_{m,0}}H_0 t\right
)^{2/3},\qquad \frac{\rho}{\rho_{0}}\sim \frac{1}{\left
(\frac{3}{2}H_0 t\right )^2}.
\label{gr21b}
\end{equation}
In the dark energy era ($a\gg a'_2$):
\begin{equation}
a\sim \left(\frac{\Omega_{m,0}}{4\Omega_{\Lambda,0}}\right
)^{1/3}e^{\sqrt{\Omega_{\Lambda,0}}H_0 t},\qquad 
\rho\simeq \rho_{\Lambda}.
\label{gr21d}
\end{equation}

\subsection{Numerical applications}
\label{sec_nv}

\begin{figure}[!h]
\begin{center}
\includegraphics[clip,scale=0.3]{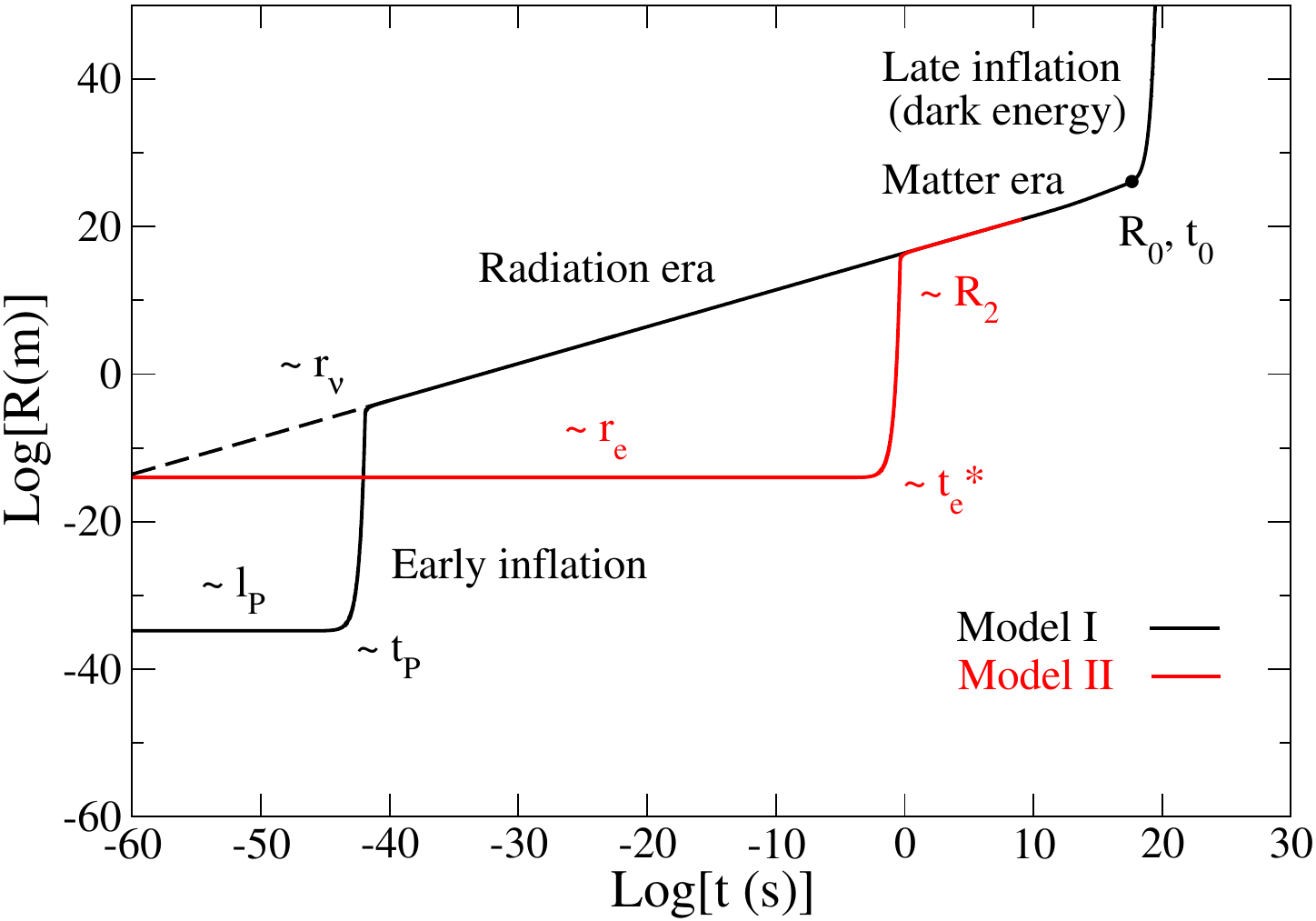}
\caption{Temporal evolution of the radius of the
universe  in logarithmic scales in Model I (black) and Model II (red). 
The universe exists at all times in the past and
in the future.  There is no singularity (aioniotic universe). The early universe
undergoes a phase of exponential inflation. During the early
inflation, the scale factor increases by $\sim 30$
orders of magnitude in $\sim 10^{-42}\, {\rm s}$ (Model I) or in $\sim
1\, {\rm s}$ (Model II). This is followed by the
radiation era ($a\propto t^{1/2}$), by the matter era ($a\propto t^{2/3}$), and
by the dark energy era responsible for the accelerated expansion of the universe
observed at present. The universe exhibits two types of inflation: An early
inflation corresponding to the Planck density $\rho_P$ (Model I) or to the
electron density (Model II) and a late inflation corresponding to the
cosmological density  $\rho_{\Lambda}$ (dark energy or cosmological
constant). The evolution of the early and late universe is
remarkably symmetric. In our model it is described by two polytropic equations
of state with index $n=1$ and $n=-1$, respectively.  The
dashed line corresponds to the $\Lambda$CDM model which presents a
big bang singularity at $t=0$. We have also represented the location of the
present universe
that is just at the transition between the matter era and the dark energy era
(cosmic coincidence).}
\label{taLOGLOGcomplet}
\end{center}
\end{figure}

\begin{figure}[!h]
\begin{center}
\includegraphics[clip,scale=0.3]{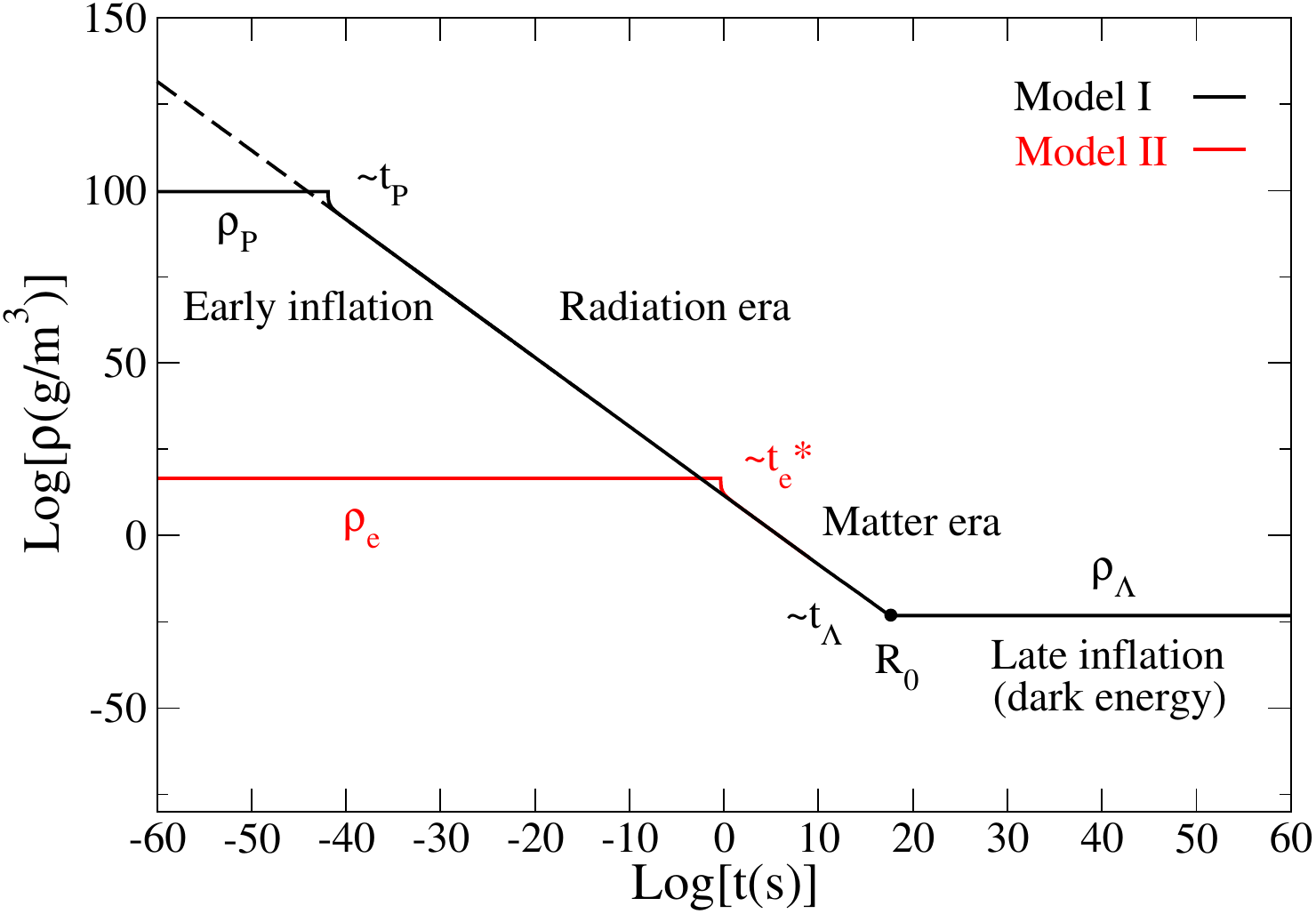}
\caption{Temporal evolution of the density of the
universe  in logarithmic scales in Model I (black) and Model II (red). The
density goes from a maximum value $\rho_{\rm max}$ equal to the Planck density
$\rho_P$ (Model I) or to the electron density $\rho_e$ (Model II) to a
minimum value $\rho_{\rm min}=\rho_{\Lambda}$ equal to the cosmological density.
These two bounds, which are fixed by fundamental
constants of physics, are responsible for the early and late inflation of the
universe. In between, the density decreases as $t^{-2}$. The
dashed line corresponds to the $\Lambda$CDM model with a
big bang singularity at $t=0$.}
\label{trhoLOGLOGcomplet}
\end{center}
\end{figure}

In order to describe quantitatively the
physical evolution of the universe from the early inflation to the late
acceleration that we observe today we use the ``radius of the universe''
$R(t)=a(t)R_{\Lambda}$ with
$R_{\Lambda}=ct_\Lambda=(8\pi/\Lambda)^{1/2}=4.77\times 10^{26}\, {\rm
m}$ defined in Appendix \ref{sec_massuniv}. We also use the
results of Appendix \ref{sec_intre}. We consider the two
models of
Sec. \ref{sec_kcdur} which differ from each other only in the early universe.
The temporal evolution of the radius and density of the universe in the two
models are represented in Figs. \ref{taLOGLOGcomplet} and
\ref{trhoLOGLOGcomplet}.

{\it Model I:} In the first model, the density of the primordial universe
(maximum density) is equal to the Planck density $\rho_I=\rho_P=5.16\times
10^{99}\, {\rm g\, m^{-3}}$. In that model, the
universe starts at $t=0$ (begining of the
inflation)  with a size equal to the Planck length $R(t=0)=l_P=1.62\times
10^{-35}\, {\rm m}$ (i.e.
$a(t=0)=3.39\times 10^{-62}$) and reaches a
size $R_c=R_1\sim \lambda_\Lambda^*=3.91\times 10^{-5}\, {\rm m}$ of the order
of the neutrino's Compton wavelength (i.e.
$a_1=a_c=1.98\times 10^{-32}$)  at
the end of the inflation  which occurs on a timescale $t_1=t_c=23.8\, t_P$ of
the order
of a few Planck times $t_P=5.39\times 10^{-44}\, {\rm
s}$. The $e$-folding number is $N_0=68.5$ (the size of the universe
increases by a factor $1/\epsilon\sim 10^{30}$ during the inflation). At
$t=0$ the density and the  magnetic field are $\rho(t=0)\simeq
\rho_P$ and $B(t=0)=3.69\times
10^{113}\, {\rm T}$. At $t=t_1=t_c$ the density and the  magnetic
field are $\rho_1=\rho_c=\rho_P/2$ and $B_1=B_c=1.08\times 10^{54}\, {\rm
T}$.\footnote{We have used Eqs. (\ref{bdim}), (\ref{tsw1a}) and (\ref{tsw1b}) to
compute
the magnetic field.} We also note that the mass
of the universe at $t=0$ is equal to the Planck mass $M_P=\rho_P
l_P^3=2.18\times 10^{-5}\, {\rm g}$
(the primordial universe has the same characteristics as a Planck black hole or
a planckion particle) while
its mass at the end of the inflation is $\rho_P R_c^3\sim
10^{90}M_P$. This suggests that $10^{90}$ particles of mass
$M_P$ have been created during the inflation, implying   that, after the
radiation era, the mass of the
universe is equal to
$10^{62}M_P$ (see
Appendix \ref{sec_massuniv}).

{\it Model II:} In the second model, the density of the primordial universe
(maximum density) is 
equal to the electron density $\rho_I=\rho_e=4.07\times 10^{16}\, {\rm g\,
m^{-3}}$.  This value is justified by applying the
nonlinear electrodynamics with the Lagrangian from Eq. (\ref{kc1}) to the
electron (see Sec. \ref{sec_nee}). In that model, the universe starts at $t=0$
(begining of the
inflation)  with a size of the order of the electron radius
$R(t=0)\sim r_e=2.82\times 10^{-15}\, {\rm m}$ (i.e.
$a(t=0)\sim 5.91\times
10^{-42}$) and reaches a size $R_c=R_1\sim {\tilde
R}_2=7.07\times 10^{15}\,
{\rm m}$ of the order of the radius of a dark energy star of the stellar mass
(i.e. $a_1=a_c=1.18\times 10^{-11}$) at
the end of the inflation  which occurs on a timescale $t_1=t_c=23.8\, t_e^*$
of the order of a few gravitoelectronic times $t_e^*=0.0192\, {\rm s}$.
The $e$-folding number is $N_0=68.5$ (the size of the universe
increases by a factor $1/\epsilon\sim 10^{30}$ during the inflation). At
$t=0$ the density and the  magnetic field are $\rho(t=0)\simeq
\rho_e$ and  $B(t=0)=1.04\times
10^{72}\, {\rm T}$. At $t_1=t_c$ the density and the  magnetic
field are $\rho_1=\rho_c=\rho_e/2$ and $B_c=3.03\times 10^{12}\, {\rm
T}$ (see footnote 14).  We also note that the mass
of the universe at $t=0$ is equal to the electron mass $m_e=\rho_e
r_e^3=9.11\times 10^{-28}\, {\rm
g}$
(the
primordial universe has the same characteristics as an electron) while
its mass at the end of the inflation is $\rho_e R_c^3\sim
10^{90}m_e$. This suggests that $10^{90}$ particles of mass
$m_e$ have been created during the inflation, implying that, after the radiation
era, the universe is made of $10^{83}$ electrons or $10^{80}$ protons, which is
Eddington's number (see Appendix \ref{sec_massuniv}).

After the inflation, the evolution of the universe is the same in the two
models. The universe undergoes a radiation era, then enters in the matter era at
$t_{\rm eq}=5.25\times 10^4\, {\rm yrs}$ (i.e.
$a_{\rm eq}=3.00\times
10^{-4}$) and in the dark energy era at $t_2'=10.2\, {\rm Gyrs}$ (i.e.
$a'_{2}=0.7634$). The  age of the universe is $t_0=0.956\,
H_0^{-1}=13.8\, {\rm Gyrs}$ (i.e. $a_0=1$). Its size and density  are
$R_0=R_{\Lambda}=4.77\times 10^{26}\, {\rm m}$ and $\rho_0=8.62\times
10^{-24}{\rm g\, m^{-3}}$. The present magnetic field is $B_0=4.23\times
10^{-10}\, {\rm T}$ (see Sec. \ref{sec_magnu}).

\section{Nonlinear electrodynamics based on the Lagrangian
${\cal L}=-{\cal F}/(1+{\cal F}/{\cal F}_*)$: Electrostatics}
\label{sec_nee}

In this section, we apply the nonlinear electrodynamics considered previously to
the electron in the spirit of the Born-Infeld \cite{born1933,borninfeld} 
theory. A similar approach has been developed by Kruglov
\cite{kruglov2015a}. We consider a
purely electrostatic situation. We compute the electric field created by a
pointlike charge $\sqrt{4\pi}e$  and determine the total electric  energy that
it carries. We then identify this electric energy with the mass-energy of the
electron and obtain an estimate of its classical radius and density. 

We consider a nonlinear electrodynamics based on the Lagrangian
\begin{equation}
{\cal L}=\frac{-{\cal F}}{1+
\frac{{\cal F}}{{\cal F}_*}}. 
\label{kc1b}
\end{equation}
This is the Lagrangian (\ref{gr8}) of the generalized radiation without
the constant term (vacuum energy) that yields an infinite total energy. We
leave the constant ${\cal
F}_*=\rho_* c^2$ undetermined for the moment. It will be determined at the end
by applying this model to the electron. The linear (Maxwell)
electrodynamics is recovered in the limit ${\cal
F}_*\rightarrow +\infty$. According to Eqs. (\ref{re1}), (\ref{re2}) and (\ref{kc1b}), the general expressions of the energy density and pressure are
\begin{equation}
\rho c^2=\frac{{\cal F}}{1+
\frac{{\cal F}}{{\cal F}_*}}+\frac{E^2}{\left (1+
\frac{{\cal F}}{{\cal F}_*}\right )^2},
\label{last1}
\end{equation}
\begin{equation}
P=-\frac{{\cal F}}{1+
\frac{{\cal F}}{{\cal F}_*}}-\frac{E^2-2B^2}{3}\frac{1}{\left (1+
\frac{{\cal F}}{{\cal F}_*}\right )^2}.
\label{last2}
\end{equation}

\subsection{Electric field}

\begin{figure}[!h]
\begin{center}
\includegraphics[clip,scale=0.3]{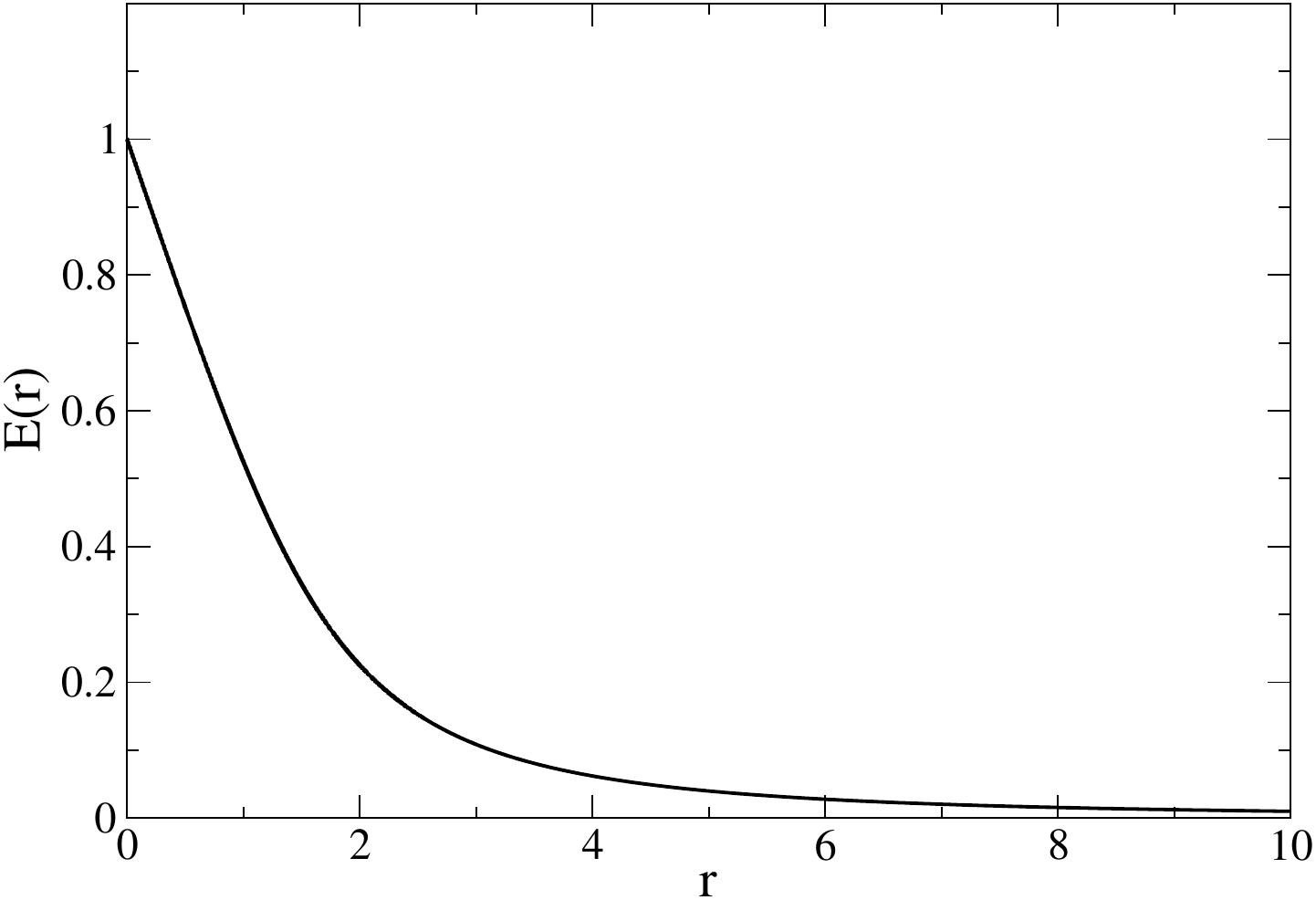}
\caption{Electric field $E$
(normalized by $\sqrt{2{\cal F}_*}$) as a function
of the distance $r$ (normalized by $r_*$).}
\label{xyC}
\end{center}
\end{figure}

In electrostatics, using Eq.
(\ref{el3}), the Lagrangian (\ref{kc1b}) takes the form
\begin{equation}
{\cal L}=\frac{E^2}{2\left (1-
\frac{E^2}{2{\cal F}_*}\right )}. 
\label{kc6}
\end{equation}
According to Eqs. (\ref{el8}) and Eq.
(\ref{kc1b}) the electric field created by a pointlike charge $\sqrt{4\pi}e$ is
given
by 
\begin{equation}
\frac{E(r)}{\left \lbrack 1-\frac{E^2(r)}{2{\cal
F}_*}\right\rbrack^2}=\frac{e}{\sqrt{4\pi}r^2}.
\label{kc7}
\end{equation}
In this nonlinear electrodynamics, the electric field is finite at the
origin $r=0$ (unlike in Maxwell's electrodynamics) with the maximum value
$E(0)=E_*=\sqrt{2{\cal F}_*}$. It decreases
monotonically with the distance. At small distances ($r\ll r_*$) it decreases as
$E(r)/\sqrt{2{\cal F}_*}=1-r/2r_*+...$,\footnote{By contrast, in the Born-Infeld
\cite{born1933,borninfeld} model the electric field decreases as
$E(r)/\sqrt{2{\cal F}_*}=1-\frac{1}{2}(r/r_*)^4+...$.}
 where 
\begin{equation}
r_*=\left (\frac{e}{\sqrt{8\pi{\cal
F}_*}}\right )^{1/2}
\label{bi18}
\end{equation}
is a characteristic radius determined by the finite
value of ${\cal F_*}$ in the nonlinear electrodynamics based on Eq.
(\ref{kc1b}). It can be interpreted as the radius of the electron in this
model. It reduces to zero in Maxwell's
electrodynanics ${\cal
F_*}\rightarrow +\infty$. At
large distances ($r\gg r_*$)  we recover
the
Coulomb law $E(r)\sim e/(\sqrt{4\pi}r^2)$. The electric field
is plotted as a function of
$r$ in Fig. \ref{xyC}.

{\it Remark:} Conversely, Eq. (\ref{bi18}) can be written as
\begin{equation}
{\cal F}_*=\rho_* c^2=\frac{e^2}{8\pi r_*^4}.
\label{nap}
\end{equation}

{\it Remark:} Let us call $\rho_e^*$ the extended charge density in the usual
(Maxwell) electrodynamics that produces the same electric field as above. It
can be determined by the usual Gauss law
\begin{equation}
\nabla\cdot {\bf E}=\rho_e^*,
\label{gauss1}
\end{equation}
where $E(r)$ is given by Eq. (\ref{kc7}). Using the Gauss theorem, we have
$E(r)=Q(r)/(4\pi r^2)$ where $Q(r)=\int_0^r \rho_e^*(r') 4\pi {r'}^2\, dr'$ is
the charge contained within the sphere of radius $r$. Since $E(r)\sim
e/(\sqrt{4\pi}r^2)$ for $r\rightarrow +\infty$, we have $Q=\lim_{r\rightarrow
+\infty}E(r)4\pi r^2=\sqrt{4\pi}e$. Therefore, the total charge $Q$ associated
with the extended distribution $\rho_e^*$ in linear electrodynamics coincides
with the charge $\sqrt{4\pi}e$ of the singular point-charge in nonlinear
electrodynamics. This is a general result that was first made in
connection to the Born-Infeld model \cite{born1933,borninfeld}.

\subsection{Energy density and pressure}

\begin{figure}[!h]
\begin{center}
\includegraphics[clip,scale=0.3]{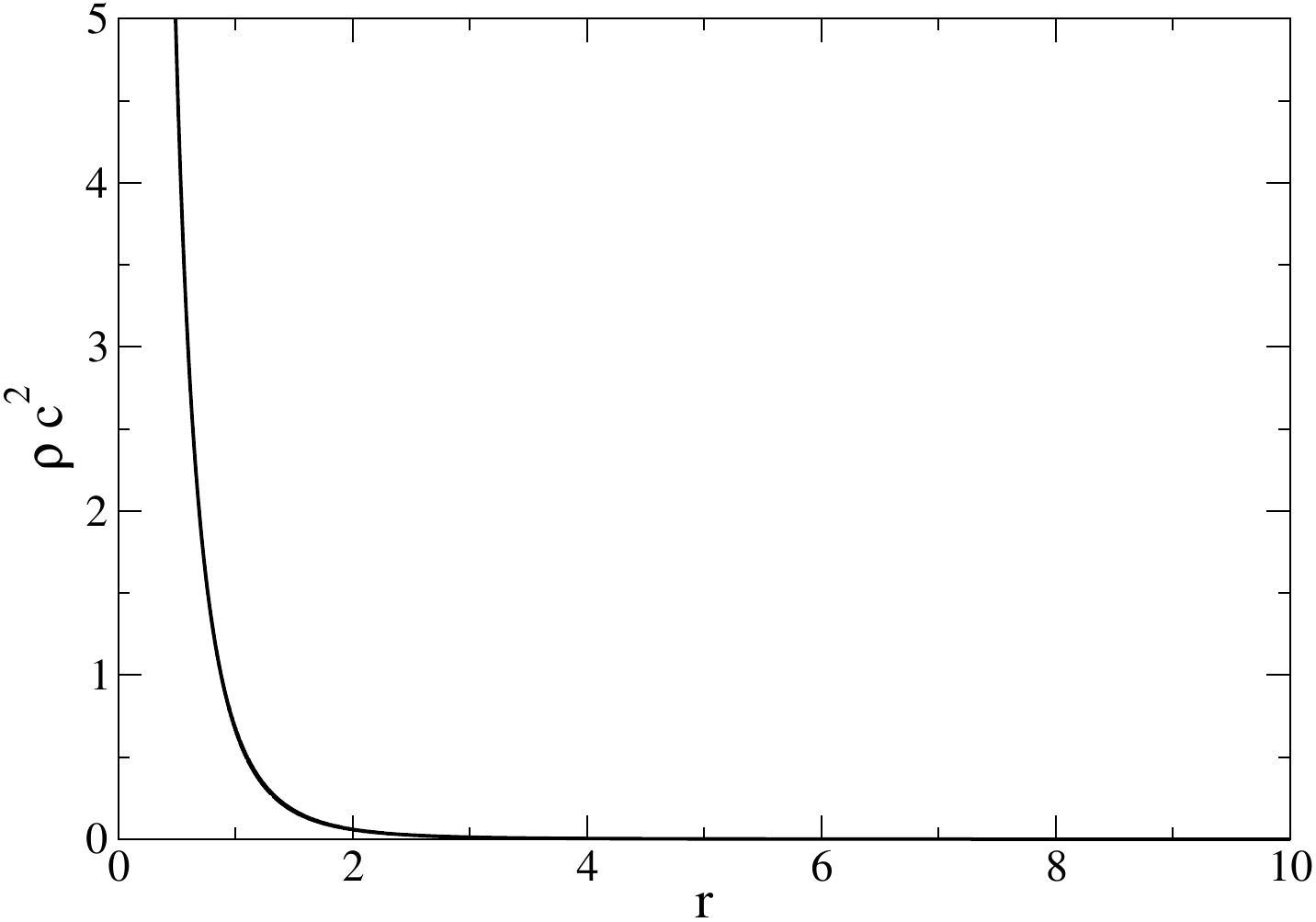}
\includegraphics[clip,scale=0.3]{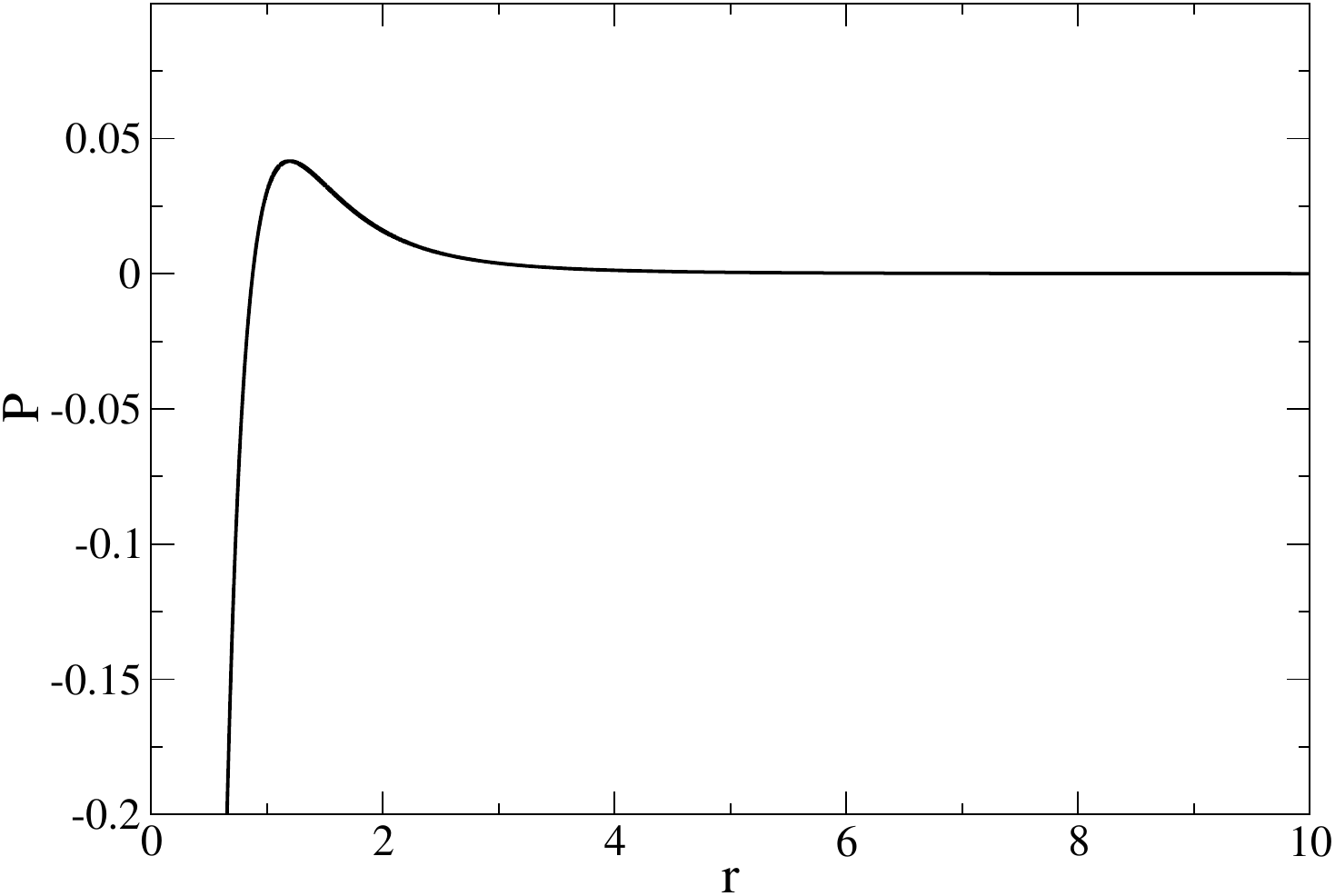}
\caption{Energy density $\rho c^2$
(normalized by $\rho_* c^2$) and pressure $P$
(normalized by $\rho_* c^2$) as a function
of the distance $r$ (normalized by $r_*$). }
\label{xrC}
\end{center}
\end{figure}

The energy density and the pressure are given by Eqs. (\ref{el9}) and
(\ref{el10}) with Eq. (\ref{kc1b})
yielding
\begin{equation}
\rho c^2=\frac{E^2(r)\left \lbrack 1+\frac{E^2(r)}{2{\cal F}_*}\right
\rbrack}{2\left
\lbrack 1-\frac{E^2(r)}{2{\cal F}_*}\right \rbrack^2},
\label{kc8}
\end{equation}
\begin{equation}
P=\frac{E^2(r)\left \lbrack 1-\frac{3E^2(r)}{2{\cal F}_*}\right
\rbrack}{6\left
\lbrack 1-\frac{E^2(r)}{2{\cal F}_*}\right \rbrack^2}.
\label{kc8b}
\end{equation}
The energy density and the pressure diverge at the
origin $r=0$ as $\rho/\rho_*\sim 2(r_*/r)^{2}$ and $P/\rho_* c^2\sim -
(2/3)(r_*/r)^{2}$ respectively (this is the same behavior as
in the
Born-Infeld \cite{born1933,borninfeld} model). The energy density
starts from $+\infty$ at  $r=0$ and decreases to $0^+$ as $\rho/\rho_*\sim
(r_*/r)^{4}$  for $r\rightarrow
+\infty$. The pressure starts from $-\infty$ at  $r=0$, increases, vanishes at
$r_0/r_*=2/3^{3/4}$ (corresponding to $E_0/\sqrt{2{\cal F}_*}=1/\sqrt{3}$ and
$\rho_0/\rho_*=1$),
becomes positive,
reaches a maximum $P_{\rm max}/\rho_*c^2=1/24$ at $r_m/r_*=4/5^{3/4}$
(corresponding to $E_m/\sqrt{2{\cal F}_*}=1/\sqrt{5}$ and $\rho_m/\rho_*=3/8$)
and decreases  to
$0^+$ as $P/\rho_* c^2\sim 
(1/3)(r_*/r)^{4}$  for $r\rightarrow
+\infty$.  The energy density
and the pressure are plotted as a function of $r$ in Fig.
\ref{xrC}.

\subsection{Equation of state}

\begin{figure}[!h]
\begin{center}
\includegraphics[clip,scale=0.3]{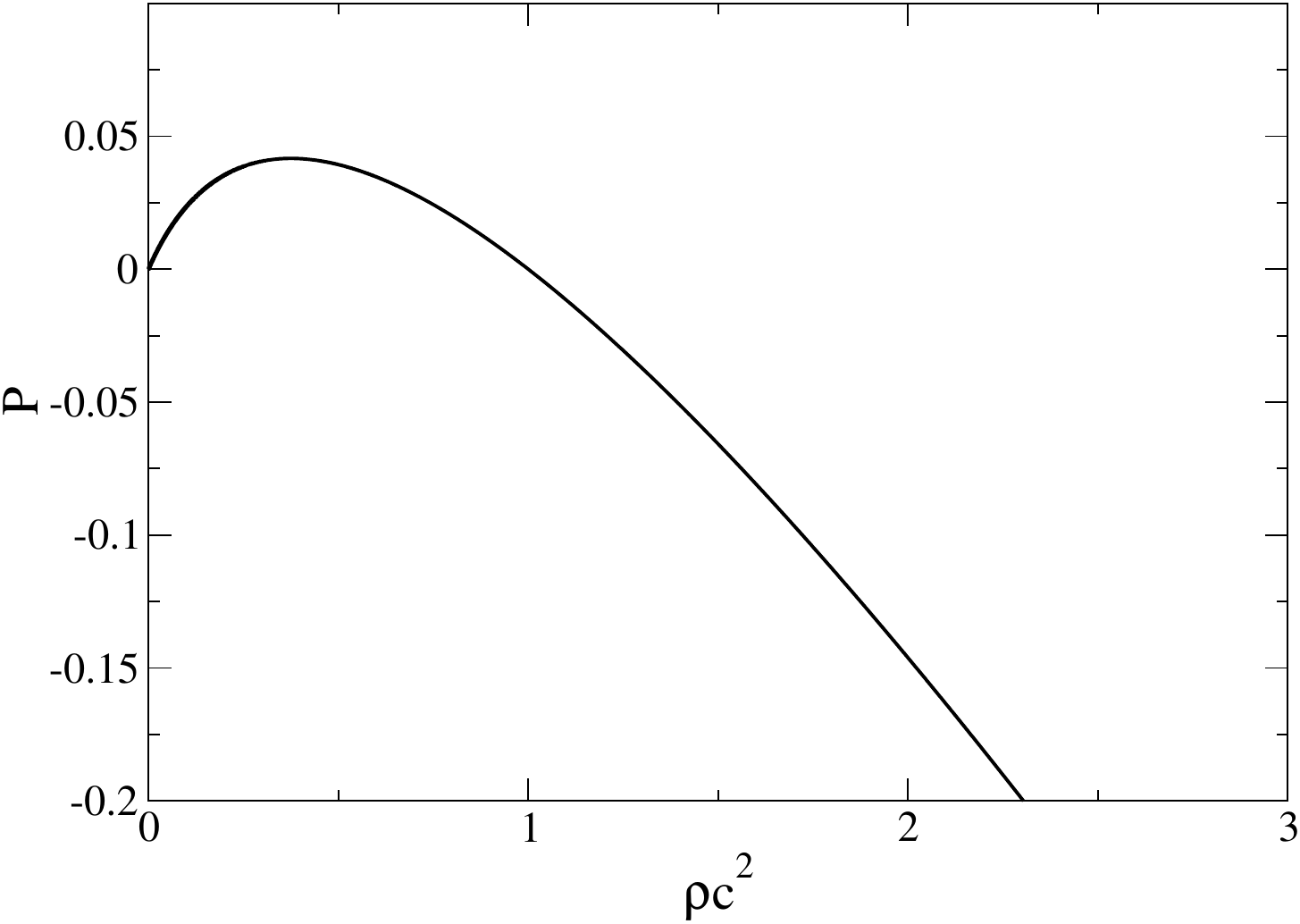}
\caption{Pressure $P$ (normalized
by $\rho_* c^2$) as a function of the
energy density $\rho$ (normalized by $\rho_*$). This is the electrostatic
equation of state associated with the Lagrangian (\ref{kc1b}).}
\label{rpC}
\end{center}
\end{figure}

Eliminating the electric field $E(r)$ between Eqs. (\ref{kc8}) and
(\ref{kc8b})
we obtain the equation
of state 
\begin{equation}
\frac{P}{\rho_* c^2}=\frac{1}{6}\left (-1-2\frac{\rho}{\rho_*}+\sqrt{1+8
\frac{\rho}{\rho_*}}\right ).
\label{nkc11a}
\end{equation}
For small energy densities $\rho c^2\rightarrow 0$ (large
distances) we recover
the usual equation of state of the radiation
\begin{equation}
P\sim \frac{1}{3}\rho c^2,
\label{nkc11b}
\end{equation}
corresponding to Maxwell's linear electrodynamics. For large energy densities
$\rho c^2\rightarrow +\infty$ (short distances) we obtain the equation of state
\begin{equation}
P\sim -\frac{1}{3}\rho c^2.
\label{nkc11c}
\end{equation}
This is a linear equation of state of the form $P=\alpha\rho c^2$ with
$\alpha=-1/3$. The same equation of state is obtained in the Born-Infeld
\cite{born1933,borninfeld}  model
at high densities (both for the
electric field of the electron and for the magnetic
universe).\footnote{Interestingly, this
equation of state occurs in
cosmology in relation to the Milne model of the universe \cite{milne}. This is
also the equation of state of a gas of cosmic string. Therefore, it may
describe a cosmic stringlike era (see, e.g., \cite{becas}).}
We note that the pressure in the core ($r<r_0$) is negative. If we view the
electron as an extended charge of typical radius $r_*$ (see below), a negative
pressure is
necessary to counteract
the ordinary electrostatic repulsion and ensure its cohesion. This is similar to
the Poincar\'e stress
\cite{poincare1905,poincareE} introduced in the
Abraham-Lorentz \cite{abraham,lorentz} electromagnetic model of the
electron to stabilize the particle (see Appendix F of \cite{massmaxrel}).
The equation of state (\ref{nkc11a}) is plotted in Fig. \ref{rpC}. We note
that this ``electric'' equation of state differs from the ``magnetic''
equation of state (\ref{gen25b}) even though the electromagnetic Lagrangian is
the same.

\subsection{Classical radius of the electron and value of ${\cal F}_*=\rho_*
c^2$}

The electric energy density $\rho(r)c^2$ diverges at the
origin $r=0$ as $r^{-2}$. However, the total electric energy
\begin{equation}
{\cal E}=\int_0^{+\infty} \rho(r) c^2 4\pi r^2\,
dr 
\label{bi16}
\end{equation}
is finite. This is like in the Born-Infeld \cite{born1933,borninfeld} model.
Following Born \cite{born1933} we
shall identify the electrostatic energy with the
mass of the
electron via the relation\footnote{This relation has a
long history in physics  even before Einstein's theory of relativity
(see Appendix F of \cite{massmaxrel}). It appeared in relation to the 
Abraham-Lorentz model
of the electron where it was believed that the mass of a particle had an
electromagnetic origin \cite{thomson1881}. The Born-Infeld
\cite{born1933,borninfeld}  theory of the
electron can be considered as a revival of the old
idea of the electromagnetic origin of mass; namely, that the
electron is a singularity in the electromagnetic field and
that its mass is purely electromagnetic.
By contrast, it is not possible
to identify ${\cal E}=m_e c^2$ in Maxwell's electrodynamics because the
electromagnetic energy ${\cal E}$ is infinite whereas $m_e$ is finite.}
\begin{equation}
{\cal E}=m_e c^2.
\label{bi17}
\end{equation}
This relation gives an electromagnetic origin to the mass of the electron and
allows one to determine
${\cal F}_*$ (or $\rho_*$ or $r_*$).
Using Eq. (\ref{kc7}), (\ref{bi18}), (\ref{kc8}) and (\ref{bi16}) 
we find that the electrostatic
energy
of the electron is
\begin{equation}
{\cal E}=\frac{e^2}{r_*} \int_0^1
\frac{(1+y^2)(3y^2+1)}{4\sqrt{y}}\,
dy=\frac{16}{15}\, \frac{e^2}{r_*},
\label{kc13}
\end{equation}
where $y=E/\sqrt{2{\cal F}_*}$.
Together with Eq. (\ref{bi17}) this gives 
\begin{equation}
m_e c^2=\frac{16}{15}\, \frac{e^2}{r_*}.
\label{nbi25}
\end{equation}
Comparing Eq. (\ref{nbi25}) with Eq. (\ref{mee1}) we get 
\begin{equation}
r_*=\frac{16}{15} r_e=3.00\times 10^{-15}\, {\rm
m},
\label{bi27}
\end{equation}
which may be interpreted as the electron radius
in the present model (it turns out that $r_*$ is very close to $r_e$ since
$16/15\simeq 1.07$). In this
sense, the present model (similarly to the
Born-Infeld \cite{born1933,borninfeld} 
model) {\it justifies} the relation from
Eq. (\ref{mee1}) defining the classical radius of the electron. 
In the present model (as in the Born-Infeld \cite{born1933,borninfeld} 
electrodynamics), the electron has a finite effective radius because the
electric field
and
the electrostatic energy of a point charge are finite. This is basically due to
the finite
value
of ${\cal F}_*$ in the Lagrangian.
According to Eqs.
(\ref{mee1}), (\ref{nap}) and
(\ref{bi27}) we have
\begin{equation}
{\cal F}_*=\frac{e^2}{8\pi
r_*^4}=\frac{1}{8\pi(\frac{16}{15})^4}\frac{e^2}{r_e^4}=\frac{1}{8\pi(\frac{
16}{15})^4}\frac{m_e c^2}{r_e^3}.
\label{bi28}
\end{equation}
Using Eqs. (\ref{gen23}) and (\ref{mee4}), this
gives
\begin{equation}
\rho_*=\frac{1}{8\pi(\frac{16}{15})^4}\frac{m_e}{r_e^3}
=\frac{1}{8\pi(\frac{16}{15})^4}\rho_e=1.25\times 10^{15}\, {\rm g\,
m^{-3}}.
\label{bi29}
\end{equation}
Therefore, as could have been
expected, we find that $\rho_*$ is of the order of the density of the electron
$\rho_e$ (up to a factor $0.0307$). This is much smaller than the Planck density
$\rho_P$. These densities differ by $80$ orders of magnitude (see Appendix
\ref{sec_ele}).

{\it Remark:} We must, however, point out a difficulty with this
model. If we assume that ${\cal F}_*$ is a universal constant and if we apply
the same argument to the proton (which has a charge opposite to that of the
electron) we would find the same
radius and the same mass as the electron, which is obviously incorrect. This
suggests that the mass of the proton has not an electromagnetic origin, even in
a context of nonlinear electrodynamics.

\subsection{Heisenberg-Euler Lagrangian}
\label{sec_hea}

Following Kruglov \cite{kruglov2020} we can compare the
previous Lagrangian model at the weak field limit 
with the  Heisenberg-Euler \cite{he} Lagrangian, which is the QED Lagrangian
with one loop correction. The
Heisenberg-Euler Lagrangian reads
\begin{equation}
{\cal L}_{\rm HE}\simeq -{\cal F}+\lambda{\cal F}^2+...
\label{bi4}
\end{equation}
with
\begin{equation}
\lambda=\frac{8\alpha^2}{45}\frac{\hbar^3}{m_e^4
c^5},
\label{bi5}
\end{equation}
where $\alpha=e^2/\hbar c$ is the fine-structure constant.  When ${{\cal
F}}/{{\cal
F}_*}\ll 1$, we can expand Eq. (\ref{kc1b}) to first order and we obtain
\begin{equation}
{\cal L}\simeq -{\cal F}+\frac{{\cal F}^2}{{\cal F}_*}.
\label{kc3}
\end{equation}
Comparing this expression with the Heisenberg-Euler Lagrangian (\ref{bi4})
we get
\begin{equation}
{\cal F}_*=\frac{45}{8\alpha^2}\frac{m_e^4
c^5}{\hbar^3}.
\label{kc3b}
\end{equation}
Recalling that ${\cal F}_{*}=\rho_* c^2$ [see Eq. (\ref{gen23})] we find that
\begin{equation}
\rho_*=\frac{45}{8\alpha^2}\frac{m_e^4
c^3}{\hbar^3}=\frac{45}{8}\alpha\rho_e=1.67\times 10^{15}\, {\rm
g\, m^{-3}},
\label{kc3c}
\end{equation}
where $\rho_e$ is the electron density from Eq. (\ref{bi10}).
Therefore, this argument confirms that 
$\rho_*$ is of the order of the electronic
density (up to a factor $0.0410$). The
comparison between Eqs. (\ref{bi29}) and (\ref{kc3c})
gives $\alpha=1/[45\pi(16/15)^4]=1/183$.

\subsection{Fundamental length}

In the present model of nonlinear electrodynamics, the electric field created by
a
point like charge (electron) is nonsingular at the origin and the electrostatic
energy is finite even though the central energy density diverges. On the other
hand, the mass of a charged particle like the electron has a purely
electromagnetic nature as in
the Abraham-Lorentz model \cite{abraham,lorentz}. If we identify the
electromagnetic energy
${\cal E}$ with the mass-energy $m_e c^2$  of the electron we can define the
classical radius $r_e$ of the electron and its density $\rho_e$.

These results are similar to the Born-Infeld \cite{born1933,borninfeld}
 electrodynamics and very
different
from Maxwell's electrodynamics where the electric field  created by a
point like charge is singular at the origin and the electrostatic energy is
infinite. In that case, the electron has a vanishing radius and an infinite
density. Therefore, its mass (which is finite) cannot have an electromagnetic
origin.

The regularization of the divergences (infinities) is due to the finite value of
${\cal F}_*$ which plays a role similar to the speed of light $c$ in the theory
of relativity (this analogy is at the basis of the Born-Infeld
\cite{born1933,borninfeld} model who
adopted a Lagrangian of nonlinear electrodynamics similar to the Lagrangian of a
relativistic particle). The
finiteness of
${\cal F}_*$ prevents the electric field to be larger than
$E(0)=E_*=\sqrt{2{\cal F}_*}$. It gives an upper bound on the possible
electric field. Similarly, the finiteness of the speed of light $c$
imposes $v<c$ in
relativistic mechanics.  When ${\cal F}_*\rightarrow +\infty$ we recover
Maxwell's electrodynamics. Similarly, when $c\rightarrow+\infty$, we recover
Newton's mechanics.

If we assume that the nonlinear Lagrangian (\ref{kc1b}) applies both to the
magnetic
universe like in Sec. \ref{sec_kc} and to the electron like in the present
section we
come to the conclusion that $\rho_I$ in Sec. \ref{sec_kc} should be
identified to the
electron density $\rho_e$, not to the Planck density $\rho_P$. This argument
selects Model II with respect to Model I in Sec. \ref{sec_kcdur}.
Therefore, in this model, the primordial density of the
universe (maximum density)  is equal to the
electron density, not to the Planck density. This implies (see Sec.
\ref{sec_kcdur}) that
the duration of the inflation ($\sim 1\, {\rm s}$) is much longer than  usually
believed ($\sim 10^{-42}\, {\rm s}$).

Following Kruglov \cite{kruglov2015a}, one can interprete
the quantity
$r_*\sim r_e$ as a
fundamental length due to quantum gravity effects. Indeed,  for strong
electromagnetic fields, nonlinear electrodynamics may arise from possible
quantum gravity corrections to linear electrodynamics. This is how a new
parameter $r_*$ with the dimension of a length is introduced in the
model [see Eq. (\ref{bi18})]. We can
substantiate this claim with the following arguments. We have already mentioned
that $r_e$ represents in Model II the initial size of the universe at time
$t=0$ (see Appendix \ref{sec_intre} and Sec. \ref{sec_nv}). On the other hand,
we will 
show in Sec. \ref{sec_heur} that the electron classical radius determines the
correct value of the cosmological constant (vacuum energy)
through
Zeldovich's second formula \cite{zeldovich,zeldovichA}. This justifies the
Eddington
formula relating the cosmological constant to the radius (or to the mass) of the
electron and the other fundamental constants of physics
\cite{eddington1931lambda,ouf}. The
electron radius can also be interpreted as a minimum scale in quantum gravity in
the sense of Karolyhazy \cite{karolyhazy}. All these arguments suggest that
$r_e$ should
be interpreted as a fundamental minimum length. Since this minimum length is
much larger than the Planck length ($r_e\gg l_P$) this shows that quantum
gravity is not a Planck scale phenomenon. This is very different
from the
results obtained in Model I where the initial size of the universe at $t=0$ is
the Planck length $l_P$.

\section{Heuristic connections between nonlinear
electrodynamics, quantum gravity, vacuum energy and dark energy}
\label{sec_heur}

In this section we relate the two cosmological models discussed in
Sec. \ref{sec_kcdur} to the two models of vacuum energy introduced by
Zeldovich \cite{zeldovich,zeldovichA} and to the two models of minimum length
introduced in quantum gravity by Karolyhazy \cite{karolyhazy} and
Amelino-Camelia \cite{gacnature}. We show
striking heuristic connections between these apparently
disconnected topics.

\subsection{Model I}

Lema\^itre \cite{lemaitre1934} was the first to understand
that the effect of the cosmological constant $\Lambda$ is
equivalent to that of a fluid with a constant density
$\rho_\Lambda=
\Lambda c^2/8\pi G$ described by an
equation of state $P=-\rho c^2$. He interpreted $\rho_{\Lambda}c^2$ as
the vacuum energy density. However, he did not connect his interpretation with
the
zero-point energy, nor relate it to quantum mechanics.
The origin of the vacuum energy was first discussed by Zeldovich
\cite{zeldovich,zeldovichA} and Sakharov \cite{sakharov}
in relation to quantum field theory. When one tries to compute the vacuum energy
density $\rho_\Lambda$  from
first principles,  one encounters a
severe problem of divergence at small scales (UV
divergence). However, following the seminal study of 
Zeldovich \cite{zeldovichA}, several procedures have been devised to deal with
these divergences (see \cite{martin} for a review). Zeldovich
\cite{zeldovich,zeldovichA} introduced two models of 
vacuum energy, each depending on a fundamental mass scale $m$.
In his first model, the regularized, finite, energy density of
the vacuum
which leads to the equation of state of
vacuum $P=-\rho c^2$ as a consequence of relativistic Lorentz
invariance reads\footnote{The energy density of vacuum
fluctuations
can be related to the 
Casimir effect \cite{casimir} and is given qualitatively by
$\rho_{\Lambda}c^2\sim (e^2/r)/r^3\sim \hbar c/r^4$ where $e^2/r$ is the
electrostatic energy and we have used
$e^2\sim \hbar c$ in order of magnitude \cite{ouf}.}
\begin{eqnarray}
\label{an15}
\rho_{\Lambda}c^2 \sim \frac{m^4c^5}{\hbar^3} \sim 
\frac{\hbar c}{\lambda_C^4},
\end{eqnarray}
where we have introduced the Compton wavelength $\lambda_C={\hbar}/{mc}$  of the
elementary particle of mass $m$. It represents the relevant small scale cut-off
$\lambda_{\rm min}$ in Zeldovich's model I. The expression
(\ref{an15}) of the vacuum energy density is commonly
adopted in the literature \cite{martin}. We note that this expression of the
vacuum energy density does not involve the gravitational constant $G$. This may
unveil a problem with this approach if vacuum energy arises from quantum
gravity (see below). In his original paper,
Zeldovich \cite{zeldovichA}  used Eq. (\ref{an15}) with the proton mass and
obtained a discrepency of $40$ orders of magnitude with the empirical
cosmological density. If we use the Planck mass and the
Planck length in Eq. (\ref{an15}), we find that $\rho_{\Lambda}\sim \rho_P$,
yielding a discrepency of $120$ orders of magnitude with the empirical value.
This is the usual
formulation of the cosmological
constant problem \cite{weinbergcosmo}. Alternatively, if we adopt the measured
value of the cosmological constant and reverse the relation from Eq.
(\ref{an15}) we get
\begin{equation}
m_{\Lambda}^*=\left (\frac{\Lambda\hbar^3}{Gc}\right )^{1/4}=
\sqrt{m_{\Lambda}M_P}=5.04\times 10^{-3}\, {\rm
eV/c^2},
\label{an16a}
\end{equation}
\begin{equation}
\lambda_{\Lambda}^*= \left (\frac{G\hbar}{\Lambda
c^3}\right )^{1/4}=
\sqrt{R_{\Lambda}^*l_P}=3.91\times 10^{-5}\, {\rm m},
\label{an16b}
\end{equation}
where $m_{\Lambda}=\hbar\sqrt{\Lambda}/c=2.08\times
10^{-33}\, {\rm eV/c^2}$ is the cosmon mass and
$R_\Lambda^*=1/\sqrt{\Lambda}=9.49\times 10^{25}\, {\rm m}$ is
the cosmological length \cite{ouf}. In Ref. \cite{ouf} we arrived at the mass
and length
scales (\ref{an16a}) and (\ref{an16b}) by different
considerations and we
identified them with the mass and Compton wavelength of the neutrino. We
conclude therefore 
that the mass
$m$ and the length $\lambda_C$ leading to the observed value of the
cosmological constant in Zeldovich's model I correspond to the mass and
Compton
wavelength of the neutrino.

Some authors have tried to determine the limit on the measurability of spacetime
distances in quantum
gravity  or
the minimum uncertainty of spacetime geodesics \cite{gacnature}. It
arises when taking into account the quantum properties of the devices used for
measurement. According
to Amelino-Camelia
\cite{gacnature}
the minimum uncertainty of the measure of the length of an object
of size $l$ due to quantum fluctuations is given by 
\begin{equation}
\delta l\sim (l l_P)^{1/2}.
\label{ac1}
\end{equation}
When applied to the
Universe as a whole ($l\sim R_{\Lambda}^*$) this gives
\begin{equation}
\lambda_{\rm
min}=\sqrt{R_{\Lambda}^*l_P}=3.91\times 10^{-5}\, {\rm m},
\label{ac2}
\end{equation}
which corresponds to
Eq. (\ref{an16b}). This minimum length is larger than
the Planck length $l_P=1.62\times
10^{-35}\, {\rm m}$ showing that
quantum
gravity is not a Planck scale phenomenon.

In Model I of Sec. \ref{sec_kcdur} we have identified the primordial (maximum)
density of
the universe with the Planck density $\rho_P$. Then, we have shown that
the universe starts with a size  $R(0)=l_P$ equal to the Planck length at $t=0$
and reaches a size $R_1=\lambda_\Lambda^*$ equal to the neutrino's Compton
wavelength at the end of the inflation of duration  $t_c=23.8\, t_P$ (see
Sec. \ref{sec_kcdur}, Sec. \ref{sec_nv} and Appendix \ref{sec_intre}).

Combining the above
results, we conclude that the minimum length $\lambda_{\rm min}$  in Zeldovich's
model I leading to the correct value of the cosmological constant
corresponds to:

(i) the Compton wavelength of the neutrino [Eq. (\ref{an16b})].

(ii) the limit on the
measurability of spacetime distances [Eq. (\ref{ac2})] in quantum gravity
according to Amelino-Camelia \cite{gacnature}.\footnote{This
lengthscale can
also be related
to  the radius of a fifth extra dimension \cite{dlb}.}

(iii) The radius of the universe at the end of the inflation in model I of Sec.
\ref{sec_kcdur} [Eq. (\ref{wn1b})]. This radius defines an
effective or physical ``minimum'' length $\lambda_{\Lambda}^*=3.91\times
10^{-5}\, {\rm
m}$ which could  replace the Planck length $l_P=1.62\times
10^{-35}\, {\rm m}$ and yield the correct value of
the vacuum energy density when substituted into Eq. (\ref{an15}).

In conclusion, if we identify $\lambda_{\rm min}$ to the Compton wavelength
of the neutrino in Zeldovich's model I, we can express the Compton wavelength
 $\lambda_\Lambda^*$ and the mass $m_\Lambda^*$ of the neutrino as a
function of the cosmological
constant [see Eqs. (\ref{an16a}) and (\ref{an16b})]. Conversely,
assuming that the neutrino has these
characteristic
parameters, we can explain the measured value of $\Lambda$ [see Eq.
(\ref{an16bintro})].

\subsection{Model II}

By considering the gravitational interaction energy
between virtual pairs of the quantum electrodynamic vacuum, Zeldovich 
\cite{zeldovich,zeldovichA} obtained another formula for the vacuum 
energy density\footnote{He wrote the vacuum energy density under the form
\begin{equation}
\rho_{\Lambda}c^2\sim \frac{Gm^2}{\lambda_C}\times \frac{1}{\lambda_C^3}.
\label{er17}
\end{equation}
This expression assumes that the vacuum
contains virtual pairs of particles with effective density $n\sim
1/\lambda_C^3$ and that these pairs have a gravitational energy of interaction
$Gm^2/\lambda_C$.}
\begin{equation}
\label{an17}
\rho_\Lambda c^2\sim \frac{G m^6c^4}{\hbar^4}\sim
\frac{G\hbar^2}{c^2\lambda_C^6},
\end{equation}
where we have introduced the Compton wavelength $\lambda_C={\hbar}/{mc}$  of the
particle of mass $m$. It represents the relevant minimum length
$\lambda_{\rm min}$ in Zeldovich's model II. We
note that this expression of the
vacuum density explicitly involves the gravitational constant $G$. Therefore,
it may be related to a theory of quantum gravity. In
his original papers, Zeldovich \cite{zeldovich,zeldovichA}  used Eq.
(\ref{an17}) with the proton
mass and obtained a discrepency of $7$ orders of magnitude with the empirical
cosmological density. If we use the Planck mass and
the Planck length in Eq. (\ref{an17}), we find that $\rho_{\Lambda}\sim \rho_P$,
yielding a discrepency of $120$ orders of magnitude with the empirical
value (cosmological
constant problem). Alternatively, if we adopt the measured
value of the cosmological constant and reverse the relation from Eq.
(\ref{an17}) we get
\begin{equation}
m_e\sim \left (\frac{\Lambda
\hbar^4}{G^2}\right )^{1/6}\sim
(m_{\Lambda}M_P^2)^{1/3}=6.77\times 10^7\, {\rm
eV/c^2},
\label{an18a}
\end{equation}
\begin{equation}
\lambda_e\sim \left (\frac{G^2\hbar^2}{\Lambda c^6}\right
)^{1/6}\sim 
(R_{\Lambda}^*l_P^2)^{1/3}=2.92\times 10^{-15}\, {\rm m}.
\label{an18b}
\end{equation}
In Ref. \cite{ouf} we arrived at the mass and length scales (\ref{an18a}) and
(\ref{an18b}) by different
considerations and we
identified them with the mass and Compton wavelength of the electron (in order
of magnitude). Actually,
we obtained accurate expressions of the electron mass and electron Compton
wavelength  under the form \cite{preouf1,preouf2,ouf}
\begin{equation}
m_e\simeq \alpha\left (\frac{\Lambda \hbar^4}{G^2}\right )^{1/6},\qquad
\lambda_e=\frac{r_e}{\alpha} \simeq \frac{1}{\alpha}\left
(\frac{G^2\hbar^2}{\Lambda c^6}\right )^{1/6},
\label{edd3}
\end{equation}
where $\alpha=e^2/(\hbar c)\simeq 1/137$ is the fine-structure constant (see
Appendix \ref{sec_ele}). We see that Eq. (\ref{an18b}) corresponds more closely
to
the classical electron radius $r_e$ than to the electron Compton
wavelength $\lambda_e$ (we note that the classical electron radius does not
depend
explicitly on $\alpha$). The formula (\ref{edd3}) provides an accurate form of
the Eddington formula \cite{eddington1931lambda,ouf} relating the mass of the
electron to the cosmological
constant (or the converse).\footnote{It would be desirable to know if this
formula is {\it exact} or just a good approximate relation (possibly the leading
term
in an expansion in powers of $\alpha$).} It can be written $m_e\simeq \alpha
(m_{\Lambda}M_P^2)^{1/3}$ and $r_e=\alpha \lambda_e\simeq
(R_{\Lambda}^*l_P^2)^{1/3}$. We
conclude therefore 
that the mass
$m$ and the length $\lambda_C$ leading to the observed value of the
cosmological constant in Zeldovich's model II correspond to the typical mass and
typical Compton
wavelength of the electron (more precisely to $m_e/\alpha$ and
$\alpha\lambda_e=r_e$).

Another formula has been obtained for  the minimum uncertainty in the
measure of the length of an
object of size $l$ due to quantum fluctuations. According
to Karolyhazy \cite{karolyhazy} it is given by
\begin{equation}
\delta l\sim (l l_P^2)^{1/3}.
\label{an18bce1}
\end{equation}
This is the
condition that the device used to make the measurement does not turn into a
black hole. This expression was later related to the holographic principle
and to the theory of quantum information. The minimum total uncertainty in the
measurement of a length equal to the
size of the universe  ($l\sim R_{\Lambda}^*$), which is a
consequence of
combining the principles of quantum mechanics and general relativity, is given
by
\begin{equation}
\lambda_{\rm min}=(R_{\Lambda}^*l_P^2)^{1/3}=2.92\times 10^{-15}\, {\rm
m},
\label{an18bce}
\end{equation}
which corresponds to Eq. (\ref{an18b}).  This minimum length is larger than
the Planck length $l_P=1.62\times
10^{-35}\, {\rm m}$ showing
that quantum
gravity is not a Planck scale phenomenon.

In Model II of Sec. \ref{sec_kcdur} we have identified the primordial
(maximum) density of
the universe with the electron density $\rho_e$. This was justified by
applying the same Lagrangian (\ref{kc1}) of nonlinear electrodynamics both to
the
magnetic universe (see Sec. \ref{sec_kcdur}) and  to the electron (see Sec.
\ref{sec_nee}). Then, we have shown that
the universe starts with a size  $R(0)=r_e$ equal to the classical radius of
the electron at $t=0$
and reaches a size  $R_1={\tilde R}_2$ of the order of the radius of a dark
energy star of the stellar mass
at the end of the inflation of duration  $t_c=23.8\, t_e^*$ (see
Sec. \ref{sec_kcdur}, Sec. \ref{sec_nv} and Appendix \ref{sec_intre}).

Combining the
above results, we conclude that the minimum length $\lambda_{\rm min}$ in
Zeldovich's model II leading to the correct value of the cosmological
constant corresponds to:

(i) the classical radius of the electron [Eqs. (\ref{an18b}) and (\ref{edd3})].

(ii) the limit on the measurability of spacetime distances [Eq. (\ref{an18bce})]
in quantum
gravity according to Karolyhazy \cite{karolyhazy}.

(iii) The radius of the universe at the begining of the inflation in model II
of Sec. \ref{sec_kcdur} [Eq. (\ref{f14})].  This radius defines
an absolute ``minimum'' length $r_e=2.82\times 10^{-15}\, {\rm
m}$ which could  replace the Planck length $l_P=1.62\times
10^{-35}\, {\rm m}$ and yield the correct value of
the vacuum energy density when substituted into Eq. (\ref{an17}).

In conclusion, if we identify $\lambda_{\rm min}$ to the classical radius
of the electron in Zeldovich's model II,  we can express the radius $r_e$, the
Compton wavelength $\lambda_e$ and the mass
$m_e$ of the electron as a
function of the cosmological constant [see Eqs. (\ref{an18a})-(\ref{edd3})] and
we justify the
mysterious Eddington relation \cite{eddington1931lambda,ouf}.
Conversely, we can express the cosmological constant in terms
of $r_e$, $\lambda_e$ or $m_e$ and, by using the empirical value of
the mass of the electron, we can explain the measured value of $\Lambda$
[see Eq. (\ref{edd3intro})]. The
physical
reason to identify  $\lambda_{\rm min}$ to the classical radius
of the electron is that it corresponds to the initial radius of the universe
at $t=0$ according to the model of nonlinear electrodynamics
based on the
Lagrangian
(\ref{kc1}) provided that this model applies both to the early magnetic universe
and to the
electron.

\section{Conclusion}

In this paper we have discussed the connection between the model of
magnetic universe based on nonlinear electrodynamics introduced by
Kruglov \cite{kruglov2015a,kruglov2015,kruglov2020}  and our model of universe
based on a quadratic (or polytropic) equation of state
\cite{chavanisAIP,jgrav,cosmopoly1,cosmopoly2,cosmopoly3,universe,stiff,mlbec,
chavanisIOP,vacuumon}.  These models both  predict a period of
early inflation followed by a radiation era. These models are essentially
equivalent since the equation of state introduced in
\cite{chavanisAIP,jgrav,cosmopoly1,cosmopoly2,cosmopoly3,universe,stiff,mlbec,
chavanisIOP,vacuumon} can be deduced
from the  nonlinear electromagnetic Lagrangian introduced in 
\cite{kruglov2015a,kruglov2015,kruglov2020}  and
{\it vice versa}. However, nonlinear electrodynamics might give a physical
interpretation to our quadratic equation of state. It may arise from possible
quantum gravity corrections to linear electrodynamics.

We have also generalized the nonlinear electromagnetic Lagrangian (\ref{kc1}) by
including
a zero-point energy [see Eq. (\ref{gr8})] so that it describes  a form of
``generalized
radiation''
\cite{chavanisAIP,jgrav,cosmopoly1,cosmopoly2,cosmopoly3,universe,stiff,mlbec,
chavanisIOP,vacuumon} that accounts simultaneously
for the early inflation, the ordinary radiation, and the dark energy. Baryonic
matter and dark matter are added as independent species. This leads to a
complete model of universe (see Sec. \ref{sec_complete}). This model essentially
coincides with the
$\Lambda$CDM model but it includes a period of early inflation.
The density decreases from a maximum density $\rho_I$ equal to the Planck
density $\rho_P$ or to the electron density $\rho_e$ up to a minimum density
$\rho_\Lambda$ equal to
the cosmological density (see Fig. \ref{trhoLOGLOGcomplet}). In this sense, it
connects two de
Sitter eras of accelerating expansion separated by a radiation era and a matter
era of decelerated expansion  (see
Fig. \ref{taLOGLOGcomplet}).\footnote{In this paper, we have considered a
model similar to the $\Lambda$CDM model where the dark energy density is
constant ($n=-1$) but, following \cite{cosmopoly2,cosmopoly3}, we could consider
more general models where the dark energy density varies with time (see Secs.
\ref{sec_gen} and \ref{sec_kcl} with $n<0$ and $n\neq -1$). There is, however, a
difficulty with such models when interpreted in terms of nonlinear
electrodynamics as pointed out in Sec. \ref{sec_genval}.}

By applying the nonlinear electrodynamics to the electron in the spirit of
the Born-Infeld model \cite{born1933,borninfeld}, following Kruglov
\cite{kruglov2015a}, we have obtained an extended model of electron and
justified the fact that the electron may have a finite radius and a finite
density like in the Abraham-Lorentz model \cite{abraham,lorentz}. Indeed, in
this model, the electric field is nonsingular at the origin and the electric
energy is finite.\footnote{When the Lagrangian (\ref{kc1b}) is
applied to the electron, we find that the electric field $E(r)$ is finite at
$r=0$. By contrast, in cosmology, the
magnetic field $B\sim 1/a^2$ diverges when $a\rightarrow 0$.} By identifying
the electrostatic energy with the mass of the
electron we can obtain the electron radius and the electron density. This
determines the fundamental density $\rho_*$, or the fundamental length $r_*$, in
the nonlinear electromagnetic Lagrangian (see Sec. \ref{sec_nee}). Then, by
applying the
same nonlinear electromagnetic Lagrangian  to cosmology we have shown that the
initial density of the universe is equal  to the electron density $\rho_e$,
its initial radius is equal to the electron radius $r_e$ and its
initial mass is equal to the electron mass $m_e$. We have then
interpreted the radius of the electron as a fundamental
minimum length and we have mentioned the connection with the result of
Karolyhazy \cite{karolyhazy} in quantum gravity. By introducing this minimum
length in the
second Zeldovich \cite{zeldovich,zeldovichA} formula of vacuum energy [see
Eq. (\ref{an17})] we have obtained a refined Eddington
relation between the cosmological constant and the mass of the electron [see Eq.
(\ref{edd3intro})]. This relation provides the exact (or almost exact) value of
the
cosmological constant. We have thus justified the Eddington relation
\cite{eddington1931lambda,ouf} and the
value of the cosmological constant from nonlinear electrodynamics. This is a
true prediction of the cosmological constant without free parameter since  the
mass of the electron is experimentally known. We have also shown that this
cosmological model produces $10^{83}$ electrons or $10^{80}$ protons in the
present universe, which is
precisely the Eddington number \cite{eddington1931lambda,ouf}.

We have considered another model in which the initial density of the universe is
equal to the Planck density $\rho_P$, its initial radius is equal to the
Planck length $l_P$ and its initial mass is equal to the Planck 
mass $M_P$. We have shown that
the radius of the universe at the end of the inflation is equal to the Compton
wavelength of the neutrino $r_\nu$ introduced in \cite{ouf}. We have
then interpreted
this radius as an effective fundamental minimum length and we have mentioned the
connection with the result of Amelino-Camelia \cite{gacnature} in quantum
gravity. Then, by
introducing this minimum length in the first Zeldovich
\cite{zeldovich,zeldovichA}  formula of vacuum
energy [see
Eq. (\ref{an15})] we have obtained the correct value of the cosmological
constant [see
Eq. (\ref{an16bintro})].
However, this is not a true prediction of $\Lambda$ since the mass of the
neutrino is not firmly known. Indeed, in  \cite{ouf} it has been
precisely determined in terms of the cosmological constant.

In the two models of inflation described above the $e$-folding number is the
same, $N_0=68.5$, corresponding to an increase of the size of the universe by
$30$ orders of magnitude between the begining and the end of the inflation.
However, 
the duration of the inflation is very different. In Model I (where the initial
density is equal to the Planck density $\rho_P=5.16\times 10^{99}\,
{\rm g\, m^{-3}}$) the size of the universe
increases from the Planck length $l_P=1.62\times
10^{-35}\, {\rm m}$ to the
Compton wavelength of the neutrino $r_\nu=3.91\times 10^{-5}\, {\rm m}$ on a
timescale $t_{c}=23.8\, t_P=1.28\times 10^{-42}\, {\rm s}$. In Model II (where
the initial density is equal to the electron density $\rho_e=4.07\times
10^{16}\, {\rm g\, m^{-3}}$) the size of the
universe increases from the classical electron radius $r_e=2.82\times
10^{-15}\, {\rm m}$ to the radius ${\tilde R}_2=7.07\times 10^{15}\,
{\rm m}$ of a dark energy star of the stellar mass on
a timescale $t_{c}=23.8\, t_e^*=0.457\, {\rm s}$.  We also note that the
fundamental minimum length, which gives the
correct value of the cosmological constant when introduced in the Zeldovich
formula of vacuum energy, is different in the two models. In Model I it
corresponds to the Compton wavelength of the neutrino which is equal to the size
of the universe at the end of the inflation and in Model II it corresponds to
the classical radius of the electron  which is equal to the size
of the universe at the begining of the inflation. In the two cases, the
fundamental minimum length is much larger than the Planck length suggesting that
quantum gravity is not a Planck scale phenomenon. It would be interesting to
know if observations favor one model over the other.

\appendix

\section{The logotropic model}
\label{sec_logomodel}

In a series of papers \cite{epjp,lettre,jcap,preouf1,action,logosf,preouf2} we
have
developed a model of unified dark matter and dark energy based on the
logotropic equation of state  
\begin{eqnarray}
\label{logomodel1}
P=A\ln \left (\frac{\rho}{\rho_P}\right ),
\end{eqnarray}
where $A$ is a new fundamental constant of physics and
$\rho_P=c^5/G^2\hbar=5.16\times 10^{99}\, {\rm
g\, m^{-3}}$ is the
Planck
density. The constant $A$ can be interpreted as a sort of
``logotropic
temperature'' in a generalized thermodynamical framework
\cite{epjp,lettre,jcap}. 
This model leads to dark matter halos with a density profile that is
flat at the center (thereby solving the core-cusp problem of the CDM model) and
that
decreases at large distances as
\begin{eqnarray}
\label{logomodel2}
\rho\sim \left (\frac{A}{8\pi G}\right )^{1/2}\frac{1}{r}.
\end{eqnarray}
At even larger distances the density falls like $r^{-2}$
(isothermal profile) or like $r^{-3}$ (NFW and Burkert profiles), or even more
rapidly in order to
have a
finite mass. This confinement may result from an incomplete relaxation
\cite{preouf2,logosf}. The logotropic model implies that the
dark matter halos (``small'' scales) have a universal surface density given by
\begin{eqnarray}
\label{logomodel3}
\Sigma_0=\rho_0 r_h=5.85\, \left (\frac{A}{4\pi G}\right )^{1/2},
\end{eqnarray}
where the prefactor is deduced from the theory (here $\rho_0$ is the central
density and $r_h$ is the halo radius where the central density is divided by
$4$). By applying the logotropic
equation of state (\ref{logomodel1}) to the universe as a whole (``large''
scales) we found
that the universal constant $A$ is related to the present value of the dark
energy density $\rho_{\Lambda}\equiv \rho_{\rm de,0}=5.96\times
10^{-24}\, {\rm
g\, m^{-3}}$ by
\begin{eqnarray}
\label{logomodel4}
{A}/{c^2}=\frac{\rho_\Lambda}{\ln \left (\frac{\rho_P}{\rho_{\Lambda}}\right
)}=2.10\times 10^{-26}\, {\rm g}\, {\rm m}^{-3}.
\end{eqnarray}
Using
$\rho_{\Lambda}=\Lambda c^2/8\pi G$ with $\Lambda=1.11\times
10^{-52}\, {\rm m^{-2}}$, we can rewrite the foregoing equation as
\begin{eqnarray}
\label{logomodel8}
A=1.40\times 10^{-4}\, \frac{c^4\Lambda}{G}.
\end{eqnarray}
This allows us to
express the
universal surface density of dark matter halos in terms of the  cosmological
constant $\Lambda$ of the $\Lambda$CDM model by
\begin{eqnarray}
\label{logomodel5}
\Sigma_0=0.01955\, \frac{c^2\sqrt{\Lambda}}{G}=133\,
M_{\odot}/{\rm pc}^2.
\end{eqnarray}
This value turns out to be in good agreement with the observational value
$\Sigma_0^{\rm
obs}=141_{-52}^{+83}\,
M_{\odot}/{\rm pc}^2$ \cite{donato}. The
average surface
density of dark matter halos is $\langle \Sigma\rangle=M_h/(\pi r_h^2)=0.474\,
\Sigma_0=63.1\,
M_{\odot}/{\rm pc}^2$ where the prefactor in the second equality is deduced from
the theory. It is in good agreement with the observational value $\langle
\Sigma\rangle_{\rm obs}=0.51\, \Sigma_0^{\rm
obs}=72_{-27}^{+42}\, M_{\odot}/{\rm pc}^2$ \cite{gentile}. The logotropic model
therefore implies a universal gravitational acceleration 
\begin{eqnarray}
\label{logomodel6}
g=\frac{GM_h}{r_h^2}=\pi G\langle\Sigma\rangle=1.49\, G\Sigma_0.
\end{eqnarray}
Using Eq. (\ref{logomodel5}) the
universal gravitational
acceleration  can be expressed in terms of the cosmological constant as
\begin{eqnarray}
\label{logomodel7}
g=0.0291\, c^2\sqrt{\Lambda}=2.76\times 10^{-11}\, {\rm m\, s^{-2}}.
\end{eqnarray}
This value is in good agreement with the observational value $g_{\rm obs}=\pi G
\langle\Sigma\rangle_{\rm obs}=3.2_{-1.2}^{+1.8}\times 10^{-11}\, {\rm
m/s^2}$ \cite{gentile}.

Using the foregoing relations, the
asymptotic behavior of the logotropic density can be rewritten as
\begin{eqnarray}
\label{logomodel9b}
\rho\sim 0.121\, \frac{\Sigma_0}{r}\sim 0.0811\, \frac{g}{Gr}\sim 
0.00236\, \frac{c^2\sqrt{\Lambda}}{Gr}. 
\end{eqnarray}
As noted in \cite{logosf} this $r^{-1}$ behavior is similar to the
density cusp in the NFW model $\rho=\rho_sr_s/[r(1+r/r_s)^2]\sim \rho_s r_s/r$.
We find that
\begin{eqnarray}
\label{logomodel10}
\rho_s r_s= 0.121\, \Sigma_0= 0.0811\, \frac{g}{G}= 
0.00236\, \frac{c^2\sqrt{\Lambda}}{G}=16.0\, M_{\odot}/{\rm pc}^2.
\end{eqnarray}

We also noticed \cite{preouf1,preouf2} that the surface density $\Sigma_0=133\,
M_{\odot}/{\rm pc}^2$ of dark matter halos (and
the surface density of the
universe $\Sigma_\Lambda=c^2\sqrt{\Lambda}/G=6800\,
M_{\odot}/{\rm pc}^2$) is of
the same order as the surface density of the electron
\begin{eqnarray}
\label{logomodel8b}
\Sigma_e=\frac{m_e}{r_e^2}=54.9\,
M_{\odot}/{\rm pc^2}.
\end{eqnarray}
Using the accurate Eddington relation (\ref{la1}) we find that
\begin{eqnarray}
\label{logomodel9}
\Sigma_e\simeq
\alpha\frac{c^2\sqrt{\Lambda}}{G}\simeq \alpha\Sigma_\Lambda\simeq 0.373\,
\Sigma_0.
\end{eqnarray}
This relation may be interpreted in terms of the holographic principle
\cite{preouf1,ouf}.

The circular velocity at the halo radius is
\begin{eqnarray}
\label{logomodel10b}
v_h^2=\frac{GM_h}{r_h}.
\end{eqnarray}
Combining the foregoing relations, we find that
\begin{eqnarray}
\label{logomodel11}
\frac{v_h^4}{M_h}=Gg=\pi\langle\Sigma\rangle G^2=1.49\, \Sigma_0
G^2.
\end{eqnarray}
This relation is connected to the Tully-Fisher relation \cite{tf,tfmcgaugh} 
which involves the baryon mass $M_b$ instead of the DM halo mass
$M_h$. Introducing the baryon fraction $f_b=M_b/M_h\sim 0.17$, we obtain
$(M_{\rm
b}/v_h^4)^{\rm th}=f_b/(1.49\, \Sigma_0 G^2)=46.4\,
M_{\odot}{\rm km}^{-4}{\rm s}^4$ which is close to the
observed value 
$(M_{\rm b}/v_h^4)^{\rm obs}=47\pm 6 \, M_{\odot}{\rm km}^{-4}{\rm s}^4$
\cite{mcgaugh}.

More generally, the rotation curve is given by
\begin{eqnarray}
\label{logomodel12}
v^2(r)=\frac{GM(r)}{r}.
\end{eqnarray}
For $r\rightarrow +\infty$, we have
\begin{eqnarray}
\label{logomodel13}
M(r)\sim 2\pi \left (\frac{A}{8\pi G}\right )^{1/2}r^2
\sim 0.760\, \Sigma_0 r^2\sim 0.510\, \frac{gr^2}{G}\sim 
0.0148\, \frac{c^2\sqrt{\Lambda}r^2}{G}
\end{eqnarray}
and
\begin{eqnarray}
\label{logomodel14}
v^2(r)\sim 2\pi G \left (\frac{A}{8\pi G}\right )^{1/2}r
\sim 0.760\, G\Sigma_0 r\sim 0.510\, gr\sim 
0.0148\, c^2\sqrt{\Lambda}r.
\end{eqnarray}
Asymptotically,  the gravitational acceleration (the gravitational force
$F(r)=m g(r)$ by unit of mass) produced by the
logotropic distribution tends to a constant
\begin{eqnarray}
\label{logomodel15}
g(r)=\frac{d\Phi}{dr}=\frac{GM(r)}{r^2}\rightarrow 0.510\,
g=g_{\infty}.
\end{eqnarray}
The gravitational potential behaves as $\Phi(r)\sim g_{\infty} r$. We find
$M(r)/M_{\odot}\sim 101\, (r/{\rm pc})^2$, $\log(\frac{v}{\rm km\,
s^{-1}})=1.32+\frac{1}{2}\log(\frac{r}{\rm kpc})$ and $g_{\infty}=1.41\times
10^{-11}\, {\rm m \, s^{-2}}$ to be compared with the observational expressions
$M(r)^{\rm obs}/M_{\odot}=200_{-120}^{+200}\, (r/{\rm pc})^2$,
$\log(\frac{v_{\rm obs}}{\rm km\, 
s^{-1}})=1.47_{-0.19}^{+0.15}+0.5\log(\frac{r}{\rm kpc})$ and
$g_{\infty}^{\rm obs}=3_{-2}^{+3}\times
10^{-11}\, {\rm m \, s^{-2}}$ \cite{walker}.

When the logotropic model is applied in a cosmological framework, the evolution
of the density of the generalized radiation is given
by \cite{epjp,lettre,jcap,preouf1,action,logosf,preouf2} 
\begin{eqnarray}
\label{logots1}
\rho_{\rm Rad}=\frac{\rho_I}{1+\left
(\frac{a}{a_1}\right )^4}+\rho_{\Lambda}(1+3B\ln a),
\end{eqnarray}
where $B=A/(\rho_\Lambda c^2)=1/\ln(\rho_P/\rho_\Lambda)=3.53\times 10^{-3}$.
In the logotropic model, the density of dark energy increases slowly
(logarithmically) with the scale factor. This corresponds to a phantom
\cite{caldwell,cosmopoly3} behavior leading to a little rip \cite{littlerip}
(the energy density and the scale factor become infinite in infinite time). The
model of Sec. \ref{sec_total}, where the dark energy density is
constant, is
recovered for $B=0$. Using Eqs. (\ref{f4}) and (\ref{f10}) the nonlinear
electromagnetic Lagrangian inspired by the logotropic model is 
\begin{eqnarray}
\label{logots2}
{\cal L}=\frac{-{\cal F}}{1+\frac{{\cal F}}{\rho_I
c^2}}-\rho_\Lambda c^2-\frac{3}{4}B\rho_{\Lambda}c^2\ln \left
(\frac{{\cal F}_*}{\cal F}\right ).
\end{eqnarray}

{\it Remark:} The above results can be expressed in terms of the present value
of the Hubble
parameter $H_0=(8\pi G\rho_0/3)^{1/2}=2.195\times 10^{-18}\, {\rm s}^{-1}$
instead of the cosmological constant by
using the relation $\rho_{\Lambda}=\Omega_{\Lambda,0}\rho_0$
with $\Omega_{\Lambda,0}=0.6911$ giving 
\begin{eqnarray}
\label{logomodel15b}
\Lambda c^2=3\Omega_{\Lambda,0}H_0^2=2.07\, H_0^2.
\end{eqnarray}
For example, the universal surface density can be written as
$\Sigma_0=0.02815 H_0 c/G$ and the universal gravitational acceleration can be
written as $g=0.0419\, H_0 c$. This relation explains why the fundamental
constant $a_0=g/f_b$ that appears in
the MOND (modification of Newtonian dynamics) theory \cite{mond} is of order
$a_0\simeq H_0 c/4=1.65\times
10^{-10}\, {\rm m/s^2}$ in good agreement with the observational
value $a_0^{\rm obs}=(1.3\pm 0.3)\times
10^{-10}\, {\rm m/s^2}$. Note, however, that our model is completely different
from the MOND theory.

\section{The mass of the universe}
\label{sec_massuniv}

\subsection{Cosmological scales}

The empirical value of the cosmological constant deduced from the observation
of the accelerated expansion of the universe is $\Lambda=1.11\times
10^{-52}\, {\rm m^{-2}}$. By using
general arguments based on physical considerations and dimensional analysis, we
can introduce cosmological scales. The cosmological
density $\rho_{\Lambda}^*=
\Lambda c^2/G=1.50\times 10^{-22}\, {\rm g\, m^{-3}}$ is of the order of
the density of the universe, the cosmological time $t_{\Lambda}^*=
1/(G\rho_{\Lambda}^*)^{1/2}=1/(c\sqrt{\Lambda})=3.16\times 10^{17}\, {\rm s}$ is
of the order of
the age of the universe, the cosmological length  $R_{\Lambda}^*= c
t_{\Lambda}^*=1/\sqrt{\Lambda}=9.49\times 10^{25}\, {\rm m}$ is of the
order of the size of the visible universe (the distance travelled by a photon on
a timescale $t_{\Lambda}^*$), and the cosmological mass $M_\Lambda^*=
\rho_{\Lambda}^*{R_{\Lambda}^*}^3= c^2/(G\sqrt{\Lambda})=1.28\times 10^{56}\,
{\rm g}$ is of the order of the mass of the universe.\footnote{These
quantities are just orders of magnitude. They are given without any
prefactor, and this is why they have been written with the symbol $*$ (we have
$\rho_\Lambda^*=8\pi\rho_\Lambda$, $t_\Lambda^*=t_\Lambda/\sqrt{8\pi}$,
$R_\Lambda^*=R_\Lambda/\sqrt{8\pi}$, and $M_\Lambda^*=M_\Lambda/\sqrt{8\pi}$).
These
relations
can be derived from the Friedmann equations by using the fact
that the
present
density of the universe is of the order of the cosmological density on
account of the cosmic coincidence \cite{ouf}.} In astronomical units,
$t_{\Lambda}^*=10.0\, {\rm Gyrs}$,
$R_{\Lambda}^*=3.07\, {\rm
Gpc}$ and 
$M_{\Lambda}^*=6.42\times 10^{22}\, M_{\odot}$. The typical number of electrons
in the universe is $N_e\sim {M_{\Lambda}^*}/{m_e}\sim 10^{83}$ and the typical
number of protons (Eddington's number) is $N_p\sim {M_{\Lambda}^*}/{m_p}\sim
10^{80}$.\footnote{These estimates assume that the Universe is made only of
electrons or protons, which is of course not correct. We should also take into
account dark matter and dark energy. But, because of the cosmic coincidence,
the density of baryonic matter, dark matter and dark energy is of the same order
of magnitude at present so our estimates are meaningful.} These quantities play
an
important role in
the theory of large numbers \cite{ouf}.

\subsection{Evolution of the mass of the universe during the different epochs}

\begin{figure}[!h]
\begin{center}
\includegraphics[clip,scale=0.3]{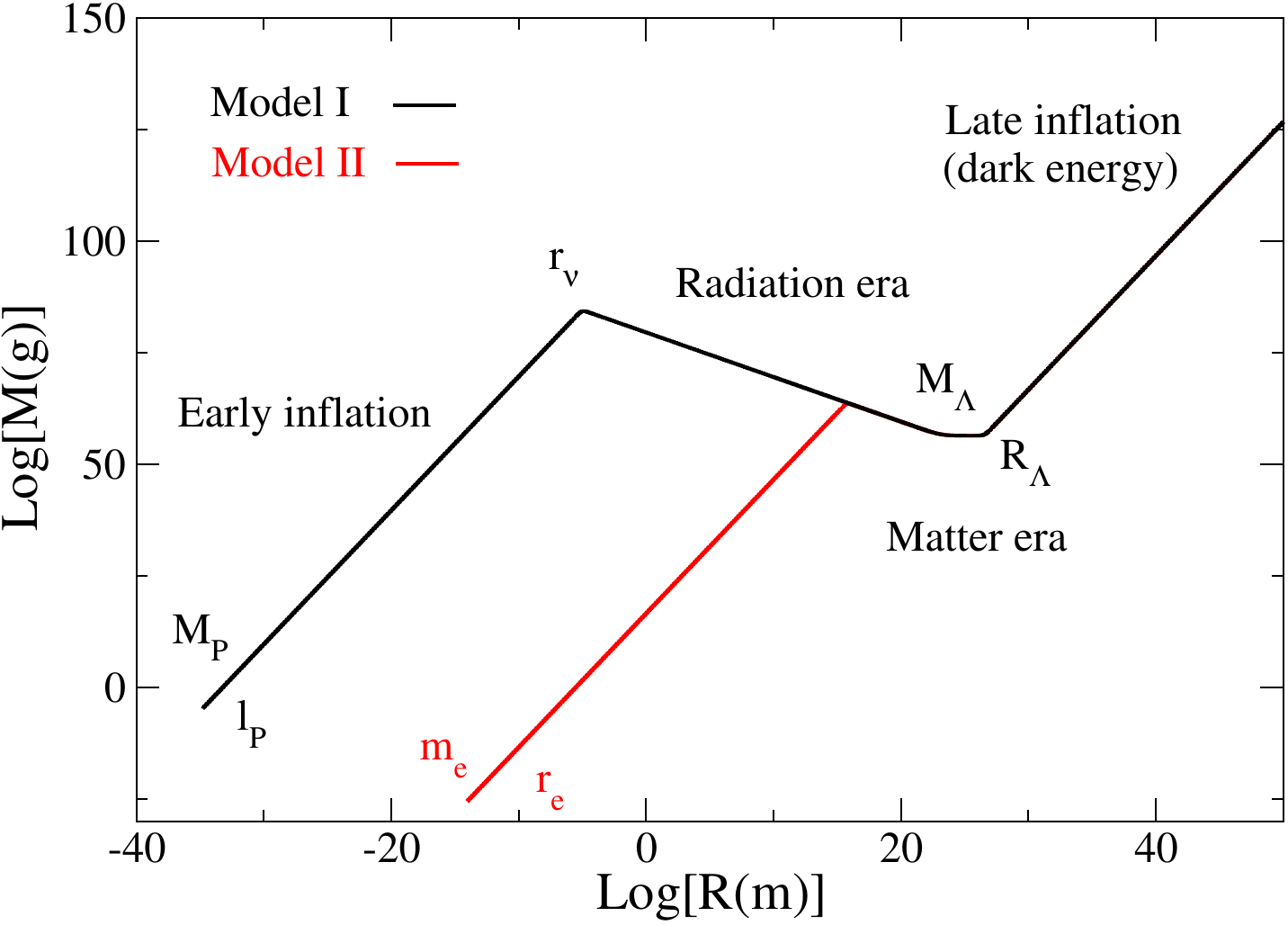}
\caption{Evolution of the mass of the universe as a function of the radius in
Model I (black) and Model II (red). }
\label{massradius}
\end{center}
\end{figure}

We define the radius and the mass of the of the universe at time $t$ by
\begin{eqnarray}
\label{mu1}
R(t)=a(t)R_{\Lambda},\qquad M(t)=\rho(t)R(t)^3.
\end{eqnarray}
During the inflation era, the density of the universe is
constant ($\rho=\rho_I$) and the mass of the universe increases as
\begin{eqnarray}
\label{mu2}
M(t)=\rho_I R(t)^3.
\end{eqnarray}
During the radiation era, the density of the universe
decreases as  $\rho\sim R^{-4}$ and the mass of the universe decreases as
\begin{eqnarray}
\label{mu3}
M(t)\sim \frac{1}{R(t)}.
\end{eqnarray}
During the matter era, the density of the universe
decreases as  $\rho\sim R^{-3}$ and the mass of the universe is constant
\begin{eqnarray}
\label{mu4}
M(t)={\rm cst}.
\end{eqnarray}
During the dark energy era, the density of the universe is
constant ($\rho=\rho_\Lambda$) and the mass of the universe increases as
\begin{eqnarray}
\label{mu5}
M(t)=\rho_\Lambda R(t)^3.
\end{eqnarray}
The evolution of the mass of the universe  as a function of the
radius in Models I and II of Sec. \ref{sec_kcdur} is plotted in Fig.
\ref{massradius}.

\subsection{Model II}

In Model II, the initial density of the universe is equal to the
density of the electron: $\rho_I=\rho_e$. The initial radius of the universe is
equal to the radius of the electron ($R(0)=r_e$) and its initial mass is
equal to the mass  $m_e$ of the electron (see Appendix \ref{sec_intre}). At
$t=0$, the primordial universe has
the same characteristics as the electron but it is ``unstable'' and
``explodes''. This picture can be viewed as a refinement of the ``primeval
atom'' of
Lema\^itre \cite{lemaitreNewton}. Let us follow its expansion accross the ages. 
Between
the begining and the end of the inflation, its radius increases by $30$ orders
of magnitude (see Appendix \ref{sec_intre}). Therefore, at the end
of the inflation, its radius is $R_1\sim 10^{30}r_e$ and its volume is
$R_1^3\sim 10^{90}r_e^3$.
Since its density is constant ($\rho\sim \rho_e$) its mass at the end of the
inflation (or at the begining of the radiation era) is $M_1\sim 10^{90}m_e$.

Between the end of the inflation (or the begining of the radiation era) and the
time of radiation-matter equality, the radius of the universe passes from $R_1$
to
$R_{\rm eq}=a_{\rm eq} R_{\Lambda}$. Using Eqs. (\ref{kc30}) and (\ref{aeq}),
we find that  $R_{\rm eq}/R_1=a_{\rm eq}/a_1\sim 10^{7}$. Therefore, its radius
increases by $7$ orders of magnitude. Using Eq. (\ref{mu3}), we find that its
mass at the time of radiation-matter equality is $M_{\rm eq}\sim 10^{-7}
M_1\sim 10^{83}m_e$.

This mass remains constant during the matter era (see Eq. (\ref{mu4}))
and may be identified with the present mass of the universe: $M_\Lambda\sim
10^{83}m_e$. This shows that the mass of the universe is equal to $N_e\sim
10^{83}$ electrons of mass $m_e$. Since $m_p/m_e=1836$, we find that the
mass of the universe  is equal to $N_p\sim
10^{80}$ protons of mass $m_p$. This justifies the Eddington number
\cite{eddington1931lambda,ouf}. These results are in good agreement with the
observations.

\subsection{Model I}

In Model I, the initial density of the universe is equal to the Planck
density: $\rho_I=\rho_P$. The initial radius of the universe is
equal to the Planck length ($R(0)=l_P$) and its initial mass is
equal to the Planck mass  $M_P$ (see Appendix \ref{sec_intre}). At $t=0$, the
primordial
universe has
the same characteristics as a Planck black hole (or a ``planckion'' particle)
but it is ``unstable'' and ``explodes''. Let us follow its expansion accross
the ages. Between
the begining and the end of the inflation, its radius increases by $30$ orders
of magnitude (see Appendix \ref{sec_intre}). Therefore, at the end of the
inflation, its radius
is $R_1\sim 10^{30}l_P$, which is of the order of the Compton wavelength of the
neutrino,
and its volume is $R_1^3\sim 10^{90}l_P^3$.
Since its density is constant ($\rho\sim \rho_P$) its mass at the end of the
inflation (or at the begining of the radiation era) is $M_1\sim 10^{90}M_P$.

Between the end of the inflation (or the begining of the radiation era) and the
time of radiation-matter equality, the radius of the universe passes from $R_1$
to
$R_{\rm eq}=a_{\rm eq} R_{\Lambda}$. Using Eqs. (\ref{kc30p}) and
(\ref{aeq}),
we find that  $R_{\rm eq}/R_1=a_{\rm eq}/a_1\sim 10^{28}$. Therefore, its
radius increases by $28$ orders of magnitude. Using Eq. (\ref{mu3}), we find
that its mass at
the time of radiation-matter equality is $M_{\rm eq}\sim 10^{-28} M_1\sim
10^{62}M_P$.

This mass remains constant during the matter era (see Eq. (\ref{mu4})) and may
be identified with the present mass of the universe: $M_\Lambda\sim
10^{62}M_P$. This
shows that the mass of the universe is equal to $N_P\sim
10^{62}$ particles of mass $M_P$. Although this result is quantitatively
correct, no such particles of mass $M_P$ exist in abundance in the universe.
Therefore, this  picture is not in agreement
with the observations. This suggests that Model II may be more relevant than
Model I.

\section{The electron}
\label{sec_ele}

The classical radius $r_e$ of the electron is defined  through the relation
\begin{eqnarray}
\label{mee1}
E=m_e c^2=\frac{e^2}{r_e}.
\end{eqnarray}
This equation expresses the equality (in order of magnitude) between the
rest-mass energy of the
electron and its electrostatic energy, assuming that the electron has a certain
size. This is a convenient manner to define the
``radius'' of the electron. This relation first appeared in the Abraham-Lorentz
\cite{abraham,lorentz} model of the extended
electron with an electromagnetic mass and later in the Born-Infeld
\cite{born1933,borninfeld} theory
of
nonlinear electrodynamics (see Appendix F of \cite{massmaxrel} for a short
review of these old theories).
Recalling the value of the charge of the electron
$e=4.80\times 10^{-13}\, {\rm
g^{1/2}\, m^{3/2}\, s^{-1}}$ and its mass $m_e=9.11\times 10^{-28}\, {\rm
g}=0.511\, {\rm MeV/c^2}$,
we obtain
\begin{eqnarray}
\label{mee2}
r_e=\frac{e^2}{m_e c^2}=2.82\times 10^{-15}\, {\rm m}.
\end{eqnarray}
The Compton wavelength of the electron is
$\lambda_e=\hbar/(m_e c)=3.86\times 10^{-13}\, {\rm m}$. It is related to the
classical radius of the electron by
\begin{eqnarray}
\label{mee3}
\lambda_e=\frac{r_e}{\alpha}\simeq 137\, r_e,
\end{eqnarray}
where
\begin{eqnarray}
\label{alpha}
\alpha=\frac{e^2}{\hbar c}\simeq
\frac{1}{137}\simeq
7.30\times 10^{-3}
\end{eqnarray}
is Sommerfeld's fine-structure constant.\footnote{Since quantum effects enter
at a distance of the order $\lambda_e$ which is much larger than $r_e$, a
purely classical electromagnetic model of the electron is not relevant.} The
typical electron density is
\begin{eqnarray}
\label{mee4}
\rho_e=\frac{m_e}{r_e^3}=4.07\times 10^{16}\, {\rm g\, m^{-3}}.
\end{eqnarray}
It can also be written as
\begin{eqnarray}
\label{bi10}
\rho_e=\frac{m_e^4c^6}{e^6}=\frac{m_e^4c^3}{\alpha^3\hbar^3}
\end{eqnarray}
or as 
\begin{equation}
\label{nf13}
\rho_e=\frac{\alpha\hbar}{c r_e^4}.
\end{equation}
The characteristic (dynamical) time 
associated with the electron is
\begin{eqnarray}
\label{mee5}
t_e=\left (\frac{m_e r_e^3}{e^2}\right
)^{1/2}=\frac{e^2}{m_e c^3}=\frac{r_e}{c}=9.40\times
10^{-24}\, {\rm
s}.
\end{eqnarray}
This is the time it takes for a light
wave to travel accross the
``size'' of an electron. This timescale first appeared in the Abraham-Lorentz
\cite{abraham,lorentz} theory of the extended electron when they tried to
calculate the recoil force on an accelerated charged particle caused by the
particle emitting electromagnetic radiation.  It can be written as
$t_e=\alpha\hbar/(m_e c^2)$. It is connected to the flight time
in the relativistic extension of Nelson's stochastic quantum mechanics
\cite{nelson} developed by  Lehr and Park \cite{lp}. This is what Caldirola
\cite{chronon} called the
``chronon'', which is a sort of ``quantum of time''. 
This is also the unit of time provided by the atomic constants that 
Dirac used in his cosmological theory based on a
large number hypothesis \cite{dirac1,dirac2,ouf}.

In \cite{ouf} we have obtained an accurate formula that relates the mass of the
electron to the cosmological constant: 
\begin{equation}
\label{la1}
m_e\simeq \alpha\left (\frac{\Lambda \hbar^4}{G^2}\right )^{1/6}.
\end{equation}
This equation can be viewed as an accurate form of  Eddington formula
\cite{eddington1931lambda,ouf}. The classical
radius of the electron is then given in good approximation by 
\begin{equation}
\label{la2}
r_e\simeq \left (\frac{G^2\hbar^2}{\Lambda c^6}\right )^{1/6}.
\end{equation}
With the empirical value $\Lambda=1.11\times 10^{-52}\, {\rm m^{-2}}$ of the
cosmological constant deduced from the observations of the accelerated
expansion of the universe we obtain the approximate value $8.80\times
10^{-28}\, {\rm g}$ for the mass of the electron  which is very close to the
measured value $m_e=9.11\times
10^{-28}\, {\rm g}$ (similarly we get $2.92\times 10^{-15}\,
{\rm m}$ for the classical radius of the electron which is very close to 
$r_e=2.82\times 10^{-15}\, {\rm m}$).  Using Eq. (\ref{la1}) we find that 
\begin{equation}
\label{la3}
\lambda_e\simeq \frac{1}{\alpha}\left (\frac{G^2\hbar^2}{\Lambda c^6}\right
)^{1/6},\qquad \rho_e\simeq \alpha\left (\frac{\Lambda^2c^{9}}{G^4\hbar}\right
)^{1/3},\qquad t_e\simeq \left (\frac{G^2\hbar^2}{\Lambda c^{12}}\right )^{1/6}.
\end{equation}
In \cite{ouf} we have developed the theory of large numbers pioneered by
Weyl, Eddington and Dirac. We have introduced the
``largest large number'' \cite{ouf}
\begin{eqnarray}
\label{la4}
\chi= \frac{\rho_P}{\rho_{\Lambda}^*}=\frac{c^3}{G\hbar\Lambda}=3.44\times
10^{121}\sim 10^{120},
\end{eqnarray}
which is the ratio between the Planck density and the cosmological density. In
terms of this number, we have
\begin{equation}
\label{la5}
\frac{m_e}{M_P}\sim \chi^{-1/6}\sim 10^{-20},\quad \frac{r_e}{l_P}\sim
\chi^{1/6}\sim 10^{20},\quad  \frac{\rho_e}{\rho_P}\sim \chi^{-2/3}\sim
10^{-80},\quad  \frac{t_e}{t_P}\sim\chi^{1/6}\sim 10^{20}. 
\end{equation}

{\it Remark:} If we assume that Eqs. (\ref{la1}) and  (\ref{la2})  are {\it
exact} (see footnote 21), then the ``theoretical'' values of the cosmological
constant and of the ``largest large number'' are
\begin{equation}
\Lambda_{\rm th}=\frac{G^2m_e^6}{\alpha^6\hbar^4}=1.36\times
10^{-52}\, {\rm m^{-2}},\qquad \chi_{\rm
th}=\left (\frac{c\hbar\alpha^2}{Gm_e^2}\right )^3=2.81\times 10^{121}.
\end{equation}

\section{The neutrino}
\label{sec_neutrino}

In \cite{ouf} we have suggested that the mass of the neutrino is related to the
cosmological constant by the relation
\begin{equation}
m_{\Lambda}^*=\left (\frac{\Lambda\hbar^3}{Gc}\right )^{1/4}=5.04\times
10^{-3}\, {\rm
eV/c^2}.
\label{la7}
\end{equation}
The Compton wavelength $\lambda_C=\hbar/mc$ of the neutrino is then given  by
\begin{equation}
\lambda_{\Lambda}^*= \left (\frac{G\hbar}{\Lambda
c^3}\right )^{1/4}=3.91\times 10^{-5}\, {\rm m}.
\label{la8}
\end{equation}
To make the numerical applications we have used the empirical
value $\Lambda=1.11\times 10^{-52}\, {\rm m^{-2}}$ of the cosmological constant.
In terms of the ``largest large number'' defined by Eq. (\ref{la4}) we
have
\begin{equation}
\label{la9}
\frac{m_{\Lambda}^*}{M_P}\sim \chi^{-1/4}\sim 10^{-30},\quad 
\frac{\lambda_{\Lambda}^*}{l_P}\sim \chi^{1/4}\sim
10^{30}.
\end{equation}

{\it Remark:} Eliminating the cosmological constant between Eqs. (\ref{la1})
and  (\ref{la7}) we obtain the following relation
\begin{equation}
\frac{m_e^6}{(m_{\Lambda}^*)^4}=\alpha^6M_P^2
\label{la7b}
\end{equation}
between the mass of the neutrino and the mass of the electron. This allows us to
determine the ``exact'' value of the neutrino mass, independently of the
uncertainty on the value of the cosmological constant. We find
$(m_{\Lambda}^*)_{\rm th}=5.30\times 10^{-3}\, {\rm eV/c^2}$.

\section{Transition between the inflation era and the radiation era}
\label{sec_intre}

As discussed in Sec. \ref{sec_kc} the transition between the inflation era and
the radiation era corresponds to a value of the scale factor $a_1$ given by Eq.
(\ref{esf6}).\footnote{Basically, this relation can be obtained
as follows. The density of radiation evolves with the scale factor as $\rho_{\rm
rad}=\rho_{\rm rad,0}/a^4$. Owing to the fact that $\rho_{\rm rad}\sim \rho_I$
at the end of the inflation, we get $a_1\sim
(\rho_{\rm rad,0}/\rho_I)^{1/4}$. If we make the additional rough approximation
$\rho_{\rm rad,0}\sim\rho_{\Lambda}$ on account of the cosmic coincidence
we obtain $a_1\sim (\rho_\Lambda/\rho_I)^{1/4}$ in order of magnitude.} We want
to determine the quantity
\begin{equation}
\label{f1}
\epsilon=\frac{a(t=0)}{a_1},
\end{equation}
where $a(t=0)$ is the value of the scale factor at the ``initial'' time
$t=0$.\footnote{This time corresponds to the moment of the big bang
singularity ($\rho=\infty$ and $a=0$) in the case of a purely radiative early
universe.} We recall that
the scale factor $a$ is dimensionless and normalized such that $a=1$ at the
present epoch. In order to have dimensional lengthscales, we introduce the
cosmological length $R_\Lambda=c
t_\Lambda=c/\sqrt{G\rho_\Lambda}=(8\pi/\Lambda)^{1/2}=4.77\times 10^{26}\, {\rm
m}$ as a reference \cite{ouf}. It is of the
order of the
present radius of the visible universe on account of the cosmic coincidence. We
then
define the ``radius of the universe'' by
$R(t)=a(t)R_{\Lambda}$.

(i) {\it Model I:} We first assume that the initial density of the universe 
(upper bound) is of the order of
the Planck density: $\rho_I=\rho_P=c^5/G^2\hbar=5.16\times 10^{99}\, {\rm
g\, m^{-3}}$. According to Eq. (\ref{esf6}) we have 
\begin{equation}
\label{f3}
a_1=\left (\frac{\Omega_{\rm rad,0}}{\Omega_{\Lambda,0}}\right
)^{1/4}\left
(\frac{\rho_\Lambda}{\rho_P}\right )^{1/4}=1.98\times
10^{-32}\sim \chi^{-1/4}\sim 10^{-30},
\end{equation}
where $\chi$ is the ``largest large number'' defined by Eq.
(\ref{la4}).
Therefore, the radius of the universe at the end
of the inflation (i.e. at the transition
between the
inflation era and the radiation era) is
\begin{equation}
\label{f8n}
R_1=a_1 R_{\Lambda}= \left (\frac{\Omega_{\rm
rad,0}}{\Omega_{\Lambda,0}}\right
)^{1/4}\left (\frac{8\pi G\hbar}{\Lambda c^3}\right
)^{1/4}=9.42\times
10^{-6}\, {\rm m}.
\end{equation}
Comparing Eq. (\ref{f8n}) with Eq. (\ref{la8}) we see that the radius of
the universe at the end of the
inflation is of the order of the Compton wavelength of the neutrino:
\begin{equation}
R_1\sim \lambda_\Lambda^*.
\label{wn1b}
\end{equation}
On the other hand, it is natural to identify $R(0)$
with the Planck length:
\begin{equation}
\label{f2y}
R(0)=l_{P}=c t_P=\left (\frac{G\hbar}{c^3}\right )^{1/2}=1.62\times
10^{-35}\, {\rm m}.
\end{equation}
Therefore
\begin{equation}
\label{f7b}
a(t=0)=\frac{R(0)}{R_{\Lambda}}=\frac{l_P}{R_{\Lambda}}=\left
(\frac{G\hbar\Lambda}{8\pi
c^3}\right )^{1/2}=3.39\times 10^{-62}\sim \chi^{-1/2},
\end{equation}
where we have used 
\begin{equation}
\label{f6n}
\frac{l_P}{R_\Lambda}=\frac{t_P}{t_\Lambda}=\left
(\frac{\rho_\Lambda}{\rho_P}\right )^{1/2}=\left
(\frac{1}{8\pi\chi}\right )^{1/2}\sim \chi^{-1/2}\sim 10^{-60}
\end{equation}
to get the last estimates. We
then get
\begin{equation}
\label{f2}
\epsilon=\frac{R(0)}{R_1}=\frac{l_P}{R_1}=\frac{l_P}{a_1 R_{\Lambda}}.
\end{equation} 
Using Eq. (\ref{f3}), we can write Eq. (\ref{f2}) as
\begin{equation}
\label{f5n}
\epsilon=\left (\frac{\Omega_{\Lambda,0}}{\Omega_{\rm rad,0}}\right
)^{1/4}\left
(\frac{\rho_P}{\rho_\Lambda}\right )^{1/4}\frac{l_P}{R_\Lambda}.
\end{equation}
Substituting  Eqs. (\ref{la4}) and (\ref{f6n}) into Eq. (\ref{f5n}) we
obtain 
\begin{equation}
\label{f7}
\epsilon=\left (\frac{\Omega_{\Lambda,0}}{\Omega_{\rm rad,0}}\right
)^{1/4}(8\pi\chi)^{-1/4}=1.71\times 10^{-30}\sim \chi^{-1/4}\sim 10^{-30}.
\end{equation}
The increase in size of the universe
during
the inflation is
\begin{equation}
\label{f8nb}
\frac{R_1}{l_P}=\frac{1}{\epsilon}= \left (\frac{\Omega_{\rm
rad,0}}{\Omega_{\Lambda,0}}\right
)^{1/4}(8\pi\chi)^{1/4}=5.83\times 10^{29}\sim \chi^{1/4}\sim 10^{30}.
\end{equation}
Therefore, during the inflation, the size of the universe increases by about
$30$ orders of magnitude. We note that the value of $\epsilon$
gives an $e$-folding  number $N_0=68.5$ 
which is fully consistent with the observations (see Sec. \ref{sec_kcdur}). This
value has been obtained by assuming $R(0)=l_P$. Conversely, assuming $N\simeq
60-70$ from the
observations implies that $R(0)\sim l_P$.

(ii)  {\it Model II:} We now assume that the initial density of the universe is
of the order of
the density of the electron: $\rho_I=\rho_e=4.07\times 10^{16}\, {\rm g\,
m^{-3}}$. According to
Eq.
(\ref{esf6}) we have
\begin{equation}
\label{f9}
a_1=\left (\frac{\Omega_{\rm rad,0}}{\Omega_{\Lambda,0}}\right
)^{1/4}\left
(\frac{\rho_\Lambda}{\rho_e}\right )^{1/4}=1.18\times
10^{-11}\sim \chi^{-1/12}\sim 10^{-10},
\end{equation}
where we have used Eqs. (\ref{la4}) and (\ref{la5}) to get the
last estimates. Therefore, the radius of the universe at the end of the
inflation (i.e. at the
transition between the
inflation era and the radiation era) is
\begin{equation}
\label{f17}
R_1=a_1R_{\Lambda}\simeq \frac{1}{\alpha^{1/4}} \left (\frac{\Omega_{\rm
rad,0}}{\Omega_{\Lambda,0}}\right
)^{1/4}\left (\frac{512\pi^3 G\hbar}{c^3\Lambda^5}\right )^{1/12}=5.82\times
10^{15}\, {\rm m},
\end{equation}
where we have used Eq. (\ref{la3}). In Ref. \cite{ouf} we 
have introduced the lengthscale 
\begin{equation}
{\tilde R}_2= \left (\frac{\hbar
G}{c^3\Lambda^5}\right )^{1/12}=
\chi^{5/12}\,
l_P= 7.07\times 10^{15}\, {\rm m}
\label{wn1hj}
\end{equation}
that we have interpreted as being the radius of a dark energy star of the
stellar mass. Therefore, we see that the radius of the universe at the end of
the inflation is of the order of  the radius of a dark energy star of the
stellar mass:
\begin{equation}
R_1\sim {\tilde R}_2.
\label{aaa}
\end{equation}
We note that this radius is
gigantic as compared
to the Compton wavelength of the neutrino obtained when we take $\rho_I=\rho_P$
[see Eqs.
(\ref{la8}) and (\ref{wn1b})]. They differ by $20$ orders of
magnitude ($\chi^{1/6}\sim 10^{20}$). On the other hand, we shall determine
$R(0)$ in order to get
the same value of $\epsilon$ as above (see Sec. \ref{sec_kcdur}). First, we
note that 
\begin{equation}
\label{f11}
\epsilon=\frac{R(0)}{R_1}=\frac{R(0)}{a_1 R_{\Lambda}}=\left
(\frac{\Omega_{\Lambda,0}}{\Omega_{\rm
rad,0}}\right
)^{1/4}\left
(\frac{\rho_e}{\rho_\Lambda}\right )^{1/4}\frac{R(0)}{R_{\Lambda}},
\end{equation}
where we have used Eq. (\ref{f9}). Equating  Eqs. (\ref{f11}) and (\ref{f5n}) we
find that
\begin{equation}
\label{nf12}
R(0)=\left (\frac{\rho_P}{\rho_e}\right )^{1/4}l_P.
\end{equation}
Using Eq. (\ref{nf13}), we get
\begin{equation}
\label{f14}
R(0)=\frac{r_e}{\alpha^{1/4}}=9.65\times 10^{-15}\, {\rm m}.
\end{equation}
In this model, the initial radius of the universe is of the order of the size
of the electron instead of being of the order of the Planck length [see Eq.
(\ref{f2y})]. This
ensures that  $\epsilon$ is given by Eq. (\ref{f7}) in
agreement with the observations (the size of the universe
increases by about $30$ orders of magnitude during the
inflation). Conversely,
assuming that the
initial radius of the universe is of the order of the size
of the electron, we find that $\epsilon$ is given by Eq. (\ref{f7}). Finally,
we find that 
\begin{equation}
\label{f18}
a(t=0)=\frac{R(0)}{R_{\Lambda}}=\frac{1}{\alpha^{1/4}
}\frac{r_e}{R_{\Lambda}}
=2.10\times
10^{-41}\sim \chi^{-1/3}\sim 10^{-40},
\end{equation}
where we have used Eqs. (\ref{la5}) and (\ref{f6n}) to get the
last estimates. The increase in size of the universe
during
the inflation is
\begin{equation}
\label{f15}
\frac{R_1}{R(0)}=\frac{\alpha^{1/4}
R_1}{r_e}=\frac{1}{\epsilon}=\left
(\frac{\Omega_{\rm
rad,0}}{\Omega_{\Lambda,0}}\right
)^{1/4}(8\pi\chi)^{1/4}=5.83\times 10^{29}\sim \chi^{1/4}\sim 10^{30}.
\end{equation}

\end{document}